Johannes Preiser-Kapeller

A Collapse of the Eastern Mediterranean?
*New results and theories on the interplay between climate and societies in Byzantium and the Near East, ca. 1000–1200 AD*[1]
With seven appendices, including three tables and 33 figures

Abstract: This paper discusses a recently proposed scenario of a climate-induced "Collapse of the Eastern Mediterranean" in the 11th century AD. It demonstrates that such a scenario cannot be maintained when confronted with proxy data from various regions. On the other hand, data on the interplay between environment and economy in the Komnenian period (1081–1185) and evidence for a change of climatic conditions in the period of the Angeloi (1185–1204) is presented, arguing that climatic parameters should be taken into consideration when comparing socio-economic dynamics in the Eastern Mediterranean with those in Western Europe. The necessity of further research on the regional as well as over-regional level for many aspects of the interaction between human society and environment in the medieval Eastern Mediterranean is highlighted.

The reflection about the impact of climate on human society goes back to antiquity. It has gained renewed intensity with the discussion about climate change and its possible anthropogenic causes in the last decades. This lead also to an increase of research on climate history and the contribution of climate to crises of past societies, popularised in books such as Jared Diamond´s "Collapse" (2005) for cases such as the Maya in Yucatan, the Vikings on Greenland or the Easter Island (Rapa Nui).[2] For the history of the Mediterranean in the Roman, post-Roman and Byzantine period, recent path-breaking papers by Michael McCormick et alii and by John F. Haldon et alii have synthesised new findings from historical, archaeological and especially natural scientific evidence.[3] As the

---

[1] Research for this paper was mostly undertaken during a research stay at the Institute of Historical Research of the National Hellenic Research Foundation in Athens financed by the Alexander S. Onassis Public Benefit Foundation in Spring 2014. My thanks go to the Alexander S. Onassis Public Benefit Foundation and its staff, especially to Frederique Hadgiantoniou and Dr. Niki Tsironi, and to all colleagues at the Institute of Historical Research of the National Hellenic Research Foundation, especially its Director Prof. Taxiarchis Kolias, as well as Dr. Stylianos Lampakis and Prof. Athina Kolia-Dermitzaki (National and Kapodistrian University Athens) for the opportunity to present a first draft of the present paper in the "Seminar Nikos Oikonomides" in March 2014. Furthermore, I thank the two anonymous reviewers as well as the editor, Prof. Ewald Kislinger (University of Vienna), for their valuable suggestions.

[2] J. DIAMOND, Collapse: How Societies choose to fail or succeed. New York 2005. For new data on the Maya cf. for instance D. J. KENNETT et alii, Development and Disintegration of Maya Political Systems in Response to Climate Change. *Science* 338 (2012) 788–791; The Great Maya Droughts in Cultural Context. Case Studies in Resilience and Vulnerability, ed. G. Iannone. Boulder, Colorado 2014. For criticism on the theoretical framework of Diamond cf. Questioning Collapse: Human Resilience, Ecological Vulnerability, and the Aftermath of Empire, ed. P. A. McAnany – N. Yoffee. Cambridge 2009; After Collapse: The Regeneration of Complex Societies, ed. G. M. Schwartz – J. J. Nichols. Tucson 2010.

[3] M. MCCORMICK et alii, Climate Change during and after the Roman Empire: Reconstructing the Past from Scientific and Historical Evidence. *Journal of Interdisciplinary History* 43,2 (2012) 169–220; J. F. HALDON et alii, The Climate and Environment of Byzantine Anatolia: Integrating Science, History, and Archaeology. *Journal of Interdisciplinary History* 45,2 (2014) 113–161. These and earlier studies have a strong focus on the transition from Late Antiquity to the middle Byzantine period, cf. for instance J. KODER, Climate Change in the Fifth and Sixth Centuries?, in: The Sixth Century – End or Beginning?, ed. P. Allen – E. Jeffreys (*Byzantina Australiensia* 10). Brisbane 1996, 270–285. Cf. also J. LUTERBACHER et alii, A Review of 2000 Years of Paleoclimatic Evidence in the Mediterranean, in: The Climate of the Mediterranean region: from

present contribution is not based on such a large scale cooperation of specialists in history, archaeology, palaeo-climatology and other sciences, its aims are more modest: to discuss aspects of the interaction between climate change and medieval Mediterranean history in the 11th–13th century against a combination of historical and natural scientific evidence. I first present theories, data and sources for the climatic and environmental history of the pre-modern period, which are especially of interest for historians not familiar with current methods and debates in this growing field. I try to demonstrate that linear or mono-causal ("deterministic") models of the impact of climatic conditions on political and economic developments are insufficient to capture the complex interplay between environmental parameters and social structures. I discuss the climatic prelude to the Komnenian period with regard to the recent hypothesis of a climate caused "Collapse of the Eastern Mediterranean" in the 11th century. As I argue, such a scenario of general "Collapse" cannot be maintained when confronted with data from various regions. I then reflect on the interplay between environment and economy in the Komnenian period and finally present evidence for another change of climatic conditions in the period of the Angeloi (1185–1204); here I argue that divergent developments of climatic parameters should be taken into consideration when comparing socio-economic dynamics in the Eastern Mediterranean (or other regions) with those in Western Europe in the 12th to 13th cent.[4] In many cases, findings are preliminary due to the lack of (conclusive) data and demand further research.

## Methodology

Recently, older concepts of climate determinism were reanimated, postulating strong linear lines of causation between climate and society and proposing "climate change as ultimate cause of human crisis in preindustrial societies".[5] Ecologists, on the other hand,

---

the past to the future, ed. P. Lionello. Amsterdam 2012, 87–185, for a most useful overview on recent findings on the climatic history of the Mediterranean combining various data.

[4] For a recent discussion of climatic aspects of the period under consideration in the present paper see J. KODER, Historical Geography of the Byzantine World in the Twelfth and Thirteenth Centuries: Problems and Sources, in: Change in the Byzantine World in the Twelfth and Thirteenth Centuries, ed. A. Ödekan – E. Akyürek – N. Necipoğlu (*First International Sevgi Gönül Byzantine Studies Symposium*). Istanbul 2010, 34–38, and J. PREISER-KAPELLER, A Climate for Crusades? Weather, climate and armed pilgrimage to the Holy Land (11th–14th Century). pre-print online: http://oeaw.academia.edu/JohannesPreiserKapeller/Papers [21.04.2015]. German version in: *Karfunkel – Zeitschrift für erlebbare Geschichte. Combat-Sonderheft* 10 (2014) 46–55. On medieval environmental studies in general see R. C. HOFFMANN, An Environmental History of Medieval Europe (*Cambridge Medieval Textbooks*). Cambridge 2014. A most useful overview of approaches to environmental history in Byzantine studies in the last decades is now provided in: I. TELELIS, Environmental History and Byzantine Studies. A Survey of Topics and Results, in: Aureus. Volume dedicated to Professor Evangelos K. Chrysos, ed. T. Kolias – K. Pitsakis. Athens 2014, 737–760. Cf. also D. STATHAKOPOULOS, Reconstructing the Climate of the Byzantine World: State of the Problem and Case Studies, in: People and Nature in Historical Perspective, ed. J. Lazlovsky – P. Szabó. Budapest 2003, 247–261. For aspects of animal husbandry, whose impairment or promotion by climatic factors would also deserve further studies, see Animals and Environment in Byzantium (7th–12th c.), ed. I. Anagnostakis – T. Kolias – Ef. Papadopoulou. Athens 2011, and H. KROLL, Tiere im Byzantinischen Reich. Archäozoologische Forschungen im Überblick (*Monographien des Römisch-Germanischen Zentralmuseum* 87). Mainz 2010, esp. 149–191. For the interplay between climate, agriculture and animal husbandry in Ottoman Anatolia in the late 16th and 17th century cf. esp. S. WHITE, The Climate of Rebellion in the Early Modern Ottoman Empire (*Studies in Environment and History*). Cambridge 2011. For the interplay between military logistics and environment in the period under consideration see also: Logistics of Warfare in the Age of the Crusades. Proceedings of a Workshop held at the Centre for Medieval Studies, University of Sydney, 30 September to 4 October 2002, ed. J. H. Pryor. Aldershot – Burlington 2006.

[5] Cf. E. WEIBERG – M. FINNÉ, Mind or Matter? People-Environment Interactions and the Demise of Early Helladic II Society in the Northeastern Peloponnese. *American Journal of Archaeology* 117/1 (2013) 1–32;

have highlighted that the actual reaction of any ecosystem – including human societies – on environmental change does not only depend on the strength and frequency of these disturbances, but also on the capability of a system to resist or to adapt to such changes. Collapse is only one of the possible results of such phenomena (see also the notion of "resilience" discussed in the conclusion below).[6] For any society, one has to take into account not only the complex interplay between environment and human communities, but also the social complexity of political, economic or cultural systems, which can process such "external" stimuli in different ways – maybe with fatal consequences, but also with adaptive measures.[7]

Also Byzantine observers were aware that the short term and long term impact of any climatic extreme event depends on the political and socio-economic framework of a society and its reactions. This can be illustrated for the often-discussed "winter of famine" in 927/928.[8] The owner of large estates were not only able to cope with such a distress much easier, but also profited from the need of smaller scale land holders, who were forced to sell their property in order to survive.[9] Emperor Romanos I Lakapenos in 934 tried to reverse these shifts of economic power in the countryside in one of the first of the "Macedonian" laws on landholding.[10] Another frequently discussed example for the possible damping or (in this case) aggravation of supply shortfalls by economic and political decisions is the establishment of the so-called *phundax* by Nikephoros (or Nikephoritzes), the *logothetes tu dromu* of Emperor Michael VII Dukas (1071–1078), in the important grain market of Rhaidestos in Thrace. This according to Michael Attaleiates contributed to a scarcity of food and an increase of prices in Constantinople, in turn raising resistance against the regime of the emperor.[11]

Both cases also confirm the findings of modern-day scholarship on famine and crises of subsistence: "droughts can cause a crop failure, but man, by withholding life-supporting food from his fellow man causes famine. (…) famine is a cultural hazard – not a physical hazard".[12] Against this background, generalisations on the impact of climatic conditions on historical trajectories become problematic; as Sarah Kate Raphael makes clear: "each environmental disaster presents an almost independent case study. There is no clear pattern of behaviour or policy that rulers followed".[13] This of course makes any analysis of the interplay between environmental and political respectively socio-economic factors challenging, as Thomas F. Homer-Dixon illustrates: "causal processes are exceedingly complex, involving multiple physical and social variables, feedback loops, interactive effects, and nonlinear responses. Analysts often must trace out long and tangled chains of causation, and data on key variables and processes are rarely abundant or high quality."[14]

---

[9] On the significance of the distribution of landed property for the impact of famine on various groups in a society cf. A. SEN, On Economic Inequality. New edition, Oxford 1997; PARKER, Global Crisis 73–76 , also 17–20 for the impact of climatic stress on crops and nutrition (see also fig. 31 for the correlation of drought and crop failure in 20th cent. Syria).

[10] N. SVORONOS, Les novelles des empereurs Macédoniens concernant la terre et les stratiotes. Introduction – edition – commentaires. Édition posthume et index établis par P. GOUNARIDIS. Athens 1994, 86, lns. 88–95; E. MCGEER, The Land Legislation of the Macedonian Emperors (*Medieval Sources in Translation* 38). Toronto 2000, 49–60, esp. 56 (for the translation): "Therefore from the previous first indiction onwards (that is, since the outbreak of famine) all particularly honoured people who have come to control over hamlets and villages and there have acquired more possessions, are expelled from there for future time, whereby they will receive back the purchase price, either by the original owners or their heirs and relatives (…) or by the village community." On the impact of the winter of 927/928 on Byzantine agriculture cf. also M. KAPLAN, Les hommes et la terre à Byzance du VIe au XIe siècle: Propriété et exploitation du sol (*Byzantina Sorbonensia* 10). Paris 1995, 395–396, 416, 421–426; J. A. HARVEY, Economic Expansion in the Byzantine Empire 900–1200. Cambridge 1989, 40–41.

[11] Michael Attaleiates 156, 5–157, 26 (Eu. Th. TSOLAKIS, Michaelis Attaliatae Historia [*CFHB* 50]. Athens 2011); Michael Attaleiates, The History, transl. by A. KALDELLIS – D. KRALLIS (*Dumbarton Oaks Medieval Library*). Cambridge, Mass. 2012, 367–373: "He [Nikephoros] thereby established a monopoly over this most essential of trade, that of grain, as no one was able to purchase it except from the phoundax. (…) For from that moment on they monopolized not only the grain carts (…) but also all other goods the circulated in the vicinity. (…) He, then, farmed out the phoundax for sixty pounds of gold, and enjoyed the proceeds, while everyone else was hard-pressed by a shortage not only of grain but of every other good. For the dearth of grain causes dearth in everything else, as it is grain that allows the purchase or preparation of other goods, while those who work for wages demand higher pay to compensate for the scarcity of food. (…) As a result of the emperor´s planning or, rather, of Nikephoros´s evil designs, grain was in short supply and abundance turned into dearth. The people´s discontent increased." For interpretations of this passage: M. ANGOLD, The Byzantine Empire 1025–1204. London – New York ²1997, 122–123; The Economic History of Byzantium, ed. A. E. Laiou. 3 Vols., Washington, D.C. 2002, 741–742; A. LAIOU – C. MORRISSON, The Byzantine Economy. Cambridge 2007, 135–136 (with reference to other studies on this event in fn. 144), and esp. A. E. LAIOU, God and Mammon: Credit, Trade, Profit and the Canons, in: Byzantium in the 12th Century. Canon Law, State and Society, ed. N. Oikonomides. Athens 1991, 261–300.

[12] W. A. DANDO, The Geography of Famine. London 1980, 11–12.

[13] S. K. RAPHAEL, Climate and Political Climate. Environmental Disasters in the Medieval Levant (*Brill´s Series in the History of the Environment* 3). Leiden 2013, 55–94 (also with an excellent analysis of the impact of and reactions to droughts and famines in the Crusader states and neighbouring Muslim polities) and 189 (for the citation). Cf. also HOMER-DIXON, Environment, Scarcity, and Violence; C. Ó GRÁDA, Famine. A short History. Princeton – Oxford 2009.

Therefore, any palaeoclimatic research has to take into account both the historical and the natural scientific evidence – as has been established by the Swiss pioneer of climate history Christian Pfister. He speaks about the "Archives of Society" – (mostly) written sources – and the "Archives of Nature", the evidence for past climatic conditions accumulated in tree rings, lake sediments or dripstones (speleothems).[15] These "proxies" allow for palaeoclimatic reconstructions of different duration and chronological resp. spatial resolution from millions of years to years (or even below) resp. from the global down to the local level. Extremely important archives of nature are sediments in lakes, which can be dissolved in annual layers. Deposited therein are for example the pollens of plants from areas even further afield. Palynologists are able to identify the different species and their relative share of the vegetation of the surrounding area in order to reconstruct climatic changes and human interventions (via "anthropogenic indicators") into the landscape (see also Appendix 4).[16] For Central Europe in the High Middle Ages, for instance, one can observe a decrease of tree pollen and an increase of pollen of domesticated crops that can be linked to forest clearances.[17] In addition to pollen the composition of sediments can offer also other important information on past climatic conditions (via oxygen isotope analyses, for instance; see also Appendix 6).[18] Archives of the society are mainly text sources, which may include direct meteorological observations of anomalies such as extreme winters or flood events, but also indirect data about the beginning of plant flowering, for instance, which allow for conclusions on weather conditions (see also Appendix 7).[19] With the combination of archives of nature and of society, historical climate research is more and more able both globally and regionally to reconstruct climate history and its potential impact on human societies over centuries and millennia; methods and results are constantly refined. For Byzantine studies, most important in this regard is the pioneering work of Ioannis Telelis, who in two massive volumes in 2004 not only provided the first systematic survey of meteorological information in Byzantine and other sources for the medieval Eastern Mediterranean, but in several articles has outlined the methodological basis for a combination of the archives of society and of nature.[20]

---

[14] HOMER-DIXON, Environment, Scarcity, and Violence 9.

[15] Ch. PFISTER, Klimageschichte der Schweiz 1525–1860. Das Klima der Schweiz von 1525–1860 und seine Bedeutung in der Geschichte von Bevölkerung und Landwirtschaft. 2 Vol.s, Bern – Stuttgart ²1985; R. BRÁZDIL – Ch. PFISTER – H. WANNER – H. VON STORCH – J. LUTERBACHER, Historical Climatology in Europe – the State of the Art. *Climatic Change* 70 (2005) 363–430; F. MAUELSHAGEN, Klimageschichte der Neuzeit (*Geschichte Kompakt*). Darmstadt 2010; R. S. BRADLEY, Paleoclimatology. Reconstructing Climates of the Quaternary. Amsterdam – Waltham – San –Diego ³2014, esp. 1–11 (general introduction) and 291–318 (on speleothems); Ch.-D. SCHÖNWIESE, Klimatologie. Stuttgart ³2008, esp. 280–333; PARKER, Global Crisis xvi–xvii: LUTERBACHER et alii, A Review of 2000 Years of Paleoclimatic Evidence (with sections on the various natural scientific data); O. M. GÖKTÜRK, Climate in the Eastern Mediterranean through the Holocene Inferred from Turkish Stalagmites. Ph.D.-Thesis, University of Bern 2011 (for several case studies for the use of speleothem data from sites in modern-day Turkey); S. W. MANNING, The Roman World and Climate: Context, Relevance of Climate Change, and some Issues, in: The Ancient Mediterranean Environment between Science and History, ed. W. V. Harris (*Columbia Studies in Classical Tradition* 39). Leiden – Boston 2013, 146–153 (on speleothems). Cf. also HALDON et alii, The Climate and Environment of Byzantine Anatolia 115–120 and Table 1.

[16] A. IZDEBSKI, A Rural Economy in Transition. Asia Minor from Late Antiquity into the Early Middle Ages (*Journal of Juristic Papyrology*, Supplement vol. 18). Warsaw 2013, 109–132, on methods, potentials and problems, especially for the interpretation and dating of pollen sequences from sites in Byzantine Anatolia and the Near East. Cf. also W. J. EASTWOOD, Palaeoecology and eastern Mediterranean Landscapes: Theoretical and practical approaches, in: General Issues in the Study of Medieval Logistics: Sources, Problems and Methodologies, ed. J. Haldon. Leiden 2006, 119–158; BRADLEY, Paleoclimatology 319–343 and 405–451; A. McMILLAN, A GIS approach to palaeovegetation modelling in the Mediterranean: the case study of southwest Turkey. PhD-Thesis, University of Birmingham 2012, also for the vulnerability of various plant species to changing climatic conditions for the south-western Anatolian case.

[17] Cf. A. IZDEBSKI – G. KOLOCH – T. SŁOCZYŃSKI – M. TYCNER-WOLICKA, On the Use of Palynological Data in Economic History: New Methods and an Application to Agricultural Output in Central Europe, 0–2000 AD. *Munich Personal RePEc Archive* Paper No. 54582, posted March 2014 (online: http://mpra.ub.uni-muenchen.de/54582/ [21.04.2015]).

## A "Collapse of the Eastern Mediterranean" in the 11th century?

Already Telelis was able to identify the 11th century as "period of extremes", marked by a series of cold as well as dry decades across all climatic zones of the Near East (see also fig. 26).[21] Similar observations have been made by Richard Bulliet in 2009 for Central Asia and Iran and most recently by Ronnie Ellenblum for the entire Near East in his book "The Collapse of the Eastern Mediterranean. Climate Change and the Decline of the East, 950-1072", but with much more far reaching hypotheses on the impact of these developments.[22] Ellenblum developed (in short) the following scenario: while several

---

[18] Izdebski, A Rural Economy in Transition 133–134. Cf. N. Roberts – G. Zanchetta – M. D. Jones, Oxygen isotopes as tracers of Mediterranean climate variability: an introduction. *Global and Planetary Change* 71 (2010) 135–140. For a practical example cf. J. R. Dean et alii, Palaeo-seasonality of the last two millennia reconstructed from the oxygen isotope composition of carbonates and diatom silica from Nar Gölü, central Turkey. *Quaternary Science Reviews* 66 (2013) 35–44. For the pitfalls connected with a neglect of the uncertainties regarding the temporal resolution and spans of dating of such sediments and data for their historical interpretation cf. Haldon et alii, The Climate and Environment of Byzantine Anatolia 120–121.

[19] Bradley, Paleoclimatology 517–551 (on the use of historical documents for climate reconstructions). Cf. also H. Grotefeld, Klimageschichte des Vorderen Orients 800–1800 A. D. nach arabischen Quellen, in: Historical Climatology in Different Climatic Zones, ed. R. Glaser – R. Walsh (*Würzburger Geographische Arbeiten* 80). Würzburg 1991, 21–43; F. Dominguez-Castro et alii, How Useful Could Arab Documentary Sources be for Reconstructing Past Climate? *Weather* 47 (2012) 76–82 (esp. for the 9th–10th cent.); St. Vogt – R. Glaser – J. Luterbacher et alii, Assessing the Medieval Climate Anomaly in the Middle East: The potential of Arabic documentary sources. *PAGES news* 19/1 (2011) 28–29. For the analysis of a single author cf. M. G. Morony, Michael the Syrian as a Source for Economic History. *Hugoye: Journal of Syriac Studies* 3.2 (2000) 141–172; M. Widell, Historical Evidence for Climate Instability and Environmental Catastrophes in Northern Syria and the Jazira: The Chronicle of Michael the Syrian. *Environment and History* 13/1 (2007) 47–70.

[20] Telelis, Μετεωρολογικά φαινόμενα; I. G. Telelis, Climatic Fluctuations in the Eastern Mediterranean and the Middle East AD 300–1500 from Byzantine Documentary and Proxy Physical Paleoclimatic Evidence – a Comparison. *JÖB* 58 (2008) 167–207; Idem, Medieval Warm Periods and the Beginning of the Little Ice Age in the Eastern Mediterranean: An Approach of Physical and Anthropogenic Evidence, in: Byzanz als Raum. Zu Methoden und Inhalten der historischen Geographie des östlichen Mittelmeerraumes, ed. Kl. Belke – F. Hild – J. Koder – P. Soustal (*VTIB* 7 = *Österreichische Akademie der Wissenschaften, phil.-hist. Kl., Denkschriften* 283). Vienna 2000, 223–243.

[21] Telelis, Μετεωρολογικά φαινόμενα 858–859; Telelis, Climatic Fluctuations in the Eastern Mediterranean and the Middle East 167–207; Idem, Medieval Warm Periods 223–243. Cf. also now Haldon et alii, The Climate and Environment of Byzantine Anatolia 124 (fig. 1) and Appendix 1 (table of climate events on the basis of Telelis´ earlier study).

[22] R. W. Bulliet, Cotton, Climate, and Camels. A Moment in World History. New York 2009; R. Ellenblum, The Collapse of the Eastern Mediterranean. Climate Change and the Decline of the East, 950–1072. Cambridge 2012; cf. also Haldon et alii, The Climate and Environment of Byzantine Anatolia 115. For

polities in the Middle East (including Byzantium) in the 10[th] to 11[th] century were weakened because of recurring extreme weather events, famine and concomitant social unrest, an abnormal cold and dry spell prevailed also in the Central Asian steppes. This damaged the herds of the nomads, which led to conflicts between different tribes and to an increasing mobility of different groups directed to the southern regions.[23] Thereby, ultimately, the political map of the Near East was enormously modified, since the Seljuks were able to gain power both in Persia and in Mesopotamia and Syria. In 1071 they defeated the Byzantines at the Battle of Manzikert, which was followed by Turkish migration to Anatolia. Thus beset, the Byzantine Emperor turned with a request for military aid to the Pope. These appeals jointly with other news about the plight of Christians due to the Seljuq invasion were used as pretext for the call to arms of Pope Urban II. The First Crusade then entered a Middle Eastern world greatly weakened and destabilized by the previous political and climatic vicissitudes of the 11[th] century.[24]

Climate, therefore, is introduced by Ellenblum not only as one, but as the prime mover of developments in the 11[th] century Near East on a "global" scale. But can this generalisation be maintained? Ellenblum has based his argument mainly on written and archaeological evidence for Egypt and Palestine; his chapter on Byzantium, for instance, is relatively weakly documented.[25] Even more, as also one reviewer has commented, he made almost no use of the growing evidence from the "archives of nature".[26] Therefore, I tried to evaluate Ellenblum´s scenario on the basis of proxy data for temperature and precipitation from more than 20 sites as well as pollen analyses from more than 40 sites for the entire Near East – with a focus on the regions of the Byzantine Empire (see Appendix 1 for the original data).

The picture emerging from this research only partly confirms Ellenblum´s scenario: data from various sites from Central Asia to Western Asia Minor indicates temperatures below average over long periods in the 11[th] century, but the proxies for precipitation show strong regional variances (see fig. 4). Rainfall was very much below average in sites from Iran, Armenia and Palestine – as were the heights of the Nile flood in Egypt. But generally, despite some dry intervals, conditions were not arid in Northern Syria or

---

similar earlier considerations regarding the causes of nomadic mobility in the 11[th] cent. see J. KODER, „Zeitenwenden". Zur Periodisierung aus byzantinischer Sicht. *BZ* 84/85 (1991/1992) 409–422.

[23] On the vulnerability of nomadic groups to adverse weather conditions as well as their dependence on the surplus production of their tilling neighbours cf. RAPHAEL, Climate and Political Climate 44–51. On a similar scenario of climatic factors for nomadic mobility in the cases of the Huns and Avars in the 5[th]–6[th] cent. AD cf. E. R. COOK, Megadroughts, ENSO, and the Invasion of Late-Roman Europe by the Huns and Avars, in: The Ancient Mediterranean Environment between Science and History, ed. W. V. Harris (*Columbia Studies in Classical Tradition* 39). Leiden – Boston 2013, 89–102. Cook´s data is also used in: MCCORMICK et alii, Climate Change during and after the Roman Empire 192 (fig. 8) and 220. On the "nomadisation" or "Bedouinization" of parts of Syria and Northern Mesopotamia, accompanying the decline of central state power, in the 10[th] to 11[th] century before the the advance of the Seljuks see: S. HEIDEMANN, Die Renaissance der Städte in Nord-Syrien und Nordmesopotamien. Städtische Entwicklung und wirtschaftliche Bedingungen in ar-Raqqa und Ḥarrān von der Zeit der beduinischen Vorherrschaft bis zu den Seldschuken (*Islamic History and Civilization. Studies and Texts* 40). Leiden 2002; K. FRANZ, Vom Beutezug zur Territorialherrschaft. Das lange Jahrhundert des Aufstiegs von Nomaden zur Vormacht in Syrien und Mesopotamien 286–420/889–1029. Beduinische Gruppen in mittelislamischer Zeit I (*Nomaden und Sesshafte* 5). Wiesbaden 2007.

[24] For a similarly "global", but somehow better documented scenario for the 17[th] cent. cf. WHITE, The Climate of Rebellion, as well as PARKER, Global Crisis.

[25] ELLENBLUM, The Collapse of the Eastern Mediterranean 123–146.

[26] The review on Ellenblum´s book by S. WHITE in: *Mediterranean Historical Review* 28 (2013) 70–72. Cf. also the review by S. HARRIS in: *Journal of Historical Geography* 42 (2013) 218–219.

in Western and Central Anatolia.[27] Therefore, it is necessary to evaluate the actual evidence for each region within Ellenblum´s scenario separately. This is equally essential due to the spatial resolution of proxy data, which may be valid only for the immediate surroundings of a site. Only in the current volume of *JÖB*, Adam Izdebski, Grzegorz Koloch and Tymon Słoczyński have produced a synthetic analysis of the trajectories of vegetation and anthropogenic indicators on the basis of pollen data for larger areas in Anatolia and the Balkans (more precisely, for seven regions: Central Greece, the highland hinterlands of Macedonia, the mountains of Western Bulgaria, Eastern Bulgaria, Eastern Bithynia, Inland Pontus and South-Western Anatolia with a focus on Pisidia). Also this synthesis in turn illustrates significant regional variability.[28]

One cornerstone of Ellenblum´s scenario, the preponderance of cold and dry conditions in the steppes of Central Asia, can be attested by proxy data, as a time series for tree ring-growth from one site in Kyrgystan for instance illustrates (see fig. 15). In this data set, the longest period of adverse climatic conditions across the Middle Ages is recorded for the decades from 950 to 1030 AD. This is also confirmed by reconstructions of the extent of the Aral Sea, which indicate a more arid and cooler period in that region from 900 AD onwards. For areas further east, the tree-ring data from Dulan-Wulan in north-central China used by Edward R. Cook does not indicate as severe multi-decadal droughts in that period as he has identified for the 4th–6th cent. AD, but still some arid trends in the first half of the 10th century. The overall validity of a scenario of cold and arid conditions in the western and central steppe region in the 10th–11th cent. has also been confirmed in a synthesis of various proxy sites in the study by Bao Yang et alii in 2009.[29]

---

[27] For regional variation of conditions in Anatolia for instance cf. M. TÜRKEŞ – T. KOÇ – F. SARŞ, Spatiotemporal Variability of Precipitation Total Series over Turkey. *International Journal of Climatology* 29 (2009) 1056–1074; Y. S. UNAL – A. DENIZ – H. TOROS – S. INCECIK, Temporal and spatial patterns of precipitation variability for annual, wet, and dry seasons in Turkey. *International Journal of Climatology* 32 (2012) 392–405; HALDON et alii, The Climate and Environment of Byzantine Anatolia 142–143. Illustrative in this regard is also the highly variable extent of areas affected by drought during the relatively arid years 1930 to 1947 in Turkey as documented in: W.-D. HÜTTEROTH, Türkei (*Wissenschaftliche Länderkunden* 21). Darmstadt 1982, 124–127.

[28] A. IZDEBSKI – G. KOLOCH – T. SŁOCZYŃSKI, Exploring Byzantine and Ottoman economic history with the use of palynological data: a quantitative approach. *JÖB* 65 (2015) 67-110. For the mathematical basis for this study cf. IZDEBSKI – KOLOCH – SŁOCZYŃSKI – TYCNER-WOLICKA, On the Use of Palynological Data in Economic History. On the wider framework of circulation systems impacting the region cf. J. KODER, Der Lebensraum der Byzantiner. Historisch-geographischer Abriß ihres mittelalterlichen Staates im östlichen Mittelmeerraum (*Byzantinische Geschichtsschreiber* Ergänzungsband 1N). Vienna ²2001, 40–44; HALDON et alii, The Climate and Environment of Byzantine Anatolia 128–131; M. Ç. KARABÖRK – E. KAHYA – M. KARACA, The influences of the Southern and North Atlantic Oscillations on climatic surface variables in Turkey. *Hydrological Processes* 19 (2005) 1185–1211; Hydrological, Socioeconomic and Ecological Impacts of the North Atlantic Oscillation in the Mediterranean Region, ed. S. M. Vicente-Serrano – R. M. Trigo. Heidelberg – London – New York 2011.

[29] Kyrgystan data: PAGES 2k Network consortium, Database S1 – 11 April 2013 version: http://www.pages-igbp.org/workinggroups/2k-network (21.04.2015). Aral Sea: I. BOOMER et alii, Advances in understanding the late Holocene history of the Aral Sea region. *Quaterny International* 194 (2009) 79–90; J.-F. CRETAUX – R. LETOLLE – M. BERGÉ-NGUYEN, History of Aral Sea level variability and current scientific debates. *Global and Planetary Change* 110 (2013) 99–113; S.K. KRIVONOGOV et alii, The fluctuating Aral Sea: A multidisciplinary-based history of the last two thousand years. *Gondwana Research* 26 (2014) 284–300 (also integrating archaeological data); Dulan-Wulan: COOK, Megadroughts, ENSO, and the Invasion, esp. 89–92. In general on climatic variations in arid Central Asia in that period cf. B. YANG et alii, Late Holocene climatic and environmental changes in arid central Asia. *Quaternary International* 194 (2009) 68–78. See also IZDEBSKI, A Rural Economy in Transition 139-140. For economic and environmental developments in China, especially in the regions neighbouring the Central Asian steppes during that period, and in Tibet cf. Q. GE – W. WU, Climate during the Medieval Climate Anomaly in China. PAGES news 19/1 (2011) 24–26; J. A. HOLMES – E. R. COOK – B. YANG, Climate change over the past 2000

Byzantine sources for the invasions of the Pechenegs and the Uzes across the lower Danube in the 1040s and 1060s equally document that several nomadic groups of the Steppe very much suffered from famine and disease.[30] In addition, larger scale relocations of population in areas east of the Carpathians and north of the Danube during the 11[th] century can also be deduced from archaeological evidence, as Florin Curta has highlighted.[31] Thus, nomadic mobility from the north during this period at least to a certain degree could be connected with adverse climatic conditions. But, as A. C. S. Peacock has made clear in his recent monograph on the Seljuks (also taking into consideration the hypotheses of Bulliet and Ellenblum), such adverse conditions alone would not "necessarily have triggered nomadic migrations, for most nomads will prefer to adapt to changing conditions if possible."[32] As both the examples of the Seljuks and of the Pechenegs and the Uzes make clear, only specific combinations of political factors in the steppe as well as in neighbouring regions allowed for or even invited the large-scale southwards relocation of these groups. These groups then could cause additional stress on agricultural regions already suffering from crop failure and famine. This is also documented for provinces in Iran, Iraq and Syria at that period, partly already before the advent of the Seljuks. But as archaeological evidence from the 11[th] to 12[th] cent. for the lower Danube indicates, at the same time the presence of these nomadic groups north of the river provided mercantile opportunities for the Byzantine cities in that region. Thus, a more nuanced picture of the actual interplay between nomadic mobility and neighbouring polities is necessary (see also below for the possible impact of nomadic groups on agricultural activity).[33]

---

[30] years in Western China. *Quaternary International* 194 (2009) 91–107; M. C. CHEUNG et alii, A stable mid-late Holocene monsoon climate of the central Tibetan Plateau indicated by a pollen record. *Quaternary International* 333 (2014) 40–48; J. CHEN et alii, Hydroclimatic changes in China and surroundings during the Medieval Climate Anomaly and Little Ice Age: spatial patterns and possible mechanisms. *Quaternary Science Reviews* 107 (2015) 98–111 (for a synthesis of data across modern-day China).

[30] On these events cf. P. STEPHENSON, Byzantium's Balkan Frontier. A Political Study of the Northern Balkans, 900–1204. Cambridge 2000, 80–110; ANGOLD, The Byzantine Empire 37–40; ELLENBLUM, The Collapse of the Eastern Mediterranean 125–130. Ioannes Skylitzes 21 (Konstantinos o Monomachos), 17 (458, 56–60 THURN); transl. WORTLEY 429: "For, once [the Pechenegs] had crossed the river [Danube], they found a plentiful supply of beasts, of wine and of drinks prepared from honey which they had never even heard. These they consumed without restraint and were afflicted by a flux of the bowels; many of them perished each day." Michael Attaleiates 25, 10–11 (TSOLAKIS); transl. KALDELLIS – KRALLIS 53: "The enemies [the Pechenegs] were not yet used to these lands that were foreign to them and were afflicted by a pestilent disease." Michael Attaleiates 68, 7–13 (TSOLAKIS); transl. KALDELLIS – KRALLIS 157: "Among those [Uzes] who were left behind, however, a vast horde still, some were devastated by an epidemic disease and hunger and were only half alive, while others had been defeated by the Bulgarians and the Pechenegs (…). And so they were killed contrary to all human expectation." Michael Attaleiates 69, 11–17 (TSOLAKIS); transl. KALDELLIS – KRALLIS 159 (also on the identification of the „Myrmidons"):"As for this Skythian nation, some crossed the Danube and were destroyed by a famine against which there was no recourse, for they had no food and no hope of foraging for it, as their land had neither been plowed nor sown. Only a few of them survived and they, it is said, went over to the ruler of the Myrmidons [Rus?] and were distributed by him among several cities, thereby leaving their own land altogether empty of people."

[31] F. CURTA, Southeastern Europe in the Middle Ages 500–1250 (*Cambridge Medieval Textbooks*). Cambridge 2006, 296–309. On the impact of climatic changes on human mobility cf. R. A. McLEMAN, Climate and Human Migration: Past Experiences, Future Challenges. Cambridge 2013.

[32] A. C. S. PEACOCK, The Great Seljuk Empire (*The Edinburgh History of the Islamic Empires*). Edinburgh 2015, 287–291. Cf. also FRANZ, Vom Beutezug zur Territorialherrschaft, esp. 41–45.

[33] PEACOCK, The Great Seljuk Empire 287–291 (for the discussion of climatic factors) and 20–52 (for a synthesis of the Seljuk advance up to the capture of Baghdad in 1055). For the emergence of the Seljuks cf. also D. KOROBEINIKOV, Raiders and Neighbours: the Turks (1040–1304), in: The Cambridge History of the Byzantine Empire, c. 500–1492, ed. J. Shepard. Cambridge 2008, 696–699. On political changes in the steppes north of Byzantium and the evidence for mercantile activity at the Danube border cf. STEPHENSON, Byzantium's Balkan Frontier 81–98, who speaks about a "boom in trade on the lower Danube" (p. 87). For

The impact of changing climatic conditions on the agricultural societies to the south very much varied across regions. As already mentioned, pollen analyses allow us to reconstruct the vegetation around a site at a specific time and also the increase or decrease of the human share in the usage of landscape (documented for instance in the percentage of grain pollen in a sample). I evaluated such data for more than 40 sites from Iran to the Balkans and tried to categorise them in indicating a decrease (red), a stagnation (white) or an increase (yellow) of agricultural activity during the 11[th] century. Pronounced differences between the East and North and the South and West of the region become visible (fig. 4).

A significant decrease of cereal pollen from the 10[th] to the 11[th] century can be observed for the sites of Lake Maharlou in Southwest Iran and of Lake Almalou in Northwest Iran (in the latter case also of fruit tree pollen) (fig. 16).[34] The difficulties in the dating of the two pollen layers is decisive for the question if the decrease can be attributed mainly to military events connected with the advance of Turkish groups from the East and/or also with otherwise adverse conditions. While the decisive layer at Almalou most probably dates after the invasion of the Seljuks in the region, the Maharlou data could indicate a decline of agriculture already before these events – which would also correlate with the pronounced drier conditions reconstructed from other sites in Iran. Yet anthropogenic indicators in general are relatively weak for the Maharlou region. Thus, this aspect of Ellenblum´s respectively Bulliet´s scenario, that is the weakening of agricultural communities in (parts of) Iran (simultaneously with the climate caused mobilisation of nomadic groups in Central Asia), seems possible (also on the basis of written evidence on extreme weather, bad harvests and famines for some regions). But the small amount of sufficient proxy data from only two sites does not allow for a generalisation across larger areas regarding the short term or even long term impact of climatic conditions and/or the arrival of Seljuk groups. Further regional studies are necessary.[35]

---

mercantile activity in that area cf. M. GEROLYMATOU, Ἀγορές, ἔμποροι καὶ εμπόριο στο Βυζάντιο (9ος–12ος αι.). Athens 2008, 171–177, as well as the systematic survey of Byzantine ports along the Black Sea and in the mouth of the Danube currently undertaken by Grigori Simeonov (University of Vienna) within the project "Harbours and landing places on the Balkan coasts of the Byzantine Empire (4th to 12th centuries)" as part of the Special Research Programme (SPP-1630) funded by the Deutsche Forschungsgemeinschaft (DFG) on "Harbours from the Roman Period to the Middle Ages" (http://www.spp-haefen.de/en/projects/byzantine-harbours-on-the-balkan-coasts/ [21.04.2015]). On the possible impact of nomadic groups on agricultural areas already under stress cf. PEACOCK, The Great Seljuk Empire 34–35 (on the impact of the arrival of Seljuk groups in Chorasan, suffering from a series of bad harvests and famine in the 1030s); RAPHAEL, Climate and Political Climate 44–53, esp. 49 (for an extremely cold winter in Iraq in 1030). See also WHITE, The Climate of Rebellion 232–243, for conflicts between nomads and agriculturalists in 16th/17th cent. Anatolia

34 Cf. M. DJAMALI et alii, A late Holocene pollen record from Lake Almalou in NW Iran: evidence for changing land-use in relation to some historical events during the last 3700 years. *Journal of Archaeological Science* 36 (2009) 1363–1375; M. DJAMALI et alii, Notes on Arboricultural and Agricultural Practices in Ancient Iran based on New Pollen Evidence. *Paléorient* 36/2 (2010) 175–188.

35 BULLIET, Cotton, Climate, and Camels 69–95 (with the written and very few palaeo-climatological evidence). Cf. also IZDEBSKI, A Rural Economy in Transition 137, and P. CHRISTENSEN, The Decline of Iranshahr: Irrigation and Environments in the History of the Middle East, 500 B.C. to A.D.1500. Copenhagen 1993, 163– 177 (for the Iranian Southwest) and 205–210 (for Azerbaijan). Cf. again PEACOCK, The Great Seljuk Empire 34–35 (on Chorasan suffering from a series of bad harvests and famine in the 1030s at the time of the appearance of Seljuk groups), but also 291–293 (indicators that there was no long-term negative impact of the arrival of the Seljuks on the province of Fars in southwest Iran) as well as the generally positive picture of agricultural development in 11th–12th century Iran in: A. LAMBTON, Aspects of Saljūq-Ghuzz Settlemen in Persian, in: Islamic Civilization 950–1150. A Colloquium, ed. D. S. Richards (*Papers in Islamic History* 3). Oxford 1973, 105–125; P. FELDBAUER, Die islamische Welt 600–1250. Ein Frühfall von Unterentwicklung? Vienna 1995, 71–73, 330–337. See also now D. T. POTTS, Nomadism in Iran: From Antiquity to the Modern Era. Oxford 2014, 179–187.

If we now focus on the areas under Byzantine control in the mid-11th century (fig. 4), we detect a combination of decreasing precipitation and agricultural activity similar to the sites in Iran in the data from Lake Van. In comparison with Western Anatolia, the modern-day balance of precipitation, evaporation and temperature is more tenuous in the Armenian highlands, making agriculture more sensible to any shifts in climatic conditions (fig. 19 and 20); we can assume similar challenges of nature for the ancient and medieval period.[36] The analysis of the ratio of isotopes of oxygen in the dated sediments of Lake Van, at this time central region of the Armenian Kingdom of Vaspurakan, documents a rapid change towards arid conditions from the 10th to the 11th century (fig. 21).[37] In the same period, Vaspurakan became one of the first victims of the Turkish advance from Central Asia to the west; in 1016/1017, an army of Türkmen raiders, maybe hired by the neighbouring Emir of Khoy, defeated the troops of King Senek'erim Arcruni. These events motivated him to give in to Byzantine pressure and to hand over his kingdom to Emperor Basileios II in exchange for extensive possessions in Cappadocia in 1022/23. The annexation of Vaspurakan was accompanied by a large scale exodus of reportedly 14,000 families to the new realm of Senek'erim in the interior of Anatolia.[38] Similar migration movements accompanied the Byzantine annexations of the kingdoms of Ani in 1045 and of Kars in 1065.[39] An actual dramatic demographic decline in Vaspurakan can be detected both in the data from Lake Van (using the sedimentation of charcoal as proxy for human activity) as well as in the rapid decline of building activity from the 10[th] century (the apex of Arcruni power as manifested in the famous church of Ałtamar) to the 11[th] century (fig. 22).[40] It seems plausible to assume a contribution of the documented adverse climatic conditions to these developments. The relatively stronger affection of Armenia by the "cold spell" of the 11[th] century can also be observed in a recent temperature reconstruction by Joel Guiot, Christophe Corona et alii for various regions of Europe.[41] The impact of such a scenario on Byzantium could have been significant: Vaspurakan and the newly annexed Armenian territories definitely constituted the extreme periphery of the Empire. Yet their relevance within the military framework of 11[th] century Anatolia, which focused on the new border commands at the cost of the traditional thematic "defence in depth" ), was considerable (as became manifest in the frequently discussed "dissolution" of thematic troops in the East by Emperor Konstantinos IX Monomachos in ca. 1049/1052).[42] Any collapse of defences here would open up the core regions to invaders – as it did for the Seljuks from the 1050s onwards, who advanced mainly through a corridor north of Lake Van and through the valley of the Araxes-river.[43] At the same time, the region was very much suitable for the nomadic lifestyle of core elements of the Seljuk retinue, as A. C. S. Peacock has demonstrated. Accordingly, Emperor Romanos IV Diogenes tried to get hold of the cities of Manzikert and Akhlat to the north of Lake Van in order to regain control over this corridor in summer 1071.[44]

---

[36] P. E. ZIMANSKY, Ecology and Empire: The Structure of the Urartian State. Chicago 1985; J. STADELBAUER, Studien zur Agrargeographie Transkaukasiens. Berlin 1983. For climatic conditions esp. in the Lake Van region cf. R. HEWSEN, "Van in this World, Paradise in the next". The historical geography of Van/Vaspurakan, in: Armenian Van/Vaspurakan, ed. R. G. Hovannisian. Costa Mesa, CA 2000, 13–42; Th. LITT et alii, A 600,000 year long continental pollen record from Lake Van, eastern Anatolia (Turkey). *Quaternary Science Reviews* 104 (2014) 30–41; O. KWIECIEN et alii, Dynamics of the last four glacial terminations recorded in Lake Van, Turkey. *Quaternary Science Reviews* 104 (2014) 42–52; M. N. ÇAGATAY et alii, Lake level and climate records of the last 90 ka from the Northern Basin of Lake Van, eastern Turkey. *Quaternary Science Reviews* 104 (2014) 97–116; H. DUZEN – H. AYDIN, Sunshine-based estimation of global solar radiation on horizontal surface at Lake Van region (Turkey). *Energy Conversion and Management* 58 (2012) 35–46.

[37] L. WICK – G. LEMCKE – M. STURM, Evidence of Lateglacial and Holocene climatic change and human impact in eastern Anatolia: high-resolution pollen, charcoal, isotopic and geochemical records from the laminated sediments of Lake Van, Turkey. *Holocene* 13 (2003) 665–675; N. RIEDEL, Der Einfluss von Vulkanausbrüchen auf die Vegetationsentwicklung und die agrarische Nutzung seit dem Weichselspätglazial in Ostanatolien anhand von palynologischen Untersuchungen an lakustrinen Sedimenten des Vansees (Türkei). Dissertation, Rheinische Friedrich-Wilhelms-Universität Bonn 2012, esp. 14–23; N. ROBERTS et alii, Palaeolimnological evidence for an east–west climate see-saw in the Mediterranean since AD 900. *Global and Planetary Change* 84–85 (2012) 23–34; LUTERBACHER et alii, A Review of 2000 Years of Paleoclimatic Evidence 115–117; MCCORMICK et alii, Climate Change during and after the Roman Empire 184 (fig. 7) and 219–220 (evaluation of the Lake Van isotope data). Cf. also IZDEBSKI, A Rural Economy in Transition 136. For the most recent data from Lake Van focused on much earlier periods cf. Th. LITT – F. S. ANSELMETTI, Lake Van deep drilling project PALEOVAN. *Quaternary Science Reviews* 104 (2014) 1–7; LITT et alii, A 600,000 year long continental pollen record, ÇAGATAY et alii, Lake level and climate records, and several other contributions in *Quaternary Science Reviews* 104 (2014).

[38] T'ovma Arcruni cont. IV, 12 (ed. K. PATKANEAN, T'ovmayi vardapeti Arcrunwoy Patmut'iwn tann Arcruneac'. St. Petersburg 1887 [Reprint Tbilisi 1917] 307–308); Thomas Artsruni, History of the House of the Artsrunik'. Translation and Commentary by R. W. THOMSON. Detroit 1985, 370–371. Cf. W. SEIBT, Die Eingliederung von Vaspurakan in das Byzantinische Reich (etwa Anfang 1019 bzw. Anfang 1022). *Handes Amsorya* 92 (1978) 49–66; W. FELIX, Byzanz und die islamische Welt im früheren 11. Jahrhundert (*BV* 14). Vienna 1981, 137–141; HEWSEN, "Van in this World, Paradise in the next" 28–30. Based on the data in ZIMANSKY, Ecology and Empire 15–19, on the maximum extent of cultivated land, one could assume a maximum "carrying capacity" for the area of the Kingdom of Vaspurakan of ca. 250,000 people (on the basis of a minimum of 0.5 ha arable land per head). Even if we assume that the Kingdom reached this number of inhabitants, 14,000 families (maybe 60,000 to 70,000 people) would have represented a most significant demographic bleeding. Of course, the figure reported by T'ovma Arcruni cont. can only be used with caution.

---

[39] K. YUZBASHYAN, L'administration byzantine en Arménie aux Xe–XIe siècles. *Revue des études Arméniennes*. N. S. 10 (1973-1974) 139–184; G. DÉDÉYAN, L'immigration arménienne en Cappadoce au XIe siècle. *Byz* 45 (1975) 41–117; W. SEIBT, Stärken und Schwächen der byzantinischen Integrationspolitik gegenüber den neuen armenischen Staatsbürgern im 11. Jh., in: Η αυτοκρατορία σε κρίση (;) το Βυζάντιο τον 11ο αιώνα, ed. N. Vlyssidou. Athens 2003, 331–347; G. A. LEBENIOTES, Η πολιτική κατάρρευση του Βυζαντίου στην Ανατολή. Το ανατολικό σύνορο και η κεντρική Μικρά Ασία κατά το β΄ ήμισυ του 11ου αι. 2 Vols, Thessalonike 2007, 126–140.

[40] WICK – LEMCKE – STURM, Evidence of Lateglacial and Holocene climatic change. For the monumental evidence see J. M. THIERRY, Monuments arméniens du Vaspurakan. Paris 1989, and L. JONES, Between Islam and Byzantium. Aght´amar and the visual Construction of Medieval Armenian Rulership. Aldershot 2007. A general "apocalyptic" mood in the description of the developments in Armenia during the 11[th] century can be found in the contemporaneous work of Aristakēs Lastiverc'i, who speaks about depopulated and uncultivated districts across the country, see Patmut'iwn Aristakisi Lastiverc'woy, ed. K. YUZBAŠYAN. Erewan 1963; Aristakes Lastivertc'i's History, translated from Classical Armenian by R. BEDROSIAN. New York 1985.

[41] J. GUIOT – C. CORONA, and ESCARSEL members, Growing Season Temperatures in Europe and Climate Forcings Over the Past 1400 Years. *Public Library of Science ONE* 5(4) (2010): e9972. doi:10.1371/journal.pone.0009972. But also within the Armenian highlands, regional differences are plausible; cities such as Ani, Artze (Arcn) or Theodosiupolis (Erzurum) continued to prosper from a mercantile point of view well into the 11[th] century, for instance, cf. FELIX, Byzanz und die islamische Welt 166–167; HARVEY, Economic Expansion in the Byzantine Empire 212–213; GEROLYMATOU, Αγορές, έμποροι και εμπόριο 119–122, 134–136. The will of Eustathios Boilas hints at an expansion of agricultural activity in the eastern border regions of Byzantium in the mid-11[th] century (albeit it is still unclear if Boilas´ estates were located in the region of Tao/Tayk in north-western Armenia or in the area of Edessa), cf. Sp. VRYONIS, Jr., The Will of a Provincial Magnate, Eustathius Boilas (1059). *DOP* 11 (1957) 263–277; HARVEY, Economic Expansion in the Byzantine Empire 64–65. At the same time, such (possibly) marginal lands were often affected by deteriorating climatic conditions to a higher degree, cf. PARKER, Global Crisis 56–58. Further studies on these variations are necessary.

[42] Cf. H.-J. KÜHN, Die byzantinische Armee im 10. und 11. Jahrhundert. Studien zur Organisation der Tagmata (*Byzantinische Geschichtsschreiber*, Ergänzungsband 2). Vienna 1991, esp. 187–195 (on the Katepanates of Iberia and Asprakan); J. HALDON, Warfare, State and Society in the Byzantine World 565–1204 (*Warfare and History*). London 1999, 83–94; LEBENIOTES, Η πολιτική κατάρρευση του Βυζαντίου στην Ανατολή 81–90 (on the "dissolution" of thematic troops in Iberia, with sources).

[43] Cf. T'ovma Arcruni cont. IV, 12 (ed. PATKANEAN 308); Thomas Artsruni, transl. THOMSON, 371: "When news of the kings' [Senek'erim of Vaspurakan and Gagik II Bagratuni of Ani] departure from Armenia and the Roman control [of that country] reached the camp of the impious, bloodthirsty, ferocious race of Elim, then the ruler of the Elimites [= the Seljuks], who was called Sultan Tullup [Toġrïl], launched a cavalry attack like an eagle swooping on flocks of birds. Reaching the metropolis of Ani, he besieged it; having

But what about the rest of Anatolia? As already mentioned, proxy data also here indicates for most sites colder temperature conditions for the 11th century (fig. 4). Recent isotopes analyses for the Nar Gölü in Cappadocia, where extraordinary sediment conditions allow for an annual reconstruction of past parameters, hint at more severe winters for the entire period from 921–1071 AD (which would also include the extreme winter of 927/928, see above). Similarly, a cooling can be observed in speleothem-data from the Kocain-Cave in south-western Anatolia.[45] Yet in contrast to Central Asian or Armenian data, precipitation conditions seem to have remained stable or even above average in the regions for which we have data during this period (as indicated by a recent analysis of past sea level change in Lake Iznik in Bithynia or data for the Tecer Lake in Cappadocia).[46] At the same time, also the pollen data indicates a continued growth of agricultural activity as observable for Byzantine Anatolia since the 9th/10th century over most of the 11th century for most sites (fig. 4).[47] One example is the Köycegiz Gölü in Karia near ancient Kaunos, which indicates uninterrupted growth of agricultural activity up to the second half of the 12th century (although the generally captured it, he put [the inhabitants] to the sword." The Seljuk conquest of Ani actually took place in 1064, that is after the death of Sultan Toğrïl (in 1063), but the causal chain of events indicated by the author becomes clear.

[44] A. C. S. PEACOCK, Nomadic society and the Seljuk campaigns in Caucasia. *Iran and the Caucasus* 9 (2005) 205–230 (with a chronology of advances); IDEM, The Great Seljuk Empire 46–66; KOROBEINIKOV, Raiders and Neighbours 698–703; LEBENIOTES, Η πολιτική κατάρρευση του Βυζαντίου στην Ανατολή 145–177; B. TODD CAREY – J. B. ALLFREE – J. CAIRNS, Road to Manzikert. Byzantine and Islamic Warfare 527–1071. Barnsley 2012, 120–153 (with various useful maps); D. NICOLLE, Manzikert 1071: The breaking of Byzantium. Botley, Oxford 2013 (equally with several illustrative maps on the routes of Seljuk advances and the campaigns of Romanos IV).

[45] J. WOODBRIDGE – N. ROBERTS, Late Holocene climate of the Eastern Mediterranean inferred from diatom analysis of annually-laminated lake sediments. *Quaternary Science Reviews* 30 (2011) 3381–3392; DEAN et alii, Palaeo-seasonality of the last two millennia. For the Kocain-cave see GÖKTÜRK, Climate in the Eastern Mediterranean 52–59.

[46] For the Lake Iznik data cf. U. B. ÜLGEN et alii, Climatic and environmental evolution of Lake Iznik (NW Turkey) over the last ~4700 years. *Quaternary International* 274 (2012) 88–101, as well as the older study of B. GEYER – R. DALONGEVILLE – J. LEFORT, Les niveaux du Lac de Nicée au Moyen Âge, in: Société rurale et histoire du paysage à Byzance, by J. Lefort. Paris 2006, 375–393 (originally published in 2001). For Tecer Lake see C. KUZUCUOGLU et alii, Mid- to late-Holocene climate change in central Turkey: The Tecer Lake record. *The Holocene* 21/1 (2011) 173–188. For a comparison of various data across the Near East cf. J. BAKKER et alii, Climate, people, fire and vegetation: new insights into vegetation dynamics in the Eastern Mediterranean since the 1st century AD. *Climate of the Past* 9 (2013) 57–87.

[47] Cf. IZDEBSKI, A Rural Economy in Transition 145–215 (with a focus on Late Antiquity); HALDON et alii, The Climate and Environment of Byzantine Anatolia 140; IZDEBSKI – KOLOCH – SŁOCZYŃSKI, Exploring Byzantine and Ottoman economic history. For written and archaeological evidence cf. A. KAZHDAN – A. W. EPSTEIN, Change in Byzantine Culture in the Eleventh and Twelfth Centuries. Berkeley – Los Angeles – London 1985, 23–39; HARVEY, Economic Expansion in the Byzantine Empire, esp. 35–79; KAPLAN, Les hommes et la terre 531–540; ANGOLD, The Byzantine Empire 81–98; J. LEFORT, The Rural Economy, Seventh–Twelfth Centuries, in: The Economic History of Byzantium, ed. A. E. Laiou. 3 Vols., Washington, D.C. 2002, 225–304; J. LEFORT, Société rurale et histoire du paysage à Byzance. Paris 2006, 395–478; LAIOU – MORRISSON, The Byzantine Economy 90–100; Le Monde Byzantin II. L'Empire byzantin (641–1204), ed. J.-C. Cheynet. Paris 2006, 206–219, 231–232; M. WHITTOW, The Middle Byzantine Economy (600–1204), in: The Cambridge History of the Byzantine Empire, c. 500–1492, ed. J. Shepard. Cambridge 2008, 465–492; E. COOPER – M. DECKER, Life and Society in Byzantine Cappadocia. Houndmills 2012; J. KODER, Regional Networks in Asia Minor during the Middle Byzantine Period (Seventh-Eleventh Centuries). An Approach, in: Trade and Markets in Byzantium, ed. C. Morrisson. Washington, D.C. 2012, 147–175.

small percentage of cultivated pollen in these samples may limit their explanatory power) (fig. 17).[48] One counterexample is Lake Abant near Klaudiupolis in Honorias, recently analysed by Adam Izdebski; there, pollen data indicates a pronounced decrease of agricultural activity in the 11th century. Yet, as in the Iranian sample, uncertainties in the possible dating of this layer pose a problem: this decline could have taken place before as well as after the Seljuk advance. Thus we cannot assume with security that adverse conditions had affected the region already before, although we have some written evidence for such a scenario (see below the events of the years 1032–1038).[49]

Mostly "positive" is the evidence for the provinces of the Byzantine Empire in the Southern Balkans and Greece. While tree ring-data from Albania shows some shorter term cold periods (as one in the late 1080s and 1090s in the reign of Alexios I Komnenos) during the 11th century, there is no indication for a longer term "Big Chill" as in Central Asia (fig. 8 and 9). Precipitation data indicates on average humid conditions, which may have resulted in massive snow falls as reported by Anna Komnene for 1090/1091[50], but also provided sufficient moisture for agriculture most of the time. This very well correlates with the growth of agricultural activity documented in almost all pollen sites in Macedonia as well as Greece: data from Lake Voulkaria in Akarnania indicates an upwards trend for the cultivation of Olea well into the 14th century (although the relative small sample size may limit its explanatory power), pollen from Lake Orestiás at Kastoria for the growing of cereals until the end of the 11th century (fig. 18).[51] Different is the picture for the regions to the north of Mt. Haemus: sources as well as archaeological data indicate a contraction of settlement in various regions for the 11[th] and earlier 12[th] century especially due to the frequent advances of nomads from north of the Danube. Yet at the same time one can observe considerable and continuing mercantile activity in the lower Danube region (see also above on relations with the nomads along the Danube).[52] Future research may illuminate possible contributions of climatic conditions to these somehow divergent developments.

---

[48] IZDEBSKI, A Rural Economy in Transition 17–177; HALDON et alii, The Climate and Environment of Byzantine Anatolia 133–134. Cf. also MCMILLAN, A GIS approach to palaeovegetation modelling, esp. 124–158, for possible impact of changing climatic conditions on vegetation in this region, and Appendix 31–32 on the site of Köycegiz Gölü.

[49] IZDEBSKI, A Rural Economy in Transition 186–190; A. IZDEBSKI, The Changing Landscapes of Byzantine Northern Anatolia. *Archaeologia Bulgarica* 16 (2012) 47–66; HALDON et alii, The Climate and Environment of Byzantine Anatolia (Table 2); IZDEBSKI – KOLOCH – SŁOCZYŃSKI, Exploring Byzantine and Ottoman economic history.

[50] Anna Komnene VIII, 3, 3 (ed. D. R. REINSCH – A. KAMBYLIS [*CFHB* 40/1–2]. Berlin – New York 2001, 241, 78–242, 84); Anna Komnene, The Alexiad, transl. by E. R. A. SEWTER, revised by P. FRANKOPAN. London 2009, 252; cf. also TELELIS, Μετεωρολογικά φαινόμενα nr 502. For a general survey of these years cf. ANGOLD, The Byzantine Empire 132–135; E. MALAMUT, Alexis Ier Comnène. Paris 2007, 84–89; M. ANGOLD, Belle époque or Crisis? (1025–1118), in: The Cambridge History of the Byzantine Empire, c. 500–1492, ed. J. Shepard. Cambridge 2008, 611–612; P. FRANKOPAN, The First Crusade: The Call from the East. Cambridge, Mass. 2012, 37–38, 58–59, 74–78. On the chronology of these years in the work of Anna Komnene cf. K. BELKE, Byzanz und die Anfänge des rumseldschukischen Staates. Bemerkungen zur Chronologie von Anna Komnēnēs „Alexias" in den Jahren 1084 bis 1093. *JÖB* 61 (2011) 65–79; for earlier years see also E. KISLINGER, Vertauschte Notizen. Anna Komnene und die Chronologie der byzantinisch-normannischen Auseinandersetzung 1081–1085. *JÖB* 59 (2009) 127–145.

[51] S. JAHNS, The Holocene history of vegetation and settlement at the coastal site of Lake Voulkaria in Acarnania, western Greece. *Vegetation History and Archaeobotany* 14 (2005) 55–66; K. KOULI – M. D. DERMITZAKIS, Contributions to the European Pollen Database: Lake Orestiás (Kastoria, northern Greece). *Grana* 49(3) (2010) 154–156. For a more general overview on palynological data for Central Greece see IZDEBSKI – KOLOCH – SŁOCZYŃSKI, Exploring Byzantine and Ottoman economic history. Cf. also H. A. FORBES – H. A. KOSTER, Fire, Axe, and Plow: Human Influence on Local Plant Communities in the Southern Argolid. *Annals of the New York Academy of Sciences* 268 (1976) 109–126; A. DUNN, The exploitation and control of woodland and scrubland in the Byzantine world. *BMGS* 16 (1992) 235–298. For accompanying archaeological and written evidence cf. LEFORT, Société rurale, esp. 212–263 (on Macedonia); CURTA, Southeastern Europe in the Middle Ages 276–281; F. CURTA, The Edinburgh History of the Greeks, c. 500 to 1050. The Early Middle Ages. Edinburgh 2011, 166–208 (with a list of 22 churches built in Greece between 1000 and 1050); J. BINTLIFF, The Complete Archaeology of Greece: From Hunter-Gatherers to the 20th Century AD. Malden – Oxford 2012, 391–395; M. VEIKOU, Byzantine Epirus. A Topography of Transformation. Settlements of the Seventh-Twelfth Centuries in Southern Epirus and Aetoloacarnania, Greece (*The Medieval Mediterranean* 95). Leiden – Boston 2012, esp. 348–349 and 352–359.

Otherwise, as already Ioannis Telelis has highlighted and the frequency of extreme weather events as mentioned in our texts especially in the first half of the 11th century indicates, the 11[th] century could be characterised as "period of extremes" with regard to the occurrence of adverse weather conditions (fig. 26 and 27). One example for a short term change from beneficial to adverse conditions are the years 1032 to 1038 (as described especially in Ioannes Skylitzes), already mentioned as one indicator of possible climatic factors for the 11[th] century crisis by Nicolas Svoronos in 1966.[53] These events again illustrate the regional variances of such phenomena (as did our proxy data) as well as the pressure such variances could exert on the mobility of population from distressed areas into less affected ones.[54] In 1032, "famine and pestilence" affected Cappadocia, Paphlagonia, the Armeniakon and Honorias (where also the site of Lake Albant with a negative pollen record is situated, see above). This caused large scale movement of population, which both Emperor Romanos III Argyros as well as the church (the Metropolitan of Ankyra) tried to temper with "money and other necessities of life". Two years later, locusts destroyed the crops in parts of Bithynia and the Thrakesion, "compelling the inhabitants to sell their children and move into Thrace". Again, the Emperor "gave to every one of them three pieces of gold and arranged for them to return home". In 1035, a severe winter accompanied by Pecheneg raids damaged the so far unaffected European provinces. In 1037, drought and famine finally reached the region of Constantinople; yet the government was able to soften the impact of crisis in the capital by buying grain in the obviously unaffected provinces of Hellas and Peloponnese. But in 1038, famine finally also struck Europe, affecting "Thrace, Macedonia, Strymon and Thessalonike, right into Thessaly".[55] Most interesting is the spatial pattern of advance of the crisis phenomena towards the core regions of the Empire both from the East and from the North (fig. 5).

---

[52] KAZHDAN – EPSTEIN, Change in Byzantine Culture 31–34, B. BORISOV, Demographic and Ethnic Changes during XI–XII Century in Bulgaria. *Archaeologia Bulgarica* 2 (2007) 71–84; STEPHENSON, Byzantium's Balkan Frontier 84–89; GEROLYMATOU, Αγορές, έμποροι και εμπόριο 171–177; CURTA, Southeastern Europe in the Middle Ages 293–295; A. MADGEARU, Byzantine Military Organization on the Danube, 10[th]–12[th] Centuries (*East Central and Eastern Europe in the Middle Ages, 450–1450*, Vol. 22). Leiden – Boston 2013, esp. 101–144. On palynological evidence for Eastern Bulgaria cf. IZDEBSKI – KOLOCH – SŁOCZYŃSKI, Exploring Byzantine and Ottoman economic history. On the other hand, from the Typikon of Gregorios Pakurianos for the Monastery of the Mother of God Petrizonitissa in Backovo near Philippupolis from December 1083 we also have hints at the establishment of new villages and the foundation of settlements and monasteries in that region: P. GAUTIER, Le typikon du sébaste Grégoire Pakourianos, *REB* 42 (1984) 5–145; Byzantine Monastic Foundation Documents: A Complete Translation of the Surviving Founders' Typika and Testaments edited by J. Thomas – A. Constantinides Hero. Washington D. C. 2000, 507–564; HARVEY, Economic Expansion in the Byzantine Empire 64–66, 160.

[53] N. SVORONOS, Etudes sur l'organisation intérieure, la société et l'économie de l'Empire Byzantin. London 1973, esp. 348–350 (originally published in 1966). Cf. also KAZHDAN – EPSTEIN, Change in Byzantine Culture 27.

[54] Cf. PARKER, Global Crisis 70–71, for a model of "connected shallow ponds" for the links between regions and markets which are affected by environmental stress in different ways.

[55] Ioannes Skylitzes 18 (Romanos o Argyros), 11 and 17 (386, 65–73 and 389, 58–64 THURN); transl. WORTLEY 364 and 367; cf. TELELIS, Μετεωρολογικά φαινόμενα nr 443–455; HALDON et alii, The Climate and Environment of Byzantine Anatolia 127; MORONY, Michael the Syrian 160. On the impact of locusts, also in combination with severe winters or droughts, in the areas of the Near East during the 12[th]–14[th] cent. cf. RAPHAEL, Climate and Political Climate 168–177; MORONY, Michael the Syrian 151–156; WIDELL, Historical Evidence for Climate Instability.

Regional variances on a larger scale across the Eastern Mediterranean can be also illustrated with an episode reported in Arabic sources: in 1054, one of the in the 11[th] century frequent low levels of Nile flooding affected Egypt. The Fatimid government in Cairo, expecting famine and unrest after the exhaustion of the stores from the previous harvest, requested for help from Emperor Konstantinos XI Monomachos. The emperor agreed to ship 400,000 irdabb, which are 2,700 tons of grain, to Egypt. This amount would have been sufficient to supply 10,000 individuals for one year.[56] This was a modest quantity if compared with the 10,000 tons of grain Egypt provided for Constantinople in the period of Justinian, but still sufficient to dampen the effects of crisis at least in the megalopolis of Cairo (a measure thus similar to the purchase of grain for Constantinople in the drought year 1037, see above). One can also assume that such quantities could be provided from the imperial storehouses in and around Constantinople.[57] That these storehouses in contrast to the ones in Egypt were full at this time we also know on the basis of the information from Zonaras who reports an above-average harvest in the empire in this period. Equally proxy data from North-western Anatolia indicates that meteorological conditions had improved in comparison with the crisis years in the 1030s (fig. 23).[58] Yet the death of Konstantinos XI in January 1055 dashed all hopes of the Fatimid regime for support from Constantinople, since Empress Theodora decided to cancel the deal.[59] In general, besides parts of Iran and of Armenia, Egypt is the region of the Near East where Ellenblum´s scenario works best. As already earlier scholars such as Thierry Bianquis (1980) and Heinz Halm (2003) have analysed in detail, frequent low levels of the Nile Flood very much contributed to the destabilisation of the Fatimid regime in the period between 1021 and especially 1062 and 1074 AD (fig. 28).[60] Similar to Byzantium around the same period, the almost-collapse of the Empire was only prevented by the coup of a military leader from outside the capital; the "Alexios Komnenos" of Fatimid Egypt was Badr al-Ğamālī, an Armenian convert to Islam and founder of a dynasty of viziers who would rule instead of the powerless Caliphs for the next decades.[61]

---

[56] Calculations based on J. KODER, Gemüse in Byzanz. Die Frischgemüseversorgung Konstantinopels im Licht der Geoponika (*Byzantinische Geschichtsschreiber*, Ergänzungsband 3). Vienna 1993, 100–103; cf. also IDEM, Regional Networks in Asia Minor 174 (Appendix II).

[57] Μ. ΜΙΟΤΤΟ, Ο ανταγωνισμός Βυζαντίου και Χαλιφάτου των Φατιμιδών στην Εγγύς Ανατολή και η δράση των ιταλικών πόλεων στην περιοχή κατά τον 10ο και τον 11ο αιώνα. Thessaloniki 2008, 251–252 (with reference to the Arabic sources); H. HALM, Die Kalifen von Kairo: Die Fatimiden in Ägypten 973–1074. Munich 2003, 381–382; FELIX, Byzanz und die islamische Welt 119–120. Cf. in general D. JACOBY, Byzantine Trade with Egypt from the Mid-Tenth Century to the Fourth Crusade. *Thesaurismata* 30 (2000) 25–77, and for the imperial granaries J.-Cl. CHEYNET, Un aspect du ravitaillement de Constantinople aux Xe/XIe siècles d'après quelques sceaux d'horreiarioi. *SBS* 6 (1999) 1–26.

[58] Zonaras XVII 29, 4 (ed Th. BÜTTNER-WOBST, Ioannis Zonarae Epitomae Historiarum Libri XIII–XVIII. Bonn 1897, 652); TELELIS, Μετεωρολογικά φαινόμενα nr 476. Speleothem-data from the Sofular cave in northwestern Turkey: D. FLEITMANN et alii, Sofular Cave, Turkey 50KYr Stalagmite Stable Isotope Data. *IGBP PAGES/World Data Center for Paleoclimatology Data Contribution Series* # 2009-132; GÖKTÜRK, Climate in the Eastern Mediterranean 13–39.

[59] See the literature cited in fn. 57 for this measure of Empress Theodora. Cf. RAPHAEL, Climate and Political Climate 82–83 and 188–189 for similar cases of the intermixture of diplomacy and the trade of grain in times of need in the 12[th] century (Norman Sicily and Tunesia) and in the 1970s (USA and Soviet Union).

[60] Th. BIANQUIS, Une Crise frumentaire dans l'Égypte Fatimide. *Journal of the Economic and Social History of the Orient* 23, 1/2 (1980) 67–101; FELDBAUER, Die islamische Welt 66–67, 361–362; The Cambridge History of Egypt, Volume I. Islamic Egypt, 640–1517, ed. C. F. Petry. Cambridge 1998, 152–153; HALM, Die Kalifen von Kairo 68–72 (in general on the impact of the Nile flood on the stability of the Fatimid regime), 316–324, 400–420; E. CHANEY, Revolt on the Nile: Economic Shocks, Religion and Political Power. *Econometrica* 81, No. 5 (September 2013) 2033–2053. For quantitative data on the extremes of the Nile flood during these years cf. E. A. B. ELTAHIR – G. WANG, Nilometers, El Niño, and Climate Variability. *Geophysical Research Letters* 26/4 (1999) 489–492; F. A. HASSAN, Extreme Nile floods and famines in Medieval Egypt (AD 930–1500) and their climatic implications. *Quaternary International* 173/174 (2007) 101–112; F. HASSAN, Nile flood discharge during the Medieval Climate Anomaly. PAGES news 19/1 (2011)

To sum up the results of the evaluation of the scenario of a "Collapse of the Eastern Mediterranean": one can reconstruct a significant deterioration of climatic conditions in the region around Lake Van, which, combined with incipient Türkmen raids, could have contributed to migration to the west and a decline of the demographic potential in a sensitive region for the defence of Byzantine Anatolia. We equally observe a higher frequency of meteorological extremes and famines in Central and Western Asia Minor. Yet despite a general "cold trend", we detect pronounced differences in the climatic and agricultural trajectories for different regions of the Near East and especially also within the Byzantine Empire in the 11[th] century. Besides other sources, pollen data indicates a continued agro-economic growth in Anatolia and Greece in most regions. Therefore, we cannot accept a scenario of general climate-caused "collapse" of Byzantium, but may assume a contribution of environmental factors to also otherwise crisis prone socio-political and military conditions.[62] The same seems true for other regions of the Eastern Mediterranean and the Near East; a very strong correlation of climatic parameters and political unrest can be documented for Egypt with its specific ecology.

## The Byzantine economy in the 11[th]–13[th] century: some considerations on climatic conditions in comparison

The discussion about the 11[th] century crisis is part of a wider debate on the socio-economic development of Byzantium before and during the Komnenian period (1081–1185). The consensus of modern scholarship was summed up by Angeliki Laiou in 2002 in the "Economic History of Byzantium", stating that despite signs of severe political crisis "the eleventh and twelfth centuries have been recognized as periods of economic growth", when "for the first time in Byzantine history, there was a disjunction between military and territorial developments on the one hand and economic activity on the other."[63] In 2008, Mark Whittow allowed himself to disagree, arguing that "the empire

---

[30]–31; D. KONDRASHOV – Y. FELIKS – M. GHIL, Oscillatory modes of extended Nile River records (A.D. 622–1922). *Geophysical Research Letters* 32 (2005): DOI: 10.1029/2004GL022156; IZDEBSKI, A Rural Economy in Transition 140–141. In general on the special ecology of Egypt and also for later periods see: St. J. BORSCH, The Black Death in Egypt and England. A Comparative Study, Austin 2005; A. MIKHAIL, Nature and Empire in Ottoman Egypt: an Environmental History. Cambridge 2011; J. P. COOPER, The Medieval Nile. Route, Navigation, and Landscape in Islamic Egypt. New York 2014, esp. 107–123.

[61] H. HALM, Kalifen und Assassinen: Ägypten und der Vordere Orient zur Zeit der ersten Kreuzzüge 1074–1171. Munich 2014, 17–86; The Cambridge History of Egypt, ed. Petry 153; S. B. DADOYAN, The Fatimid Armenians. Cultural and political Interaction in the Near East (*Islamic History and Civilization* 18). Leiden – New York – Cologne 1997, 107–127; G. DÉDÉYAN, Les Arméniens entre Grecs, Musulmans et Croisés: étude sur les pouvoirs arméniens dans le Proche-Orient méditerranéen (1068–1150), 2 vol.s. Lisbonne 2003, 881–891.

[62] In general on the 11[th] century in Byzantium see J. C. CHEYNET, Pouvoir et contestations à Byzance (963–1210) (*Byzantina Sorbonensia* 9). Paris 1990; ANGOLD, The Byzantine Empire 15–23 (on scholarly debates on the 11[th] century); Η αυτοκρατορία σε κρίση (;) το Βυζάντιο τον 11ο αιώνα, ed. N. Vlyssidou. Athens 2003; ANGOLD, Belle époque or Crisis 583-626; D. KRALLIS, Michael Attaleiates and the Politics of Imperial Decline in Eleventh-Century Byzantium. Tempe 2012; M. C. BARTUSIS, Land and Privilege in Byzantium. The Institution of Pronoia. Cambridge 2012, 112–160, and P. FRANKOPAN, Land and Power in the Middle and Late Period, in: The Social History of Byzantium, ed. J. Haldon. Malden – Oxford 2009, 112–142 (both on the distribution of land and power in the countryside); HALDON et alii, The Climate and Environment of Byzantine Anatolia 127.

[63] The Economic History of Byzantium, ed. A. E. Laiou. 3 Vols., Washington, D.C. 2002, 1150. Cf. also P. MAGDALINO, The Empire of Manuel I Komnenos (1143–1180). Cambridge 1993, 140–171; ANGOLD, The

was not as rich as it appeared to be" and that the 11[th] and 12[th] century growth was not that impressive, especially when compared to contemporaneous developments in the "Latin West".[64] Climate proxies and pollen data cannot produce answers to questions discussed within this context such as the actual impact of the rise of great estates or the growth of Latin commerce; but they illuminate the general environmental framework within which Byzantium and other economies of the Eastern Mediterranean had to operate in the 12[th] century (also in comparison with Western Europe) and provide additional data for the increase or decrease of agricultural activity in various regions. In general, in the 12[th] century temperature and precipitation conditions – with the exception of sites in Iran and Armenia, again, as well as Egypt – seem to have been more beneficial than in the 11[th] century across the Near East, Anatolia and the Balkans (fig. 6).[65] Yet, especially for Anatolia, pollen data indicates very different trajectories for agricultural activity between the coastal areas and the interior; thus, again, regional variability dominates the picture.

A pronounced improvement of precipitation conditions as well as an increase of human activity from the mid-11[th] century onwards has been recently reconstructed on the basis of data from the site of Jableh, then within the territory of the Principality of Antioch, part-time adversary as well as vassal of Byzantium. A "peak" of viticulture has been identified for the period from 1100 to 1250 AD. A similar increase in indicators of human agricultural activity can be seen in pollen data from the Southern Bekaa Valley in Lebanon. As within the scenario of Ellenblum, one could propose that climate somehow "favoured" the crusaders in several regards. Other evidence illustrates the occurrence of severe droughts and famine in the Crusader states especially during the 12[th] century (see fig. 29); but in general, the "Latin" regimes were able to cope with these situations relatively well (in contrast to the 13[th] century, when also general political-military conditions had changed to their disadvantage).[66]

The differences between regions within Asia Minor[67] despite similarly beneficial climatic conditions can be highlighted for two sites relatively near to each other: the already

---

Byzantine Empire 173–180 (on debates on the 12[th] century); A. HARVEY, The Byzantine Economy in an International Context. *Historisch Tijdschrift Groniek* 39/171 (2006) 163–174; M. KAPLAN – C. MORRISSON, L'économie byzantine: perspectives historiographiques. *Revue historique* 630/2 (2004) 391–411; LAIOU – MORRISSON, The Byzantine Economy 90–165 (for a more recent synthesis). On the debasement of the nomisma cf. M. F. HENDY, Studies in the Byzantine Monetary Economy c. 300–1450. Cambridge 1985, esp. 506–510; KAZHDAN – EPSTEIN, Change in Byzantine Culture 25–26; C. MORRISON, Byzantine Money: Its Production and Circulation, in: The Economic History of Byzantium, ed. A. E. Laiou. 3 Vols., Washington, D.C. 2002, esp. 931–933; C. CAPLANIS, The Debasement of the "Dollar of the Middle Ages". *The Journal of Economic History* 63/3 (2003) 768–801; Le Monde Byzantin, ed. Cheynet 292–311; LAIOU – MORRISSON, The Byzantine Economy 147–155.

[64] WHITTOW, The Middle Byzantine Economy, esp. 487–491. See also the conclusion below on comparisons of socio-economic developments in Western Europe, Eastern Europe and Byzantium in the Middle Ages.

[65] Cf. RAPHAEL, Climate and Political Climate, esp. 21–27, MORONY, Michael the Syrian 147–149 and WIDELL, Historical Evidence for Climate Instability, for droughts and famines in various regions during the 12[th] century.

[66] D. KANIEWSKI et alii, Medieval coastal Syrian vegetation patterns in the principality of Antioch. *The Holocene* 21 (2011) 251–262; D. KANIEWSKI et alii, The Medieval Climate Anomaly and the Little Ice Age in Coastal Syria inferred from Pollen-derived Palaeoclimatic Patterns. *Global and Planetary Change* 78 (2011) 178–187; L. HAJAR et alii, Environmental changes in Lebanon during the Holocene: Man vs. climate impacts. *Journal of Arid Environments* 74 (2010) 746–755 (for the Southern Bekaa-Valley). Cf. also S. REDFORD, Trade and Economy in Antioch and Cilicia in the Twelfth and Thirteenth Centuries, in: Trade and Markets in Byzantium, ed. C. Morrisson. Washington, D.C. 2012, 297–309. On famine and droughts in the Crusader states and their reactions to it cf. RAPHAEL, Climate and Political Climate 21–27, 56–94 and 191–193.

[67] Cf. IZDEBSKI, A Rural Economy in Transition 204–233 on the loss of homogeneity of the "rural world" in Asia Minor in the transition from Late Antiquity to the Middle Byzantine period and the diversity ("sharp

mentioned Köycegiz Gölü in Karia, where pollen data indicates a significant increase of cereal cultivation from the 11[th] to the 12[th] century[68], and the Söğüt Gölü at the borders between Karia and Lykia (fig. 17). Here, we observe from the late 11[th] until the early 13[th] century a significant decrease of grain pollen – as in other pollen sites beyond the new borders of Byzantine power in Anatolia.[69] As John Haldon has argued for the Nar Gölü in Cappadocia and Adam Izdebski et alii for other sites, this could be connected with the occupation of these areas by various Turkish groups as well as the unrest before and after these events. Such an interpretation of the impact of the arrival of Türkmen nomads on the agricultural areas in the Near East is also influenced by contemporaneous historiography.[70]

Yet evidence from other regions as recently discussed by A. C. S. Peacock suggests that despite short-term severe crises due to invasion and warfare, the long-term impact of the arrival of Türkmen groups on agriculture was not necessarily negative.[71] In the Byzantine-Seljuk border regions of the 12[th] century, a "symbiotic" relationship can be found in the Maeander valley, where transhumant Turkish nomad groups used to winter while returning to the Anatolian plateau in summer. This happened with the consent of Byzantine authorities and was even acknowledged in a treaty between Emperor Manuel

---

[68] IZDEBSKI, A Rural Economy in Transition 172–177. On the continuing, albeit impaired urban economic activity in Western Asia Minor cf. R.-J. LILIE, Handel und Politik zwischen dem byzantinischen Reich und den italienischen Kommunen Venedig, Pisa und Genua in der Epoche der Komnenen und der Angeloi (1081–1204). Amsterdam 1984, 145–177; HENDY, Studies in the Byzantine Monetary Economy 108–131; KAZHDAN – EPSTEIN, Change in Byzantine Culture 36–39; MAGDALINO, The Empire of Manuel I Komnenos 123–132; GEROLYMATOU, Ἀγορές, ἔμποροι καὶ ἐμπόριο 129–133; J. T. ROCHE, Conrad III and the Second Crusade in the Byzantine Empire and Anatolia, 1147. PhD-Thesis, University of St. Andrews 2008, 102–123.

[69] IZDEBSKI, A Rural Economy in Transition 165–168; MCMILLAN, A GIS approach to palaeovegetation modelling, Appendix 32–33.

[70] Cf. Matthew of Edessa II, 43 (transl. A. DOSTOURIAN, Armenia and the Crusades: The Chronicle of Matthew of Edessa. Lanham 1993, 143): "At the beginning of the year 528 of the Armenian era [1079–1080], a severe famine occurred throughout all the lands of the venerators of the cross, lands which are located on this side of the Mediterranean Sea; for the bloodthirsty and ferocious Turkish nation spread over the whole country to such an extent that not one area remained untouched, rather all the Christians were subjected to the sword and enslavement. The cultivation of the land was interrupted, there was a shortage of food, the cultivators and labourers decreased due to the sword and enslavement, and so famine spread throughout the whole land. Many areas became depopulated, the Oriental peoples [Armenian and Syrian Christians] began to decline, and the country of the Romans became desolate; neither food nor security was to be found". Cf. ROCHE, Conrad III and the Second Crusade 141–152 (also citing this passage). J. F. HALDON, "Cappadocia will be given over to ruin and become a desert". Environmental evidence for historically-attested events in the 7[th]–10[th] centuries, in: Byzantina Mediterranea: Festschrift für Johannes Koder zum 65. Geburtstag, ed. K. Belke – E. Kislinger – A. Külzer – M. Stassinopoulou. Vienna 2007, 215–230; IZDEBSKI, A Rural Economy in Transition 145–215; HALDON et alii, The Climate and Environment of Byzantine Anatolia 141–142 (with fig. 4), 151; IZDEBSKI – KOLOCH – SŁOCZYŃSKI, Exploring Byzantine and Ottoman economic history

[71] PEACOCK, The Great Seljuk Empire 293–297, esp. 297: "the Türkmen thus had a relationship of mutual dependency with the settled population, to whom they sold their animal products and from whom they would have bought grain. Even if the relationship was far from unambiguously happy, in the main the evidence bears out (…) that nomads were a positive economic force"; see also D. DURAND-GUÉDY, New Trends in the Political History of Iran Under the Great Saljuqs (11th–12th Centuries). History Compass 13/7 (2015) 321–337. For a new analysis of the establishment of Turkish power in Asia Minor cf. A. D. BEIHAMMER, Politische Praxis, Ideologie und Herrschaftsbildung in der Frühphase der Türkischen Expansion in Kleinasien (Deutsche Arbeitsgemeinschaft zur Förderung Byzantinischer Studien, Sonderheft 2013). Munich 2013. For considerations on the possible size of Turkish nomad groups and conditions in Byzantine-Seljuk borderlands in the 12[th] to 13[th] cent. see now D. KOROBEINIKOV, Byzantium and the Turks in the Thirteenth Century (Oxford Studies in Byzantium). Oxford 2014, 219–245.

I Komnenos and Sultan Kiliç Arslan II in 1161, recognizing "a stable and – within limits – beneficial ecological status quo", as Peter Thonemann has stated.[72] The potential for conflict between nomads and sedentary groups was of course always there, but the mere presence of Türkmen groups is not sufficient to explain a long term decline of agricultural activity in a certain region. Again, specific (regional) combinations of environmental, socio-economic and political forces played a decisive role and demand further in-depth analysis.[73] The hinterland of Attaleia, in which Sögüt Gölü as well as neighbouring sites with similar trajectories were situated, seems to have remained a relatively insecure area, as also observed by the chroniclers of the Second Crusade in the 1140s (who mention that the fertile fields of the city were not cultivated due to fear from the Turks), and therefore of limited potential for renewed agricultural growth.[74]

These "frontier conditions" did not apply to Greece (including Crete) and the Southern Balkans. As most scholars agree, they now constituted the economically most important provinces of the empire –illustrated on the basis of the distribution of imperial domains in this period, for instance.[75] Still, also here we observe regional differences in the pollen data: while agricultural growth at sites in Eastern Macedonia or around Lake Voulkaria in Akarnania continued, data from Kastoria and nearby sites in Western Macedonia indicates a stagnation or even decrease of the growing of cereals (fig. 18). Proxy data from Lake Prespa indicates precipitation conditions below average for this region in the 12th century, but especially also the position of these sites near the Via Egnatia and the main route of advance of the Normans during the wars in the 11th as well as 12th century may have contributed to less beneficial conditions for agricultural development.[76] Yet in general, palaeo-environmental data confirms the scenario of the 12th century economic growth developed for most areas of modern-day Greece on the basis of other data, be it the increase in the number of settlements mentioned in the documents of Mount Athos for the western Chalkidike, the increase in the number of church buildings in Messenia from the 10th to the 13th century or the increase in the monetary finds in Corinth or Athens.[77] Also the increase in the number of sites of Venetian commercial activity (as documented in a comparison between the Chrysobulls of 1082 and 1198) especially in the provinces of the Southern Balkans has been interpreted as indicator of economic growth and demand – despite the long term consequences these activities may have had.[78]

---

[72] P. THONEMANN, The Maeander Valley. A Historical Geography from Antiquity to Byzantium. Cambridge 2011, 8–9 (with further references); cf. also A. F. STONE, Stemming the Turkish tide: Eustathios of Thessaloniki on the Seljuk Turks. *BSl* 62 (2004) 125–142, esp. 137–138; MAGDALINO, The Empire of Manuel I Komnenos 125–127; E. RAGIA, Η αναδιοργάνωση των θεμάτων στη Μικρά Ασία τον 12ο αι. και το θέμα Μυλάσσης και Μελανουδίου. *Byzantina Symmeikta* 17 (2005–2007) 223–238, and F. DÖLGER, Regesten der Kaiserurkunden des Oströmischen Reiches von 565–1453. 2. Teil: Regesten von 1025–1204, zweite, erweiterte und verbesserte Auflage bearbeitet von F. WIRTH. Munich 1995, nr 1441a, 1444 and 1446 (on the treaties between Byzantium and the Seljuks in 1161/1162).

[73] An example how a specific extended domain could partly fall into disarray due to a combination of political and environmental factors is provided in a charter from the Monastery of St. John Prodromos on Patmos (Βυζαντινά έγγραφα της Μονής Πάτμου: Β′ Δημοσίων Λειτουργών: διπλωματική έκδοσις, ed. M. NYSTAZOPOULOU-PELEKIDOU. Athens 1980, nr 50, lns. 69–100): in February 1073, Andronikos Dukas, cousin of Emperor Michael VII Dukas, was granted a part of the former imperial domain of Alopekai in the Maeander delta plain including five villages and ca. 5,000 modioi of land. Earlier, this estate had belonged to the Parsakutenoi family and most probably had been confiscated by imperial authorities some time before. From the praktikon drawn up for Andronikos, we learn that some buildings and infrastructure (such as a xenodocheion) of the estate had fallen into ruins under the new, obviously less careful imperial administration; equally, larger pieces of farm land had been lost to the river Maeander due to flooding. While some areas of the estate were still productive, its output was clearly below the amount which could be expected under better conditions. For a detailed reading and interpretation of this document cf. now THONEMANN, The Maeander Valley 259–270, and BARTUSIS, Land and Privilege in Byzantium 123–124. Cf. also ANGOLD, The Byzantine Empire 87, and HARVEY, Economic Expansion in the Byzantine Empire 68–69 for references to this document. On the changing landscape and the micro-climate in that area cf. G. TUTTAHS, Milet und das Wasser – ein Leben in Wohlstand und Not in Antike, Mittelalter und Gegenwart (*Schriften der Deutschen Wasserhistorischen Gesellschaft*, Sonderband 5). Bochum 2007, 16–38, 428–430; M. KNIPPIG – M. MÜLLENHOFF – H. BRÜCKNER, Human induced landscape changes around Bafa Gölü (western Turkey). *Vegetation History and Archaeobotany* 17 (2008) 365–380; B. DUSAR – G. VERSTRAETEN – B. NOTEBAERT – J. BAKKER, Holocene environmental change and its impact on sediment dynamics in the Eastern Mediterranean. *Earth-Science Reviews* 108 (2011) 137–157 (also for a general overview on the possible impact of expanding human activity on sediment dynamics for various sites across the Eastern Mediterranean).

[74] IZDEBSKI – KOLOCH – SŁOCZYŃSKI, Exploring Byzantine and Ottoman economic history (especially for data from the area of Pisidia), and BAKKER et alii, Climate, people, fire and vegetation (for a discussion of possible factors for the decline of agricultural activity as documented in pollen data for south-west Anatolia). HENDY, Studies in the Byzantine Monetary Economy 108–131; KOROBEINIKOV, Raiders and Neighbours 713–717; MAGDALINO, The Empire of Manuel I Komnenos 124–132 (also in general on conditions in western Asia Minor in the 12th century); GEROLYMATOU, Αγορές, έμποροι και εμπόριο 125–126. For Attaleia see also: H. HELLENKEMPER – F. HILD, Lykien und Pamphylien (*TIB* 8). Vienna 2004, 305–307, and F. HILD, Verkehrswege zu Lande: Die Wege der Kreuzfahrer des Ersten und Zweiten Kreuzzugs in Kleinasien, in: Handelsgüter und Verkehrswege. Aspekte der Warenversorgung im östlichen Mittelmeerraum (4. bis 15. Jahrhundert), ed. E. Kislinger – J. Koder – A. Külzer (*Veröffentlichungen zur Byzanzforschung* 18). Vienna 2010, 111–113 (both studies with detailed discussion of the sources on the Second Crusade); ROCHE, Conrad III and the Second Crusade 151–152; KOROBEINIKOV, Byzantium and the Turks 152–153.

From a climate historical point of view, the Komnenian Empire of the 12th century benefited from generally favourable conditions as they can be observed in other parts of the Eastern Mediterranean as well as in Western Europe.[79] There, demographic and economic growth was indeed more spectacular, documented for instance in the number of newly founded cities in Central Europe or the large scale colonisation movement towards Eastern Central Europe. At the same time, growth and agricultural expansion started here from a lower level of economic and demographic intensity when compared

---

[75] IZDEBSKI – KOLOCH – SŁOCZYŃSKI, Exploring Byzantine and Ottoman economic history; HENDY, Studies in the Byzantine Monetary Economy 85–90; MAGDALINO, The Empire of Manuel I Komnenos (esp. map 2); BARTUSIS, Land and Privilege in Byzantium 165–170. For data on Crete cf. M. A. ATHERDEN – J.A. HALL, Human impact on vegetation in the White Mountains of Crete since AD 500. The Holocene 9(2) (1999) 183–193.

[76] JAHNS, The Holocene history of vegetation and settlement; KOULI – DERMITZAKIS, Contributions to the European Pollen Database: Lake Orestiás; A. AUFGEBAUER et alii, Climate and environmental change in the Balkans over the last 17 ka recorded in sediments from Lake Prespa (Albania/F.Y.R. of Macedonia/Greece). Quaternary International 274 (2012) 122–135; K. PANAGIOTOPOULOS et alii, Vegetation and climate history of the Lake Prespa region since the Lateglacial. Quaternary International 293 (2013) 157–169. Cf. GEROLYMATOU, Αγορές, έμποροι και εμπόριο 141–144 (on the situation along these parts of the Via Egnatia). See also IZDEBSKI – KOLOCH – SŁOCZYŃSKI, Exploring Byzantine and Ottoman economic history, whose synthesis for the mountainous hinterland of Macedonia indicates a significant increase in cerealia-pollen for instance only from the late 12th cent. onwards.

[77] KAZHDAN – EPSTEIN, Change in Byzantine Culture 34–37; ANGOLD, The Byzantine Empire 280–286; LEFORT, Société rurale (various contributions); LAIOU –MORRISSON, The Byzantine Economy 91-96; WHITTOW, The Middle Byzantine Economy 475–476; CURTA, Southeastern Europe in the Middle Ages 323–327; GEROLYMATOU, Αγορές, έμποροι και εμπόριο 152–170; BINTLIFF, The Complete Archaeology of Greece 391–393 (with graphs); IZDEBSKI – KOLOCH – SŁOCZYŃSKI, Exploring Byzantine and Ottoman economic history.

[78] LILIE, Handel und Politik, esp. 117–221 on the Italian presence in the cities in the Byzantine provinces; LAIOU – MORRISSON, The Byzantine Economy 141–147; GEROLYMATOU, Αγορές, έμποροι και εμπόριο 102–109; WHITTOW, The Middle Byzantine Economy 476–477; D. JACOBY, Venetian commercial expansion in the eastern Mediterranean, 8th–11th centuries, in: Byzantine trade, 4th-12th centuries. The Archaeology of Local, Regional and International Exchange. Papers of the Thirty-Eighth Spring Symposium of Byzantine Studies, St. John's College, University of Oxford, March 2004, ed. M. Mundell Mango. Farnham – Burlington 2009, 371–391.

[79] Cf. already the observations in KAZHDAN – EPSTEIN, Change in Byzantine Culture 27.

with the ancient cultural lands of the Mediterranean.[80] Equally, in contrast to the Eastern Mediterranean, beneficial climatic conditions in Western Europe not only characterised most of the 11th century, but continued until the end of the 13th century (despite several severe famines, see fig. 30), well up to the onset of the so-called "Little Ice Age" and the Plague pandemics of the "calamitous" 14th century. In the Eastern Mediterranean (also in contrast to the Western Mediterranean or to Sicily, where data indicates a continuation of humid conditions until the 14th century), on the contrary, we observe another change of climatic conditions towards less favourable parameters from the end of the 12th century onwards (fig. 7). The weather across the region from the Balkans to Iran became drier – with the exception, again (and this time to its benefit) of Egypt.[81] For North-western Anatolia, proxy data (speleothems) documents a turn

[80] F. SIROCKO – K. DAVID, Das mittelalterliche Wärmeoptimum (1150–1260 AD) und der Beginn der Kleinen Eiszeit (nach 1310 AD) mit ihren kulturhistorischen Entwicklungen, in: Strategien zum Überleben. Umweltkrisen und ihre Bewältigung, ed. F. Daim – D. Gronenborn – R. Schreg. Mainz 2011, 243–254; Medieval Climate Anomaly, ed. E. Xoplaki et alii (= Past Global Changes News 19/1, March 2011; online: http://www.pages-igbp.org/download/docs/NL2011-1_lowres.pdf [21.04.2015]); HOFFMANN, An Environmental History of Medieval Europe 114–154. Cf. also P. TOUBERT, Byzantium and the Mediterranean Agrarian Civilization, in: The Economic History of Byzantium, ed. A. E. Laiou. 3 Vols., Washington, D.C. 2002, 370–380; R. BARTLETT, The Making of Europe. Conquest, Colonization and Cultural Change 950–1350. London 1993, esp. 106–166; W. BEHRINGER, Kulturgeschichte des Klimas. Von der Eiszeit bis zur globalen Erwärmung. Munich 2007, 103–115; P. MALANIMA, Europäische Wirtschaftsgeschichte. 10.–19. Jahrhundert. Vienna – Cologne – Weimar 2010, 100–106; J. D. COTTS, Europe′s Long Twelfth Century (European History in Perspective). Houndsmill 2013, 80–84. For an overview an economic growth in France in that period cf. F. MAZEL, Féodalités, 888–1180 (Histoire de France 2). Paris 2010, 493–539, and J.-Ch. CASSARD, L′âge d′or Capétien, 1180–1328 (Histoire de France 3). Paris 2011, 235–293. For the ecological impact of Latin expansion across the Mediterranean and in the Baltic, cf. A. G. PLUSKOWSKI – A. BOAS – C. GERRARD, The ecology of crusading: Investigating the environmental impact of holy war and colonisation at the frontiers of medieval Europe, Medieval Archaeology 55 (2011) 192–225, and A. PLUSKOWSKI, The Archaeology of the Prussian Crusade. Holy War and Colonisation. London – New York 2013. For an earlier approach along these lines cf. A. W. CROSBY, Ecological Imperialism. The Biological Expansion of Europe, 900–1900 (Studies in Environment and History). Cambridge ²2004.

[81] MANNING, The Roman World and Climate 143–145, with fig. 15 (comparison of precipitation data for Western Europe and the Northern Aegean); SCHÖNWIESE, Klimatologie 304–307; R. GLASER, Klimageschichte Mitteleuropas. 1200 Jahre Wetter, Klima, Katastrophen. Darmstadt ²2008; HOFFMANN, An Environmental History of Medieval Europe 169–174 (on conditions in the Western Mediterranean) and 318–329 (on the beginning of the "Little Ice Age" in Western Europe). For the data from Sicily see L. SADORI et alii, The last 7 millennia of vegetation and climate changes at Lago di Pergusa (central Sicily, Italy). Climate of the Past 9 (2013) 1969–1984. Cf. also I. G. TELELIS, Medieval Warm Periods and the Beginning of the Little Ice Age; RAPHAEL, Climate and Political Climate, esp. 95–110; On famines and crises of subsistence in Western Europe and in the Western Mediterranean in the 11th–14th cent. see esp. the contributions in: Les disettes dans la conjuncture de 1300 en Méditerranée occidentale, ed. M. Bourin – J. Drendel – F. Menant. Rome 2011. On possible effects of temporarily negative climatic conditions on the mobilisation of population for the First Crusade cf. W. Ph. SALVIN, Crusaders in Crisis: towards the Re-Assessment of the Origins and Nature of the "People′s Crusade" of 1095–1096. Imago Temporis. Medium Aevum 4 (2010) 175–199, and P. B. I MONCLÚS, Famines sans frontiers en Occident avant la "conjuncture de 1300". À propos d′une enquête en cours, in: Les disettes dans la conjuncture de 1300, ed. Bourin – Drendel – Menant 37–86, esp. 48–50 on famine in many parts of Western Europe in the years 1093–1095. On the spatial and temporal diversity of climatic conditions during this period cf. H. F. DIAZ et alii, Spatial and Temporal Characteristics of Climate in Medieval Times Revisited. Bulletin of the American Meteorological Society 92 (2011) 1487–1500; Medieval Climate Anomaly, ed. Xoplaki et alii; L. BENSON et alii, Possible impacts of early-11th-, middle-12th-, and late-13th-century droughts on western Native Americans and the Mississippian Cahokians. Quaternary Science Reviews 26 (2007) 336–350. On different trajectories of climatic conditions in the Western and Eastern Mediterranean see ROBERTS et alii, Palaeolimnological evidence for an east–west climate see-saw 23–34, LUTERBACHER et alii, A Review of 2000 Years of Paleoclimatic Evidence, and A. MAURI et alii, The climate of Europe during the Holocene: a gridded pollen-based reconstruction and its multi-proxy evaluation. Quaternary Science Reviews 112 (2015) 109–127. For a global perspective see

towards more arid conditions from the 1180s onwards (fig. 24 and 25). In a precipitation reconstruction for South-western Anatolia for the years 1097 to 2000, Ramzi Touchan and colleagues even have identified the 70 years from 1195 to 1264 as the driest period in their entire record (while the years 1098 to 1167 marked one of the most humid ones) (see also fig. 11, 12, 13 and 14 for data on precipitation in the Northern Aegean). An equal pattern can be found in recently analysed lake sediments from that region, but also in speleothems from Thrace (Uzuntarla Cave, Turkey), in sediments of the Tecer Lake from Cappadocia or in the pollen data from the area of Antioch, with a shift towards drier conditions from the later 12[th] century onwards, which also resulted in a severe famine in Syria between 1178 and 1181 (see also fig. 29). Similar observations have been made for the region of Baghdad on the basis of written sources.[82]

In general, the years of the Angeloi (1185–1204) in comparison with those of the Komnenoi started to become drier (see fig. 13 and 14) and especially on the Balkans, as the tree ring-data from Albania illustrates (fig. 10)[83], also colder. The problems Emperor Isaak II Angelos for instance faced on his campaigns against the insurgents in Bulgaria in the winter 1187/88 coincide with a general colder trend in the second half of the 1180s.[84] As for the 11[th] century one can ask how less beneficial or even adverse climatic parameters contributed to an aggravation of also otherwise crisis prone conditions within a fragmenting Byzantine polity ("programmed for destruction"?) in the period between 1180 and 1204 (see fig. 27 for the frequency of years of internal unrest).[85] Equally interesting would be an answer to the question how the impressive military as well as economic "Byzantine revival" in the exile state of Nicaea in Western Asia Minor (1204–1261) as well as the bloom of the Seljuk state of Konya until the Mongol invasion (1243) can be brought in line with this continuing "dry spell" of the 13[th] century (see fig. 13 and 24), also in comparison with competing polities to the East and to the West.[86] Both aspects are beyond the scope of the present paper and shall be discussed in a further study.

---

[82] GÖKTÜRK, Climate in the Eastern Mediterranean 13–39 (on the speleothem-data from the Sofular Cave in north-western Anatolia); for north-western Anatolia see also C. B. GRIGGS et alii, Regional Reconstruction of Precipitation in the North Aegean and Northwestern Turkey from an Oak Tree-Ring Chronology, AD 1089–1989. *Türkiye Bilimler Akademisi Arkeoloji Dergisi* IX (2006) 141–146; C. B. GRIGGS et alii, A regional high-frequency reconstruction of May–June precipitation in the north Aegean from oak tree rings, A.D. 1089–1989. *International Journal of Climatology* 27 (2007) 1075–1089. For south-western Anatolia see R. TOUCHAN – Ü. AKKEMIK – M. K. HUGHES – N. ERKAN, May–June precipitation reconstruction of southwestern Anatolia, Turkey during the last 900 years from tree rings. *Quaternary Research* 68 (2007) 196–202; I. HEINRICH – R. TOUCHAN et alii, Winter-to-spring temperature dynamics in Turkey derived from tree rings since AD 1125. *Climate Dynamics* 41 (2013) 1685–1701. For lake sediments in south-western Anatolia cf. BAKKER et alii, Climate, people, fire and vegetation. For the Uzuntarla Cave in Thrace: LUTERBACHER et alii, A Review of 2000 Years of Paleoclimatic Evidence 104–106; GÖKTÜRK, Climate in the Eastern Mediterranean 67–80. For Lake Tecer: KUZUCUOGLU et alii, Mid- to late-Holocene climate change. For Antioch and Syria: KANIEWSKI et alii, The Medieval Climate Anomaly, RAPHAEL, Climate and Political Climate 76–87 (on the drought of 1178–1181), MORONY, Michael the Syrian, and WIDELL, Historical Evidence for Climate Instability (also with statistical calculations on the frequency of extreme weather events). For Baghdad: VOGT – GLASER – LUTERBACHER et alii, Assessing the Medieval Climate Anomaly.

[83] For the data see: PAGES 2k Network consortium, Database S1 - 11 April 2013 version: http://www.pages-igbp.org/workinggroups/2k-network (21.04.2015).

[84] Niketas Choniates 398, 30–42 (ed. J. A. VAN DIETEN, Nicetae Choniatae Historia [*CFHB* 11/1–2]. Berlin 1975); O City of Byzantium: Annals of Niketas Choniatēs, transl. by H. MAGOULIAS. Detroit 1984, 219: "The emperor [Isaak II Angelos] decided once again to enter Zagora to attempt to force the Vlachs to submit. Leaving Philippopolis, he came to Triaditza [Sofia]; for he had heard that the paths from there to the Haimos were not too difficult to travel, in some places being straight and level, and that there were abundant water supplies and pasturage by the wayside for the pack animals should one pass over them in season. However, as the sun was passing the meridian of the winter solstice in its course [December 1187], rivers were freezing over, the cold north wind prevailed in that region, and so much snow had fallen that it covered the face of the earth and packed ravines and even blocked the doors of buildings, he postponed the campaign until the coming of spring. The army was left encamped in that province while the emperor returned with his light-armed troops to the imperial city, where he enjoyed himself at the horse races and delighted in the spectacles."; cf. TELELIS, Μετεωρολογικά φαινόμενα nr 591. See also STEPHENSON, Byzantium's Balkan Frontier 288–294; ANGOLD, The Byzantine Empire 304–307.

**Conclusion**

As our evaluation of Ronnie Ellenblum´s scenario has demonstrated, neither written nor archaeological nor natural scientific evidence allows us to speak of a general "Collapse" across the Eastern Mediterranean and the Near East in the 11[th] century. On the contrary, we encounter high regional variations across and within political boundaries. Where evidence indicates a significant agricultural/demographic decline and/or a change of political regime (as in parts of Iran, Vaspurakan or in Fatimid Egypt), climate-induced stress can be identified as a significant, but not as a sufficient cause of these developments.[87] It is one factor within diverse and specific combinations of environmental, socio-economic and political dynamics. At the same time, the cases of Byzantium or even of Fatimid Egypt point at a relative "resilience" of polities when faced with these dynamics. As "resilience" one can understand "the capacity to absorb sudden shocks, to adapt to longer-term changes in socioeconomic conditions, and to resolve societal disputes sustainably without catastrophic breakdown."[88] As both medieval data and data from the mid-20[th] century (for Syria and Turkey) indicate, years of weather extremes and bad harvests often tend to cluster (as in the seven years of famine in the story of Joseph in Egypt[89]), thus testing the resilience of agricultural communities and political regimes, sometimes beyond its limits (see fig. 26–30 and 31–33).[90] Both in the Byzantine Empire and the Fatimid Caliphate, the 11th century crises were characterised by severe internal conflict (leading to a replacement of parts of the ruling elites) and the loss of considerable territories (large areas of Anatolia in the Byzantine, Syria and Palestine in the Fatimid case). Yet core frameworks of political and religious power and "imperial grandeur" were maintained. Only towards the end of the period under consideration in the present paper, both empires encountered another combination of internal conflict and exterior pressure, accompanied by aspects of environmental stress (the "cold and dry years" of the Angeloi, frequent disastrous Nile floods in Egypt from 1164 onwards). This time it lead to actual "collapse" in 1171 respectively 1204, ultimately due to foreign invasions and the establishment of new "foreign" regimes in Cairo and Constantinople.[91] A more detailed comparison of these periods of crisis in the 11th and in the 12th centuries for both polities would also deserve a further study.

From the comparison of regional variability via the level of polities these phenomena also suggest global comparisons with other macro-regions and polities to evaluate the "performance" of Byzantine or Eastern Mediterranean medieval economies against the background of similar or different environmental, socio-economic and geo-political conditions[92] – as Mark Whittow does in his assessment of the Komnenian empire vis-à-vis the high medieval "boom years" of Western Europe (see above). This would contribute to the more far-reaching debate on a possible "Great Divergence" of socio-economic evolution between Western Europe and other regions on the Old World such as China, India, Eastern Europe or the Eastern Mediterranean and Near East, starting in the early modern or even in the medieval period and ultimately leading to the political and economic "pre-dominance" of Western European countries in the 19th–20th century.[93] New data on climatic and environmental parameters (which already have been included in earlier analyses) can enrich the set of factors and arguments brought forth in these debates and invite to undertake a more elaborate study, which is beyond the scope of the present paper.

---

[90] For the correlation of a series of droughts and internal unrest in Syria since 2006 cf. C. P. KELLEY et alii, Climate change in the Fertile Crescent and implications of the recent Syrian drought. *Proceedings of the National Academy of Sciences* 112/11 (2015) 3241–3246; P. H. GLEICK, Water, Drought, Climate Change, and Conflict in Syria. *Weather, Climate, and Society* 6 (2014) 331–340.

[91] On the collapse of the Fatimid Caliphate cf. HALM, Kalifen und Assassinen; A.-M. EDDÉ, Saladin. Cambridge, Mass. – London 2011, 13–55. For the Nile flood data cf. HASSAN, Extreme Nile floods and famines in Medieval Egypt; also the coming into power of the Fatimids in Egypt in 969 was accompanied by a preceding series of disastrously low Nile floods.

[92] Cf. Natural Experiments of History, ed. J. Diamond – J. A. Robinson. Cambridge, Mass. – London 2010, esp. 1–14.

[93] Cf. Mark Whittow´s upcoming book "The Feudal Revolution", looking at the transformation of Europe and the Near East between 950 and 1250 (cf. http://www.history.ox.ac.uk/faculty/staff/profile/whittow.html [21.04.2015]), and HARVEY, The Byzantine Economy in an International Context. For the concept of "divergence" cf. esp. K. POMERANZ, The Great Divergence: Europe, China, and the Making of the Modern World Economy. Princeton 2000. For criticism cf. J.-L. ROSENTHAL – R. BIN WONG, Before and beyond Divergence. The Politics of Economic Change in China and Europe. Cambridge, Mass. - London 2011. On India see T. ROY, India in the World Economy: From Antiquity to the Present. Cambridge 2012; P. PARTHASARATHI, Why Europe Grew Rich and Asia Did Not: Global Economic Divergence, 1600–1850. Cambridge 2011. For the debate on a divergence of the Islamic world see. FELDBAUER, Die islamische Welt, esp. 12–30; T. KURAN, The Long Divergence. How Islamic Law held back the Middle East. Princeton – Oxford 2011. For an early to high medieval special development of Western Europe cf. M. MITTERAUER, Warum Europa? Mittelalterliche Grundlagen eines Sonderwegs. Munich ⁵2009. On the idea of a late medieval divergence between Western Europe, Eastern Europe and the Eastern Mediterranean cf. The Origins of Backwardness in Eastern Europe. Economic and Politics from the Middle Ages until the Early Twentieth Century, ed. D. Chirot. Berkeley – Los Angeles – Oxford 1989; Ş. PAMUK, The Black Death and the Origins of the "Great Divergence" across Europe, 1300–1600. *European Review of Economic History* 11 (2007) 289–317; Europas Aufstieg. Eine Spurensuche im späten Mittelalter, ed. Th. Ertl (*Expansion. Interaktion. Akkulturation. Globalhistorische Skizzen* 23). Vienna

Maybe in contrast to what one may have expected, the increasing number of proxy data demands even more balanced, cautious and complex scenarios for the impact of changing climatic conditions on economic, social and political ones than in earlier (but also some parts of current) scholarship. Against this background historians, based on their expertise with the analysis and exploration of complex and sometimes contradictory evidence and phenomena, are able to contribute to a more nuanced and appropriate evaluation of the actual impact of environmental and climatic change on past human society beyond models of mere linear causation or general collapse.[94] The recently intensified dialogue of historians, archaeologists and natural scientists in this regard[95] therefore should be extended for the field of Byzantine Studies, in close cooperation with specialists across the entire medieval world.

## APPENDIX 1: PROXY DATA

### 1. Palaeo-environmental data sources used for the present study

## 2. Temperature proxy data sites

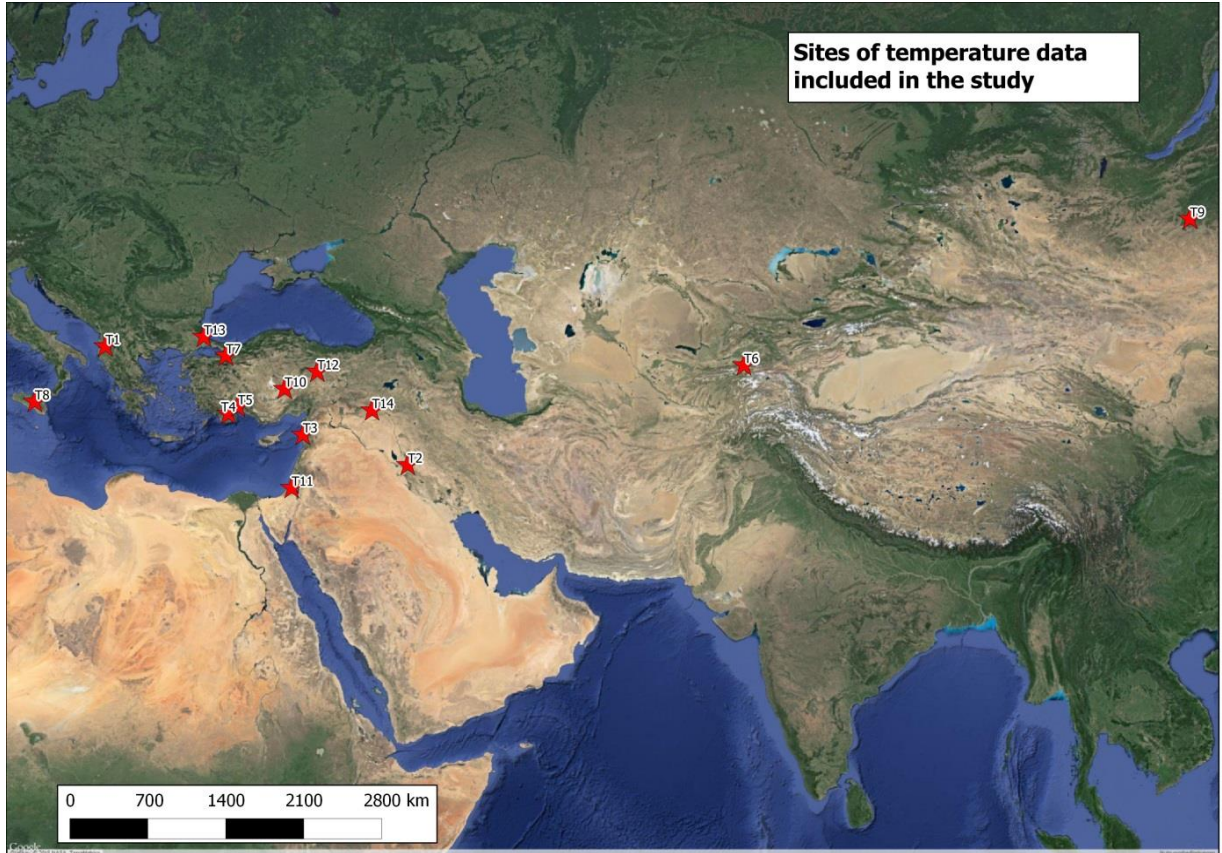

**Fig. 1:** Sites of temperature proxy data included in the study (basemap: GoogleSatellite; created with QuantumGIS*); for the identification of the sites see the table below

| ID | Locality | Latitude | Longitude | Type of Data |
|----|----------|----------|-----------|--------------|
| T1 | **Albania** | 41.00 | 20.00 | Tree-rings |
| T2 | **Baghdad** | 33.333333 | 44.383333 | Textual evidence |
| T3 | **Jableh** | 35.359167 | 35.921111 | Pollen |
| T4 | **Jsibeli** | 36.740278 | 29.916667 | Tree-rings, Isotopes |
| T5 | **Kocain Cave** | 37.202765 | 30.676059 | Speleothem, Isotopes |
| T6 | **Kyrgystan** | 39.83 | 71.50 | Tree-rings |
| T7 | **Lake Iznik** | 40.429167 | 29.721111 | Lake sediments, Isotopes, Pollen |
| T8 | **Lake Pergusa** | 37.513933 | 14.305894 | Pollen |
| T9 | **Mongolia** | 48.35 | 107.47 | Tree-rings |
| T10 | **Nar Gölü** | 38.340000 | 34.456389 | Lake sediments, Isotopes, Pollen |
| T11 | **Soreq Cave** | 31.755833 | 35.023333 | Speleothem, Isotopes |
| T12 | **Tecer Lake** | 39.430957 | 37.084702 | Lake sediments |
| T13 | **Uzuntarla Cave** | 41.583583 | 27.943056 | Speleothem, Isotopes |
| T14 | **Wadi Jarrah** | 36.959344 | 41.50365 | Pollen |

**Table 1:** Sites of temperature proxy data included in the study

## 3. Precipitation proxy data sites

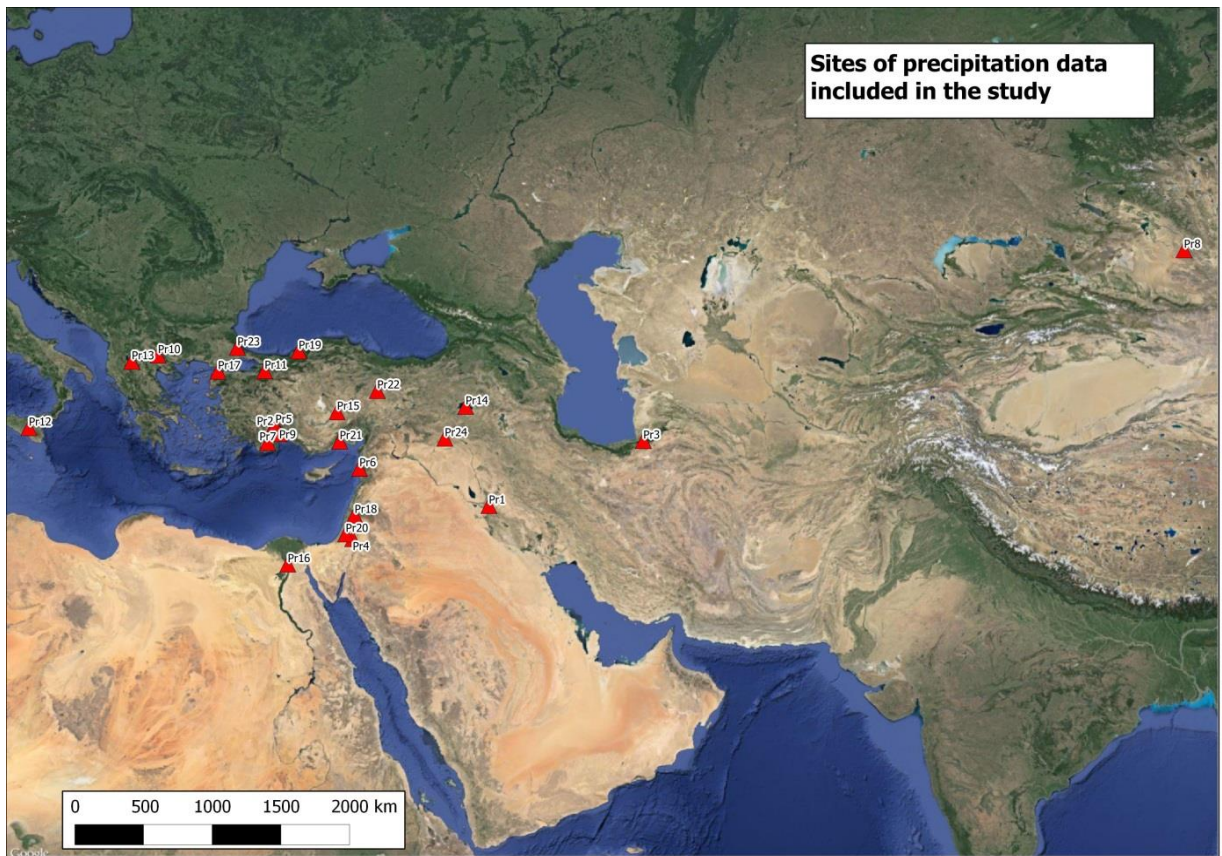

**Fig. 2:** Sites of precipitation proxy data included in the study (basemap: GoogleSatellite; created with QuantumGIS*); for the identification of the sites see the table below

| ID | Locality | Latitude | Longitude | Type of Data |
|----|----------|----------|-----------|--------------|
| Pr1 | **Baghdad** | 33.333333 | 44.383333 | Textual evidence |
| Pr2 | **Bereket** | 37.547278 | 30.282030 | Pollen |
| Pr3 | **Caspian Sea** | 36.833333 | 54.483333 | Pollen |
| Pr4 | **Dead Sea** | 31.490833 | 35.479722 | Pollen |
| Pr5 | **Gravgaz** | 37.657806 | 30.425215 | Pollen |
| Pr6 | **Jableh** | 35.359167 | 35.921111 | Pollen |
| Pr7 | **Jsibeli** | 36.740278 | 29.916667 | Tree-rings, Isotopes |
| Pr8 | **Kesang Cave** | 46.217103 | 89.838717 | Speleothem, Isotopes |
| Pr9 | **Kocain Cave** | 37.202765 | 30.676059 | Speleothem, Isotopes |
| Pr10 | **Lake Doirani** | 41.204722 | 22.7475 | Lake sediments |
| Pr11 | **Lake Iznik** | 40.429167 | 29.721111 | Lake sediments, Isotopes, Pollen |
| Pr12 | **Lake Pergusa** | 37.513933 | 14.305894 | Pollen |
| Pr13 | **Lake Prespa** | 40.897222 | 21.032222 | Lake sediments, pollen |
| Pr14 | **Lake Van** | 38.616667 | 42.866667 | Lake sediments, Pollen, Isotopes, Charcoal |
| Pr15 | **Nar Gölü** | 38.340000 | 34.456389 | Lake sediments, Isotopes, Pollen |
| Pr16 | **Nile** | 30.05 | 31.233333 | Nile flood documentation |
| Pr17 | **North Aegean** | 40.408333 | 26.673611 | Tree-rings |
| Pr18 | **Sea of Galilee** | 32.811389 | 35.604444 | Pollen, Charcoal |
| Pr19 | **Sofular Cave** | 41.416852 | 31.951094 | Speleothem, Isotopes |

| Pr20 | **Soreq Cave** | 31.755833 | 35.023333 | Speleothem, Isotopes |
|------|----------------|-----------|-----------|----------------------|
| Pr21 | **Southern Anatolia** | 36.81 | 34.629722 | Tree-rings |
| Pr22 | **Tecer Lake** | 39.430957 | 37.084702 | Lake sediments |
| Pr23 | **Uzuntarla Cave** | 41.583583 | 27.943056 | Speleothem, Isotopes |
| Pr24 | **Wadi Jarrah** | 36.959344 | 41.50365 | Pollen |

**Table 2:** Sites of precipitation proxy data included in the study

## 4. Pollen data sites

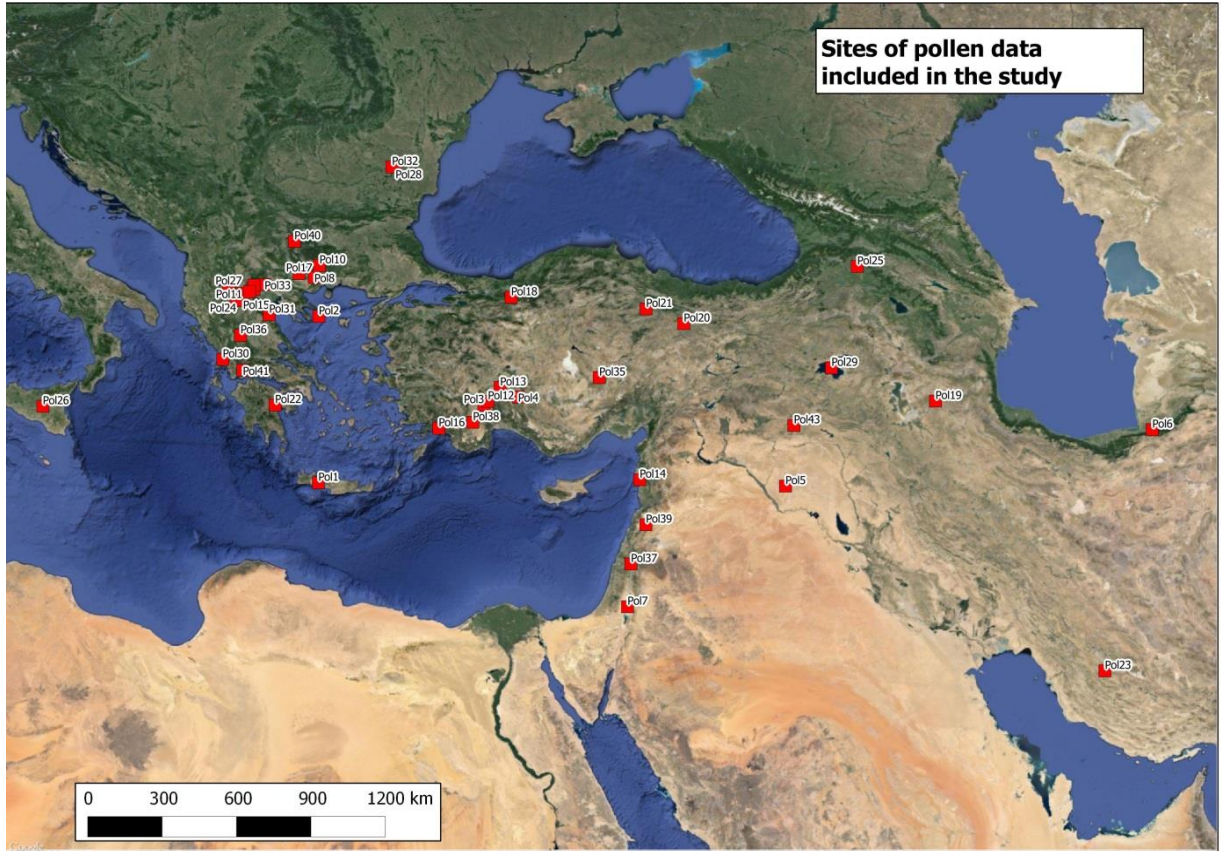

**Fig. 3:** Sites of pollen data included in the study (basemap: GoogleSatellite; created with QuantumGIS*); for the identification of the sites see the table below

| ID | Locality | Latitude | Longitude | Type of Data |
|------|--------------|-----------|-----------|--------------|
| Pol1 | **Asi Gonia** | 35.266667 | 24.283333 | Pollen |
| Pol2 | **Athos** | 40.068434 | 24.300837 | Pollen |
| Pol3 | **Bereket** | 37.547278 | 30.282030 | Pollen |
| Pol4 | **Beysehir Gölü** | 37.782222 | 31.518889 | Pollen |
| Pol5 | **Bouara** | 35.166667 | 41.2 | Pollen |
| Pol6 | **Caspian Sea** | 36.833333 | 54.483333 | Pollen |
| Pol7 | **Dead Sea** | 31.490833 | 35.479722 | Pollen |
| Pol8 | **Drama** | 41.151389 | 24.139167 | Geoarchaeology, alluvial deposits, Pollen |
| Pol9 | **Edessa** | 40.818056 | 21.9525 | Pollen |
| Pol10 | **Elatia** | 41.48 | 24.325 | Pollen |
| Pol11 | **Flampouro** | 40.71 | 21.52 | Pollen |

| Pol12 | **Gravgaz** | 37.657806 | 30.425215 | Pollen |
|---|---|---|---|---|
| Pol13 | **Hoyran Gölü** | 38.056667 | 30.866111 | Pollen |
| Pol14 | **Jableh** | 35.359167 | 35.921111 | Pollen |
| Pol15 | **Khimaditis** | 40.599410 | 21.558901 | Pollen |
| Pol16 | **Köycegiz Gölü** | 36.875 | 28.64167 | Pollen |
| Pol17 | **Lailias Mt.** | 41.255157 | 23.587311 | Polen |
| Pol18 | **Lake Abant** | 40.6 | 31.266667 | Pollen |
| Pol19 | **Lake Almalou** | 37.664037 | 46.631944 | Pollen |
| Pol20 | **Lake Demiyurt** | 39.867771 | 37.517083 | Pollen |
| Pol21 | **Lake Kaz** | 40.278927 | 36.151589 | Pollen |
| Pol22 | **Lake Lerna** | 37.55 | 22.716667 | Pollen |
| Pol23 | **Lake Maharlou** | 29.473383 | 52.767002 | Lake sediments, pollen |
| Pol24 | **Lake Orestias** | 40.515 | 21.3 | Pollen |
| Pol25 | **Lake Paravani** | 41.45 | 43.8 | Lake sediments, Pollen |
| Pol26 | **Lake Pergusa** | 37.513933 | 14.305894 | Pollen |
| Pol27 | **Lake Prespa** | 40.897222 | 21.032222 | Lake sediments, pollen |
| Pol28 | **Lake Srebarna** | 44.114444 | 27.078056 | Pollen |
| Pol29 | **Lake Van** | 38.616667 | 42.866667 | Lake sediments, Pollen, Isotopes, Charcoal |
| Pol30 | **Lake Voulkaria** | 38.860134 | 20.822186 | Pollen |
| Pol31 | **Litokhoro** | 40.118333 | 22.503611 | Pollen |
| Pol32 | **mire Garvan** | 44.11694 | 26.95 | Pollen |
| Pol33 | **Mt. Paiko** | 40.953197 | 22.335828 | Pollen |
| Pol34 | **Mt. Voras** | 40.936 | 21.95 | Pollen |
| Pol35 | **Nar Gölü** | 38.340000 | 34.456389 | Lake sediments, Isotopes, Pollen |
| Pol36 | **Pertouli** | 39.537658 | 21.465816 | Pollen |
| Pol37 | **Sea of Galilee** | 32.811389 | 35.604444 | Pollen, Charcoal |
| Pol38 | **Söğüt Gölü** | 37.05 | 29.8833 | Pollen |
| Pol39 | **Southern Bekaa Valley** | 34.008889 | 36.145278 | Pollen |
| Pol40 | **Suho Ezero** | 42.133333 | 23.416667 | Pollen |
| Pol41 | **Trikhonis** | 38.55 | 21.55 | Pollen |
| Pol42 | **Vegoritis** | 40.760556 | 21.789167 | Pollen |
| Pol43 | **Wadi Jarrah** | 36.959344 | 41.50365 | Pollen |

**Table 3:** Sites of pollen data included in the study

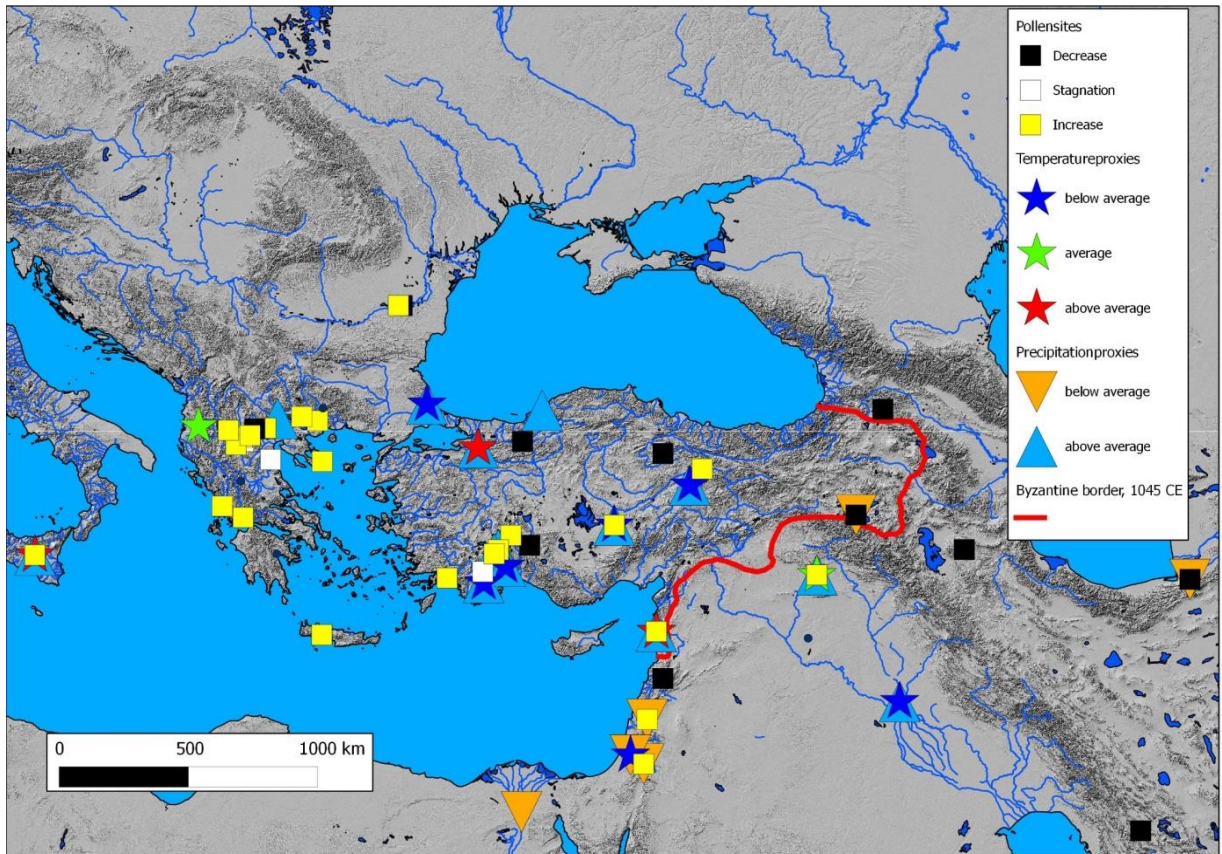

**Fig. 4**: Reconstruction of climatic conditions and general trends in agricultural production in the Byzantine Empire in the 11[th] cent. AD (red line: eastern border at ca. 1064; created with QuantumGIS*; data: see Appendix 1)

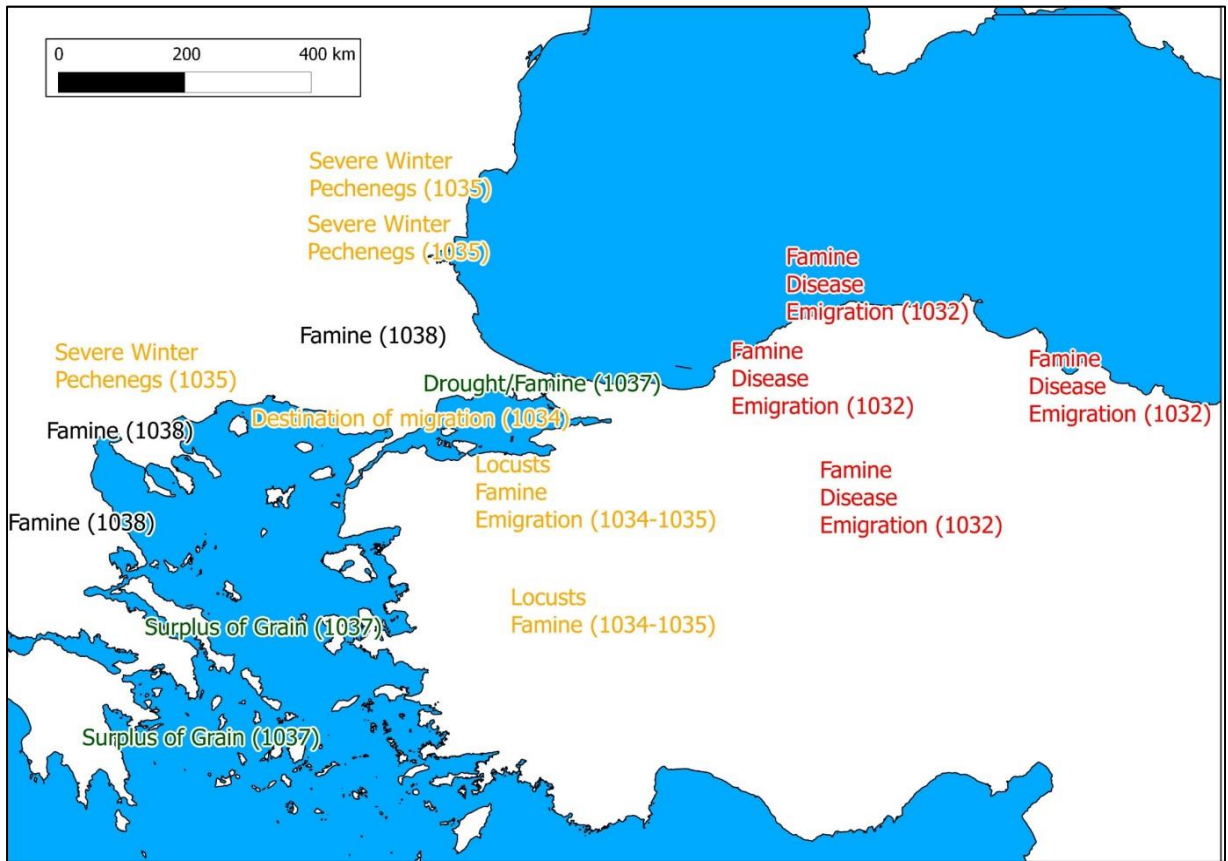

**Fig. 5:** Extreme weather events, famine, disease and invasions in the Byzantine Empire, 1032–1038 AD (created with QuantumGIS*)

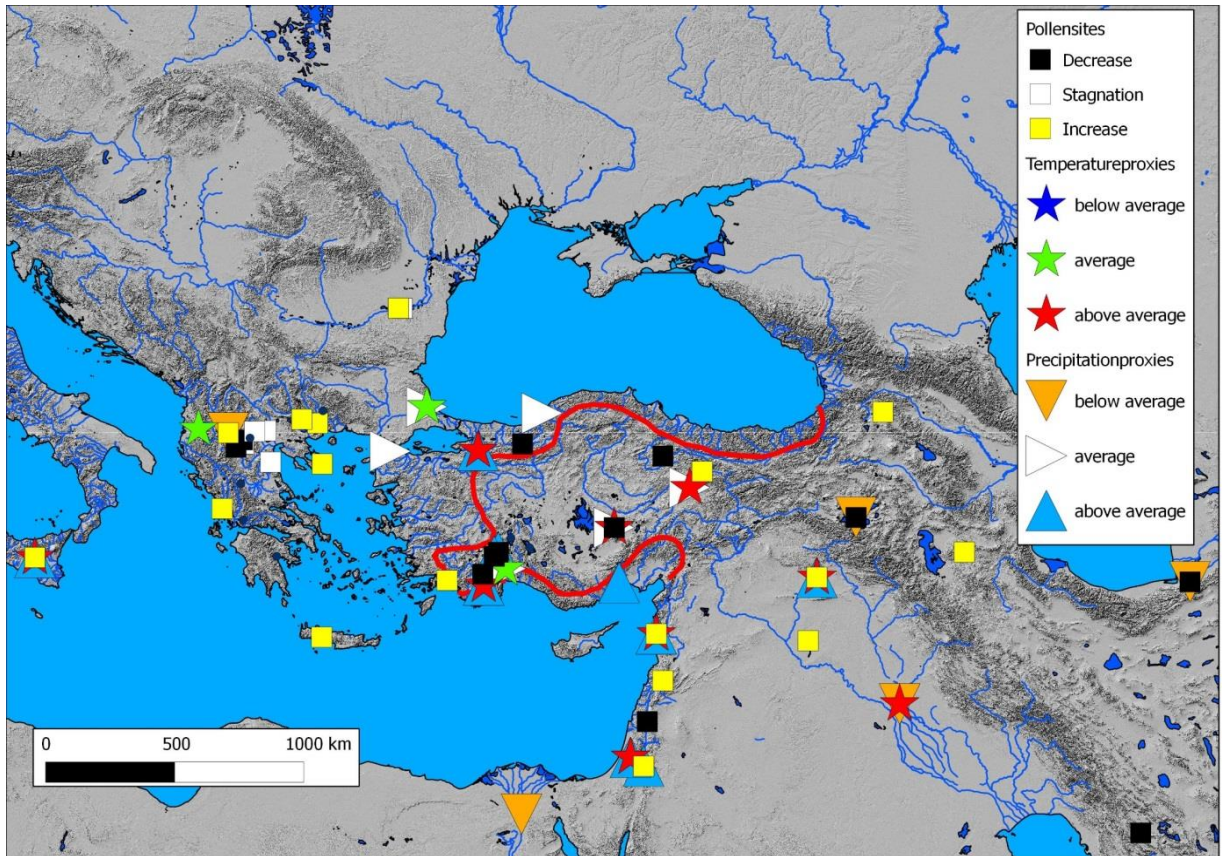

**Fig. 6**: Reconstruction of climatic conditions and general trends in agricultural production in the Near East in the 12<sup>th</sup> cent. AD (red line: eastern border of the Byzantine Empire at ca. 1176; created with QuantumGIS*; data: see Appendix 1)

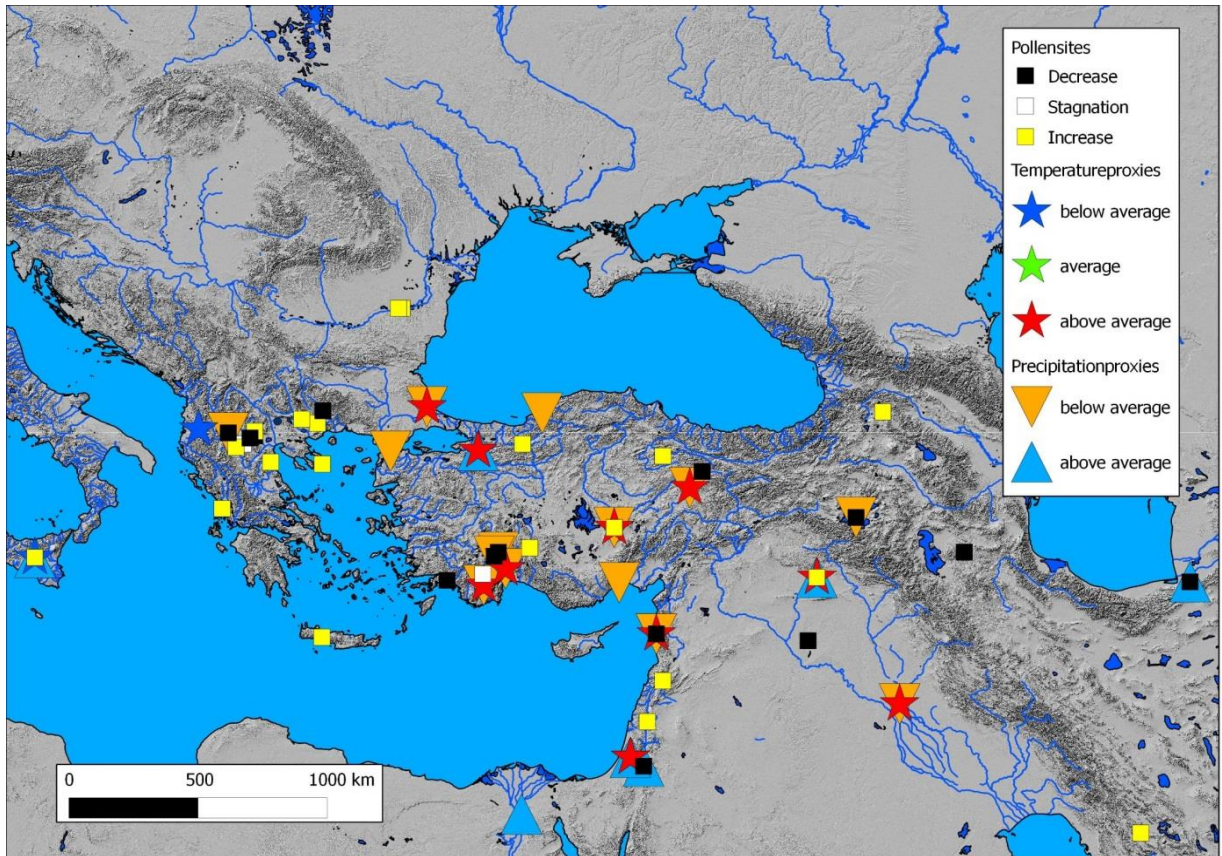

**Fig. 7**: Reconstruction of climatic conditions and general trends in agricultural production in the Near East in the 13th cent. AD (created with QuantumGIS*; data: see Appendix 1)

APPENDIX 3: SELECTED TREE RING DATA

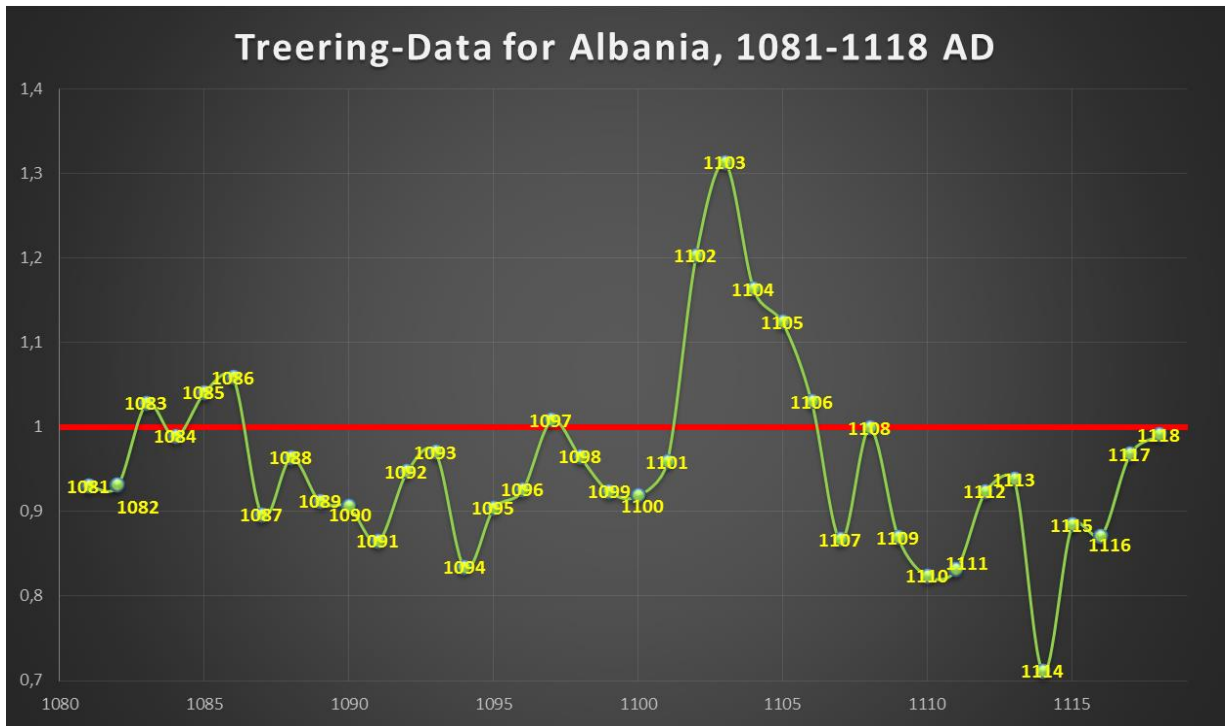

**Fig. 8:** Tree ring-data for Albania, 1081–1118 AD, as proxy for temperature conditions (data: PAGES 2k Network consortium, Database S1 – 11 April 2013 version: http://www.pages-igbp.org/workinggroups/2k-network [21.04.2015])

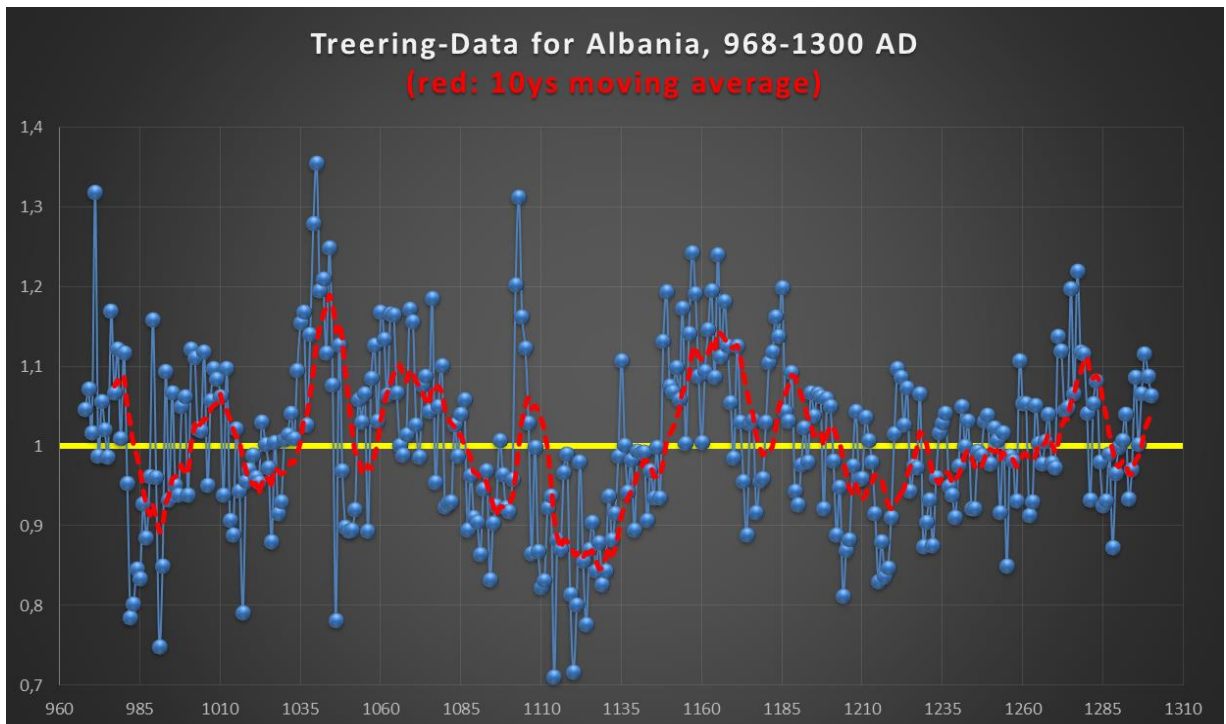

**Fig. 9:** Tree ring-data for Albania, 968–1300 AD, as proxy for temperature conditions (data: PAGES 2k Network consortium, Database S1 – 11 April 2013 version: http://www.pages-igbp.org/workinggroups/2k-network [21.04.2015])

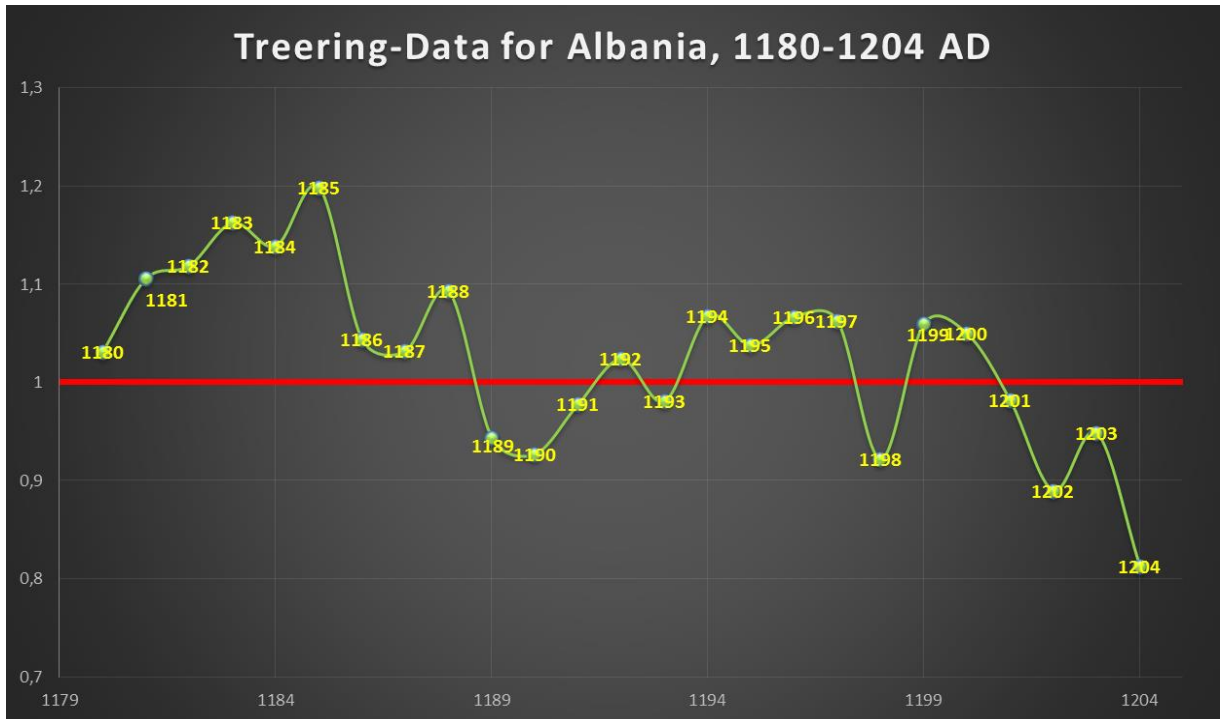

**Fig. 10:** Tree ring-data for Albania, 1180–1204 AD, as proxy for temperature conditions (data: PAGES 2k Network consortium, Database S1 – 11 April 2013 version: http://www.pages-igbp.org/workinggroups/2k-network [21.04.2015])

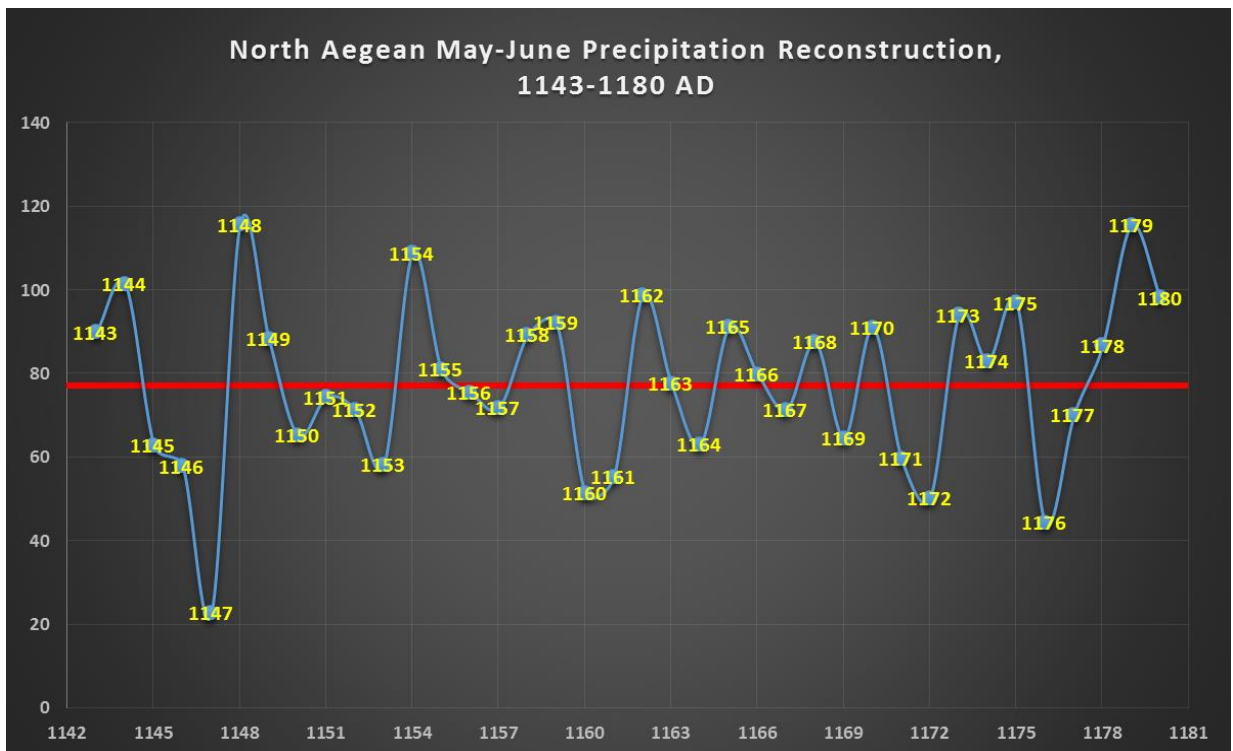

**Fig. 11:** Tree ring-based reconstruction of May-June precipitation in the Northern Aegean, 1143–1180 AD (data: Griggs *et alii*, Regional Reconstruction of Precipitation; Griggs *et alii*, A regional high-frequency reconstruction of May–June precipitation)

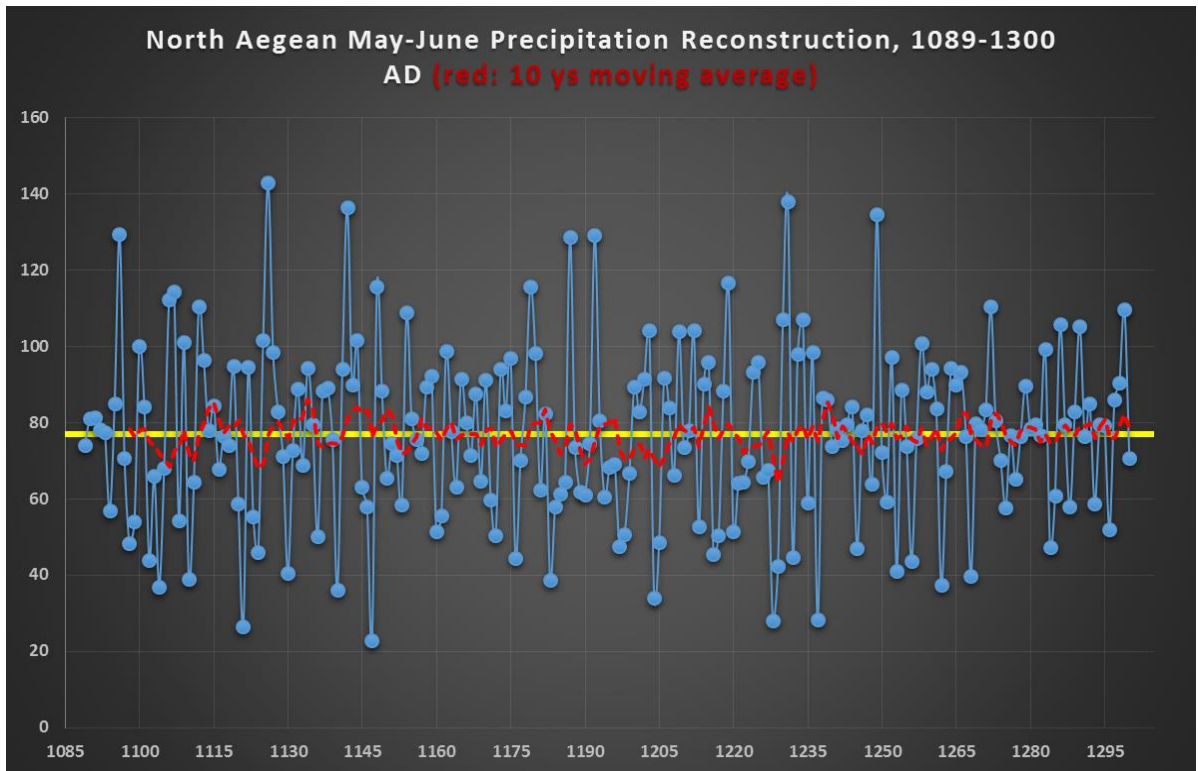

**Fig. 12:** Tree ring-based reconstruction of May-June precipitation in the Northern Aegean, 1089–1300 AD (data: GRIGGS *et alii*, Regional Reconstruction of Precipitation; GRIGGS *et alii*, A regional high-frequency reconstruction of May–June precipitation)

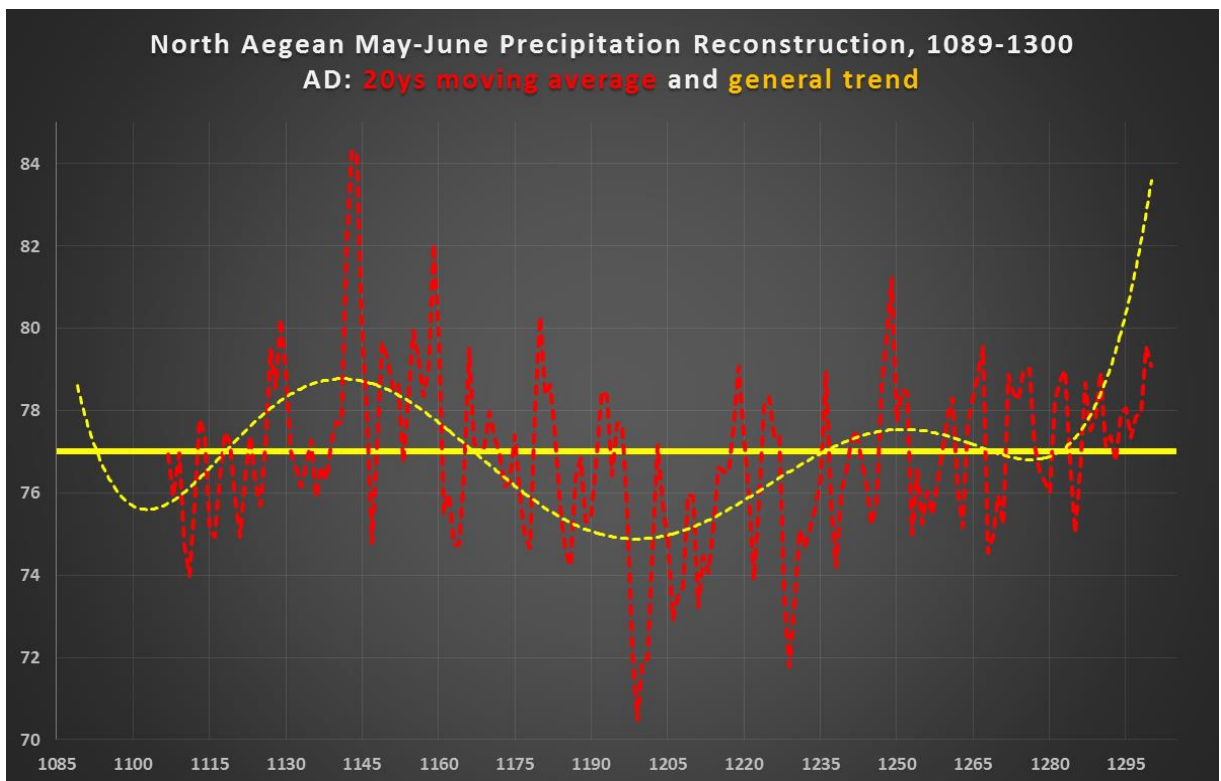

**Fig. 13:** Tree ring-based reconstruction of May-June precipitation in the Northern Aegean, 1089–1300 AD: 20 years moving average and general trend (data: GRIGGS *et alii*, Regional Reconstruction of Precipitation; GRIGGS *et alii*, A regional high-frequency reconstruction of May–June precipitation)

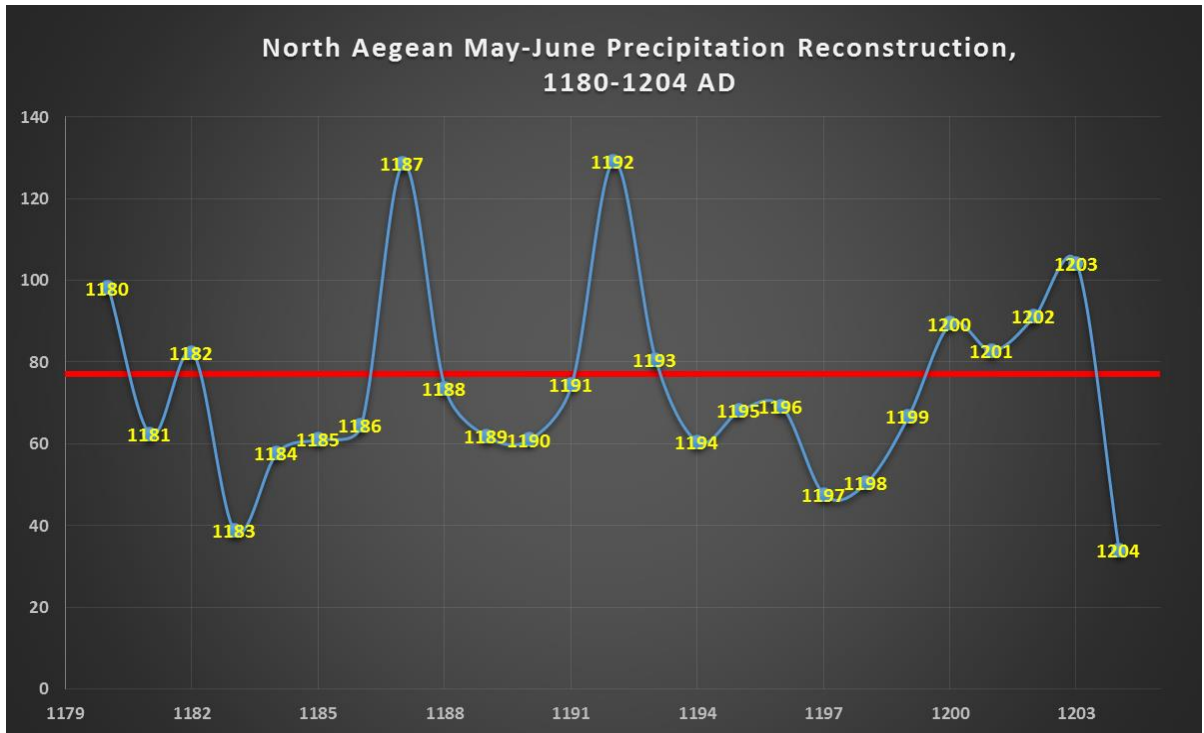

**Fig. 14:** Tree ring-based reconstruction of May-June precipitation in the Northern Aegean, 1180–1204 AD (data: GRIGGS *et alii*, Regional Reconstruction of Precipitation; GRIGGS *et alii*, A regional high-frequency reconstruction of May–June precipitation)

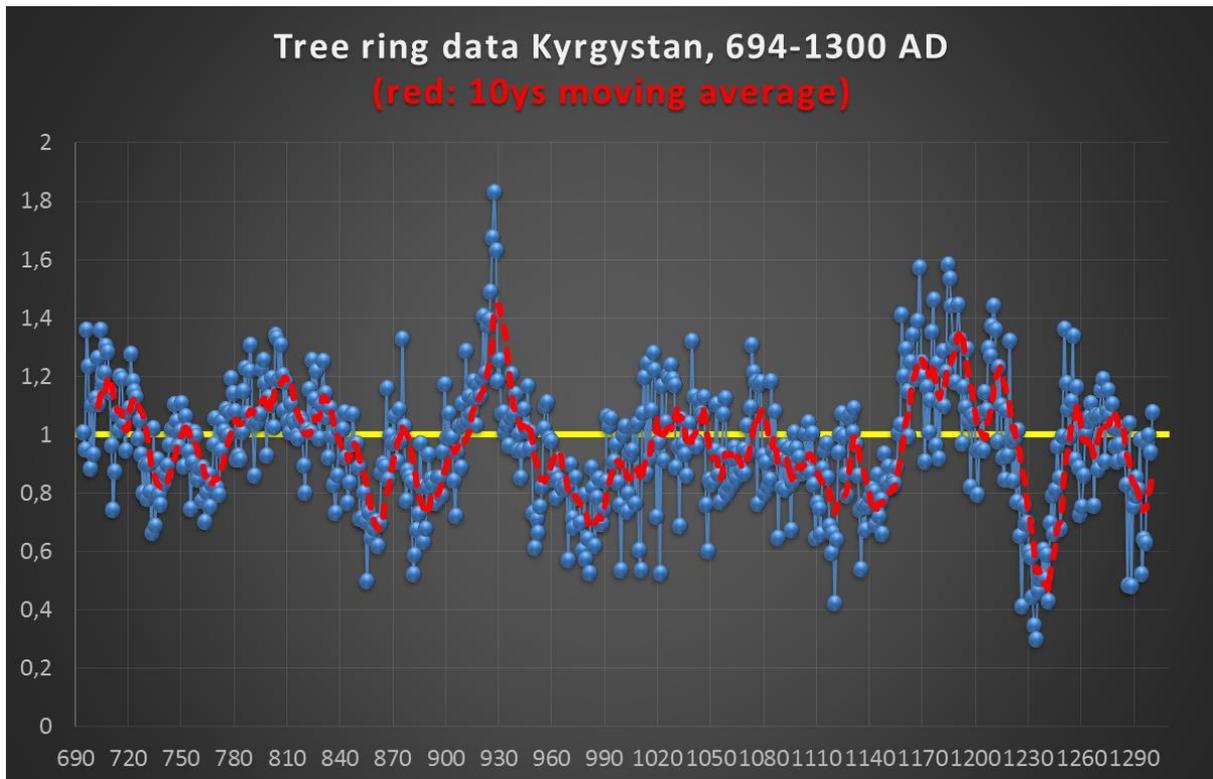

**Fig. 15:** Tree ring data from Kyrgystan, 694–1300 AD, as proxy for temperature and precipitation conditions (data: PAGES 2k Network consortium, Database S1 – 11 April 2013 version: http://www.pages-igbp.org/workinggroups/2k-network [21.04.2015])

## APPENDIX 4: SELECTED POLLEN DATA

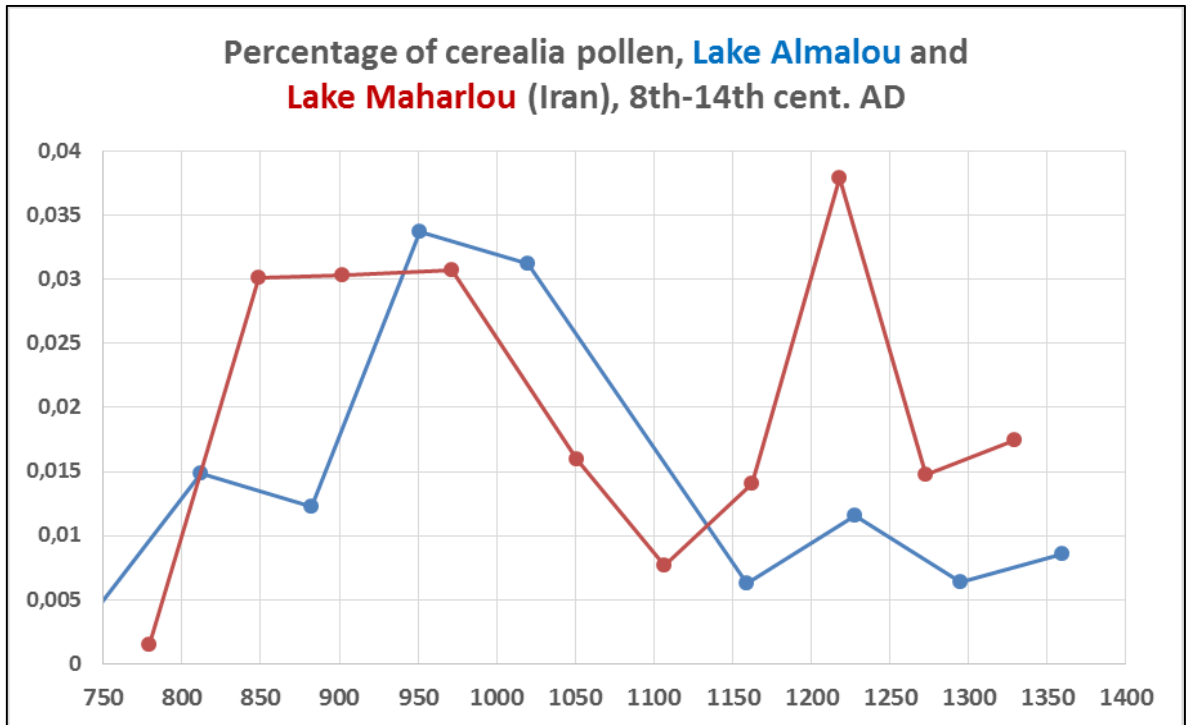

**Fig. 16:** Percentage of cerealia pollen in the samples from Lake Almalou and Lake Maharlou (both Iran), 8th–14th cent. AD (data: DJAMALI *et alii*, A late Holocene pollen record from Lake Almalou; DJAMALI *et alii*, Notes on Arboricultural and Agricultural Practices; EPD: *European Pollen Database* [http://www.europeanpollendatabase.net/] [21.04.2015])

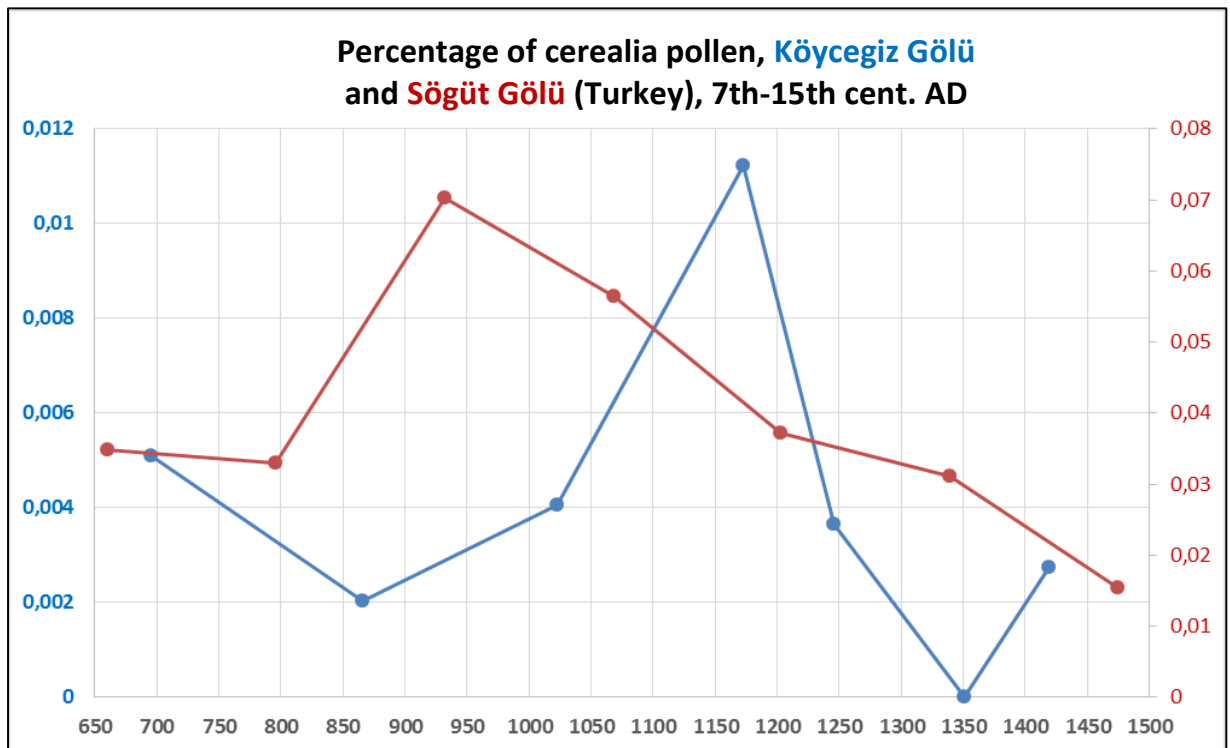

**Fig. 17:** Percentage of cerealia pollen in the samples from Köycegiz Gölü and Sögüt Gölü (both Turkey), 8th–15th cent. AD (data: EPD: *European Pollen Database* [http://www.europeanpollendatabase.net/] [21.04.2015])

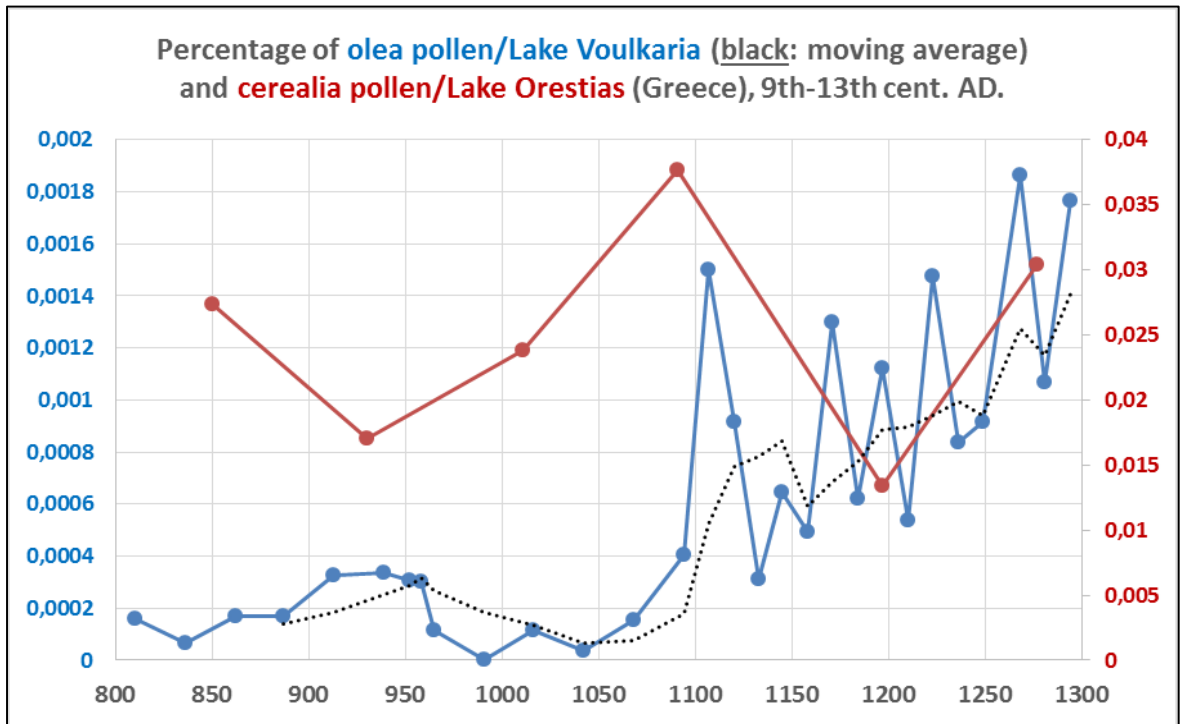

**Fig. 18:** Percentage of olea pollen in the sample from Lake Voulkaria and of cerealia pollen in the sample from Lake Orestias (both Greece), 9[th]–13[th] cent. AD (data: EPD: *European Pollen Database* [http://www.europeanpollendatabase.net/] [21.04.2015])

## APPENDIX 5: LAKE VAN/VASPURAKAN DATA

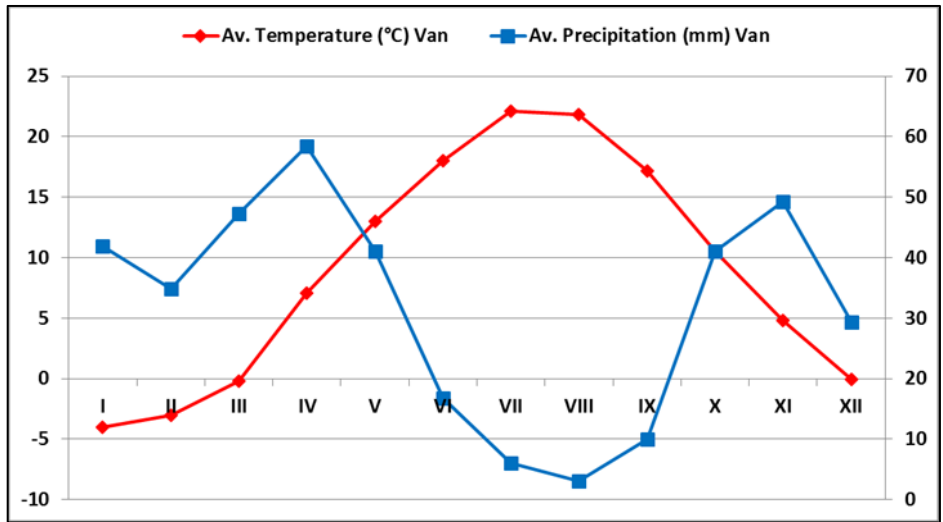

**Fig. 19:** Average monthly temperature (°C) and average monthly precipitation (mm) in the city of Van (Turkey) today (data: http://www.climate-charts.com [21.04.2015])

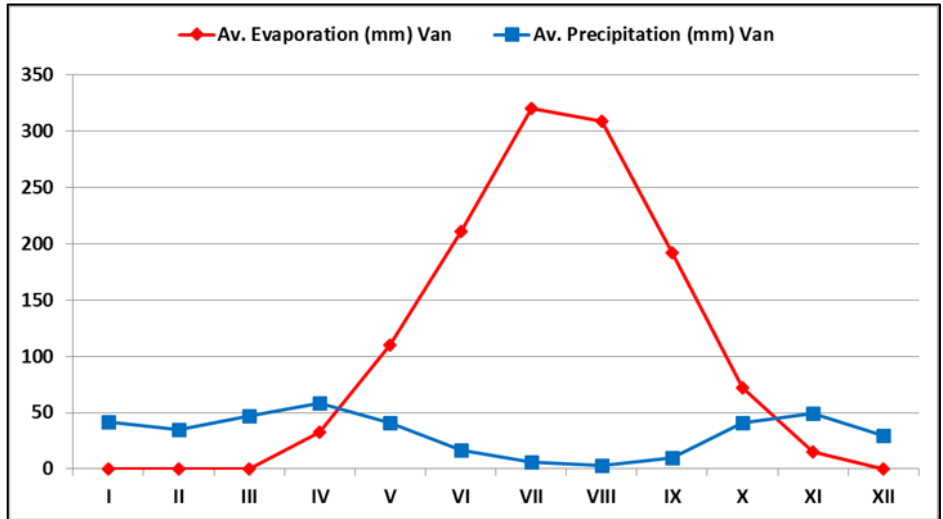

**Fig. 20:** Average monthly evaporation (mm) and average monthly precipitation (mm) in the city of Van (Turkey) today (data: http://www.climate-charts.com [21.04.2015])

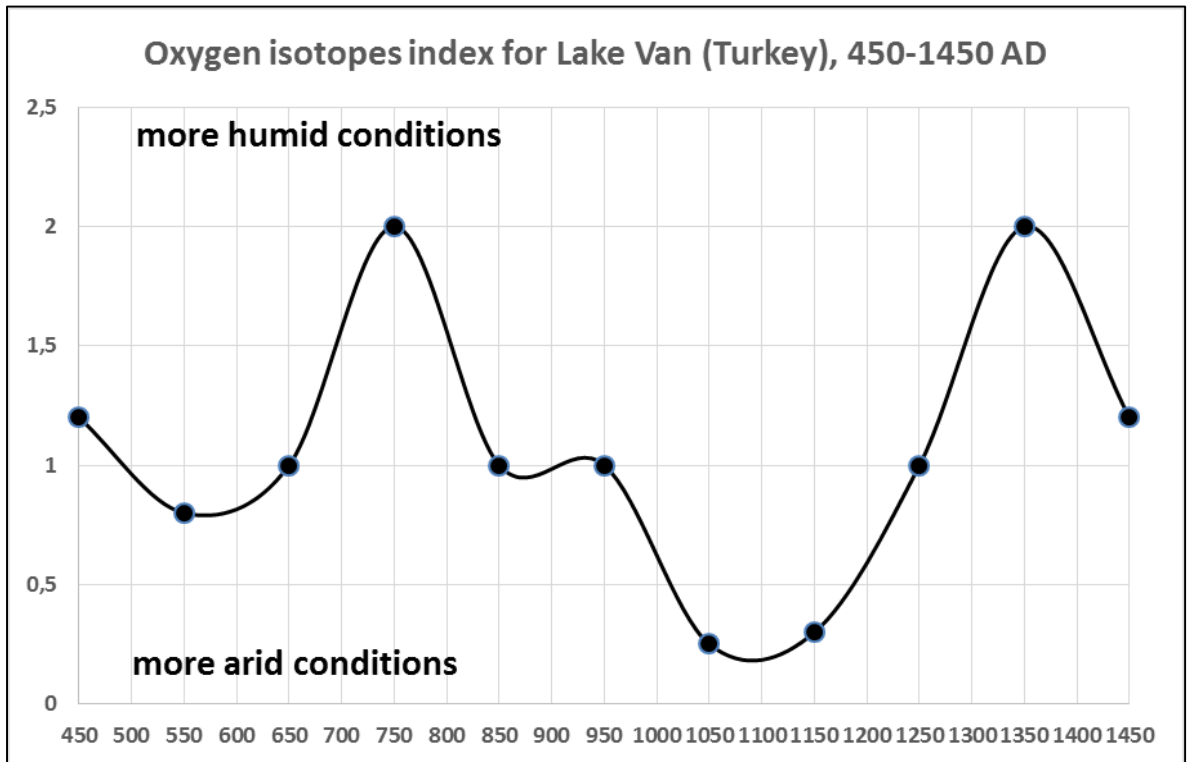

**Fig. 21:** Oxygen isotopes index for core samples from sediments in Lake Van (Turkey), 450–1450 AD (data: WICK – LEMCKE – STURM, Evidence of Lateglacial and Holocene climatic change)

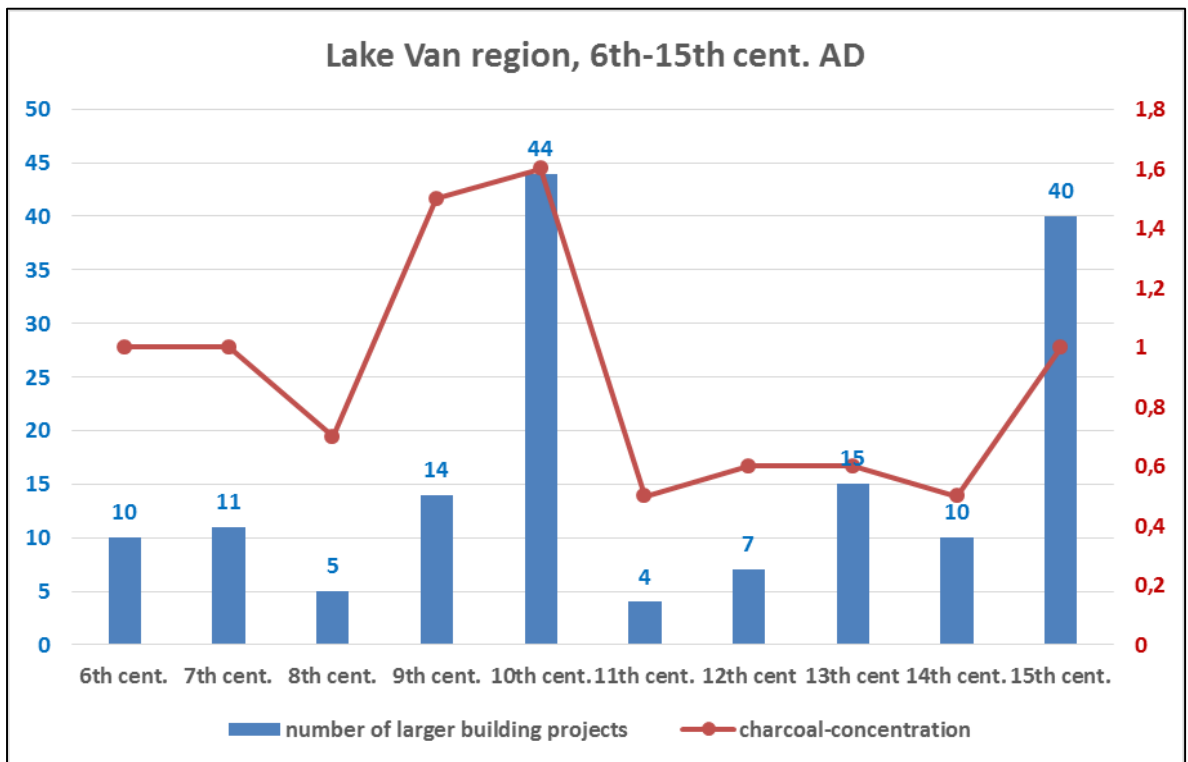

**Fig. 22:** Number of larger building projects in the Lake Van region and charcoal index for core samples from sediments in Lake Van (Turkey) as proxy for human activity, 6th–15th cent. AD (data: THIERRY, Monuments arméniens du Vaspurakan resp. WICK – LEMCKE – STURM, Evidence of Lateglacial and Holocene climatic change)

APPENDIX 6: CARBONATE ISOTOPES DATA FROM SPELEOTHEMS
IN THE SOFULAR CAVE (TURKEY)

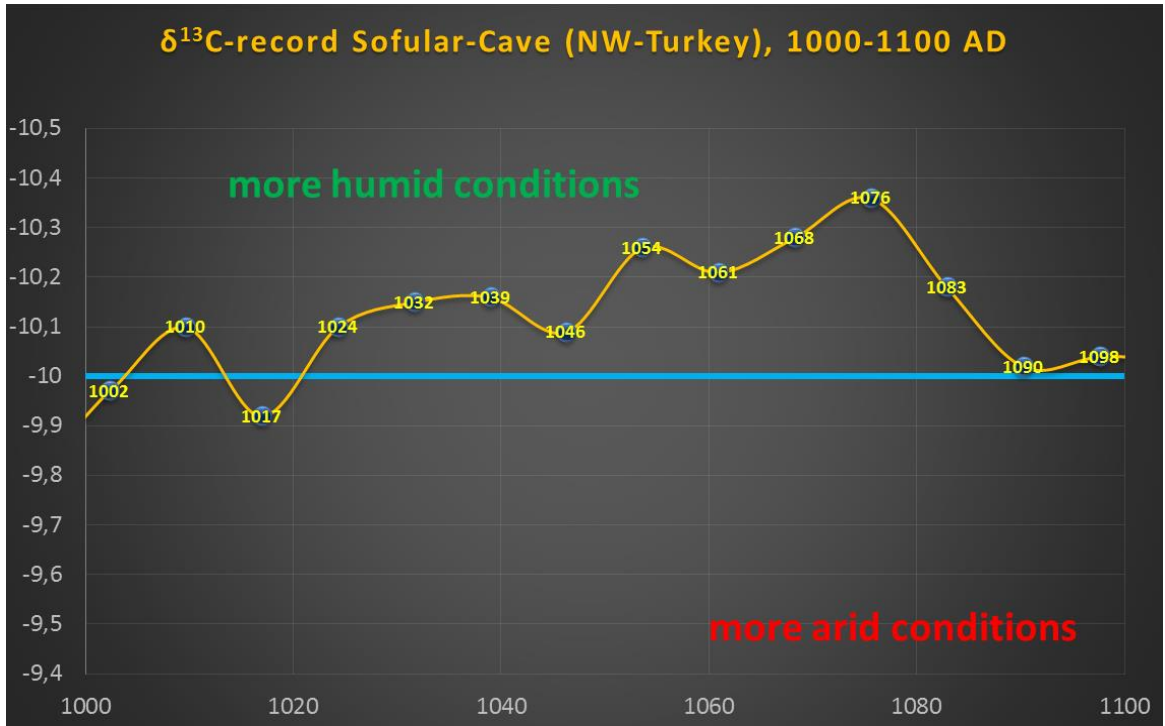

**Fig. 23:** Carbonate isotopes data from speleothems in the Sofular Cave (Turkey) as precipitation proxy, 1000–1100 AD (data: FLEITMANN *et alii*, Sofular Cave, Turkey 50KYr Stalagmite Stable Isotope Data)

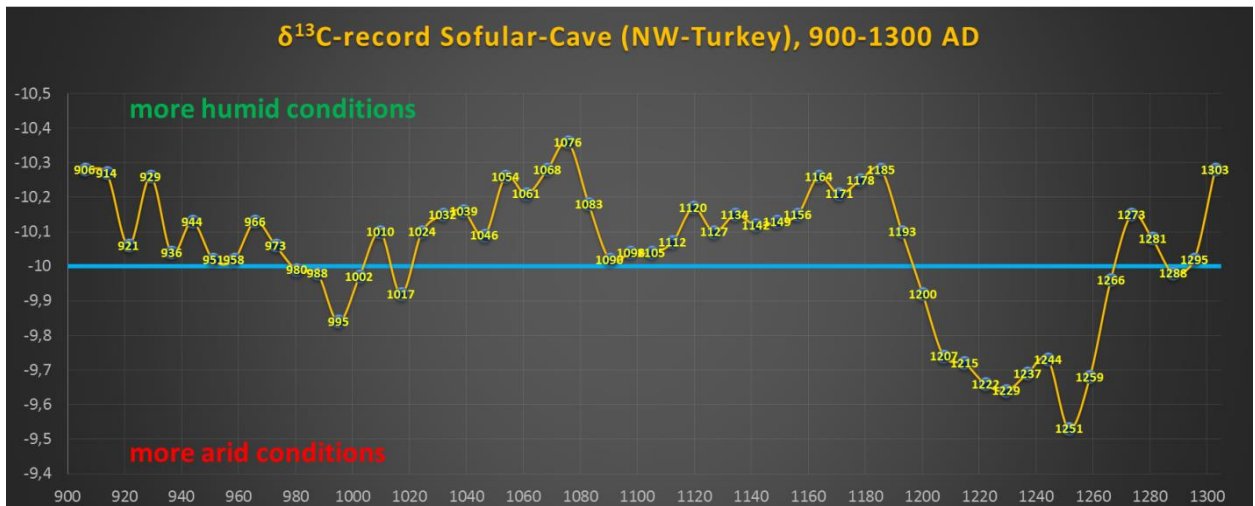

**Fig. 24:** Carbonate isotopes data from speleothems in the Sofular Cave (Turkey) as precipitation proxy, 900–1300 AD (data: FLEITMANN *et alii*, Sofular Cave, Turkey 50KYr Stalagmite Stable Isotope Data)

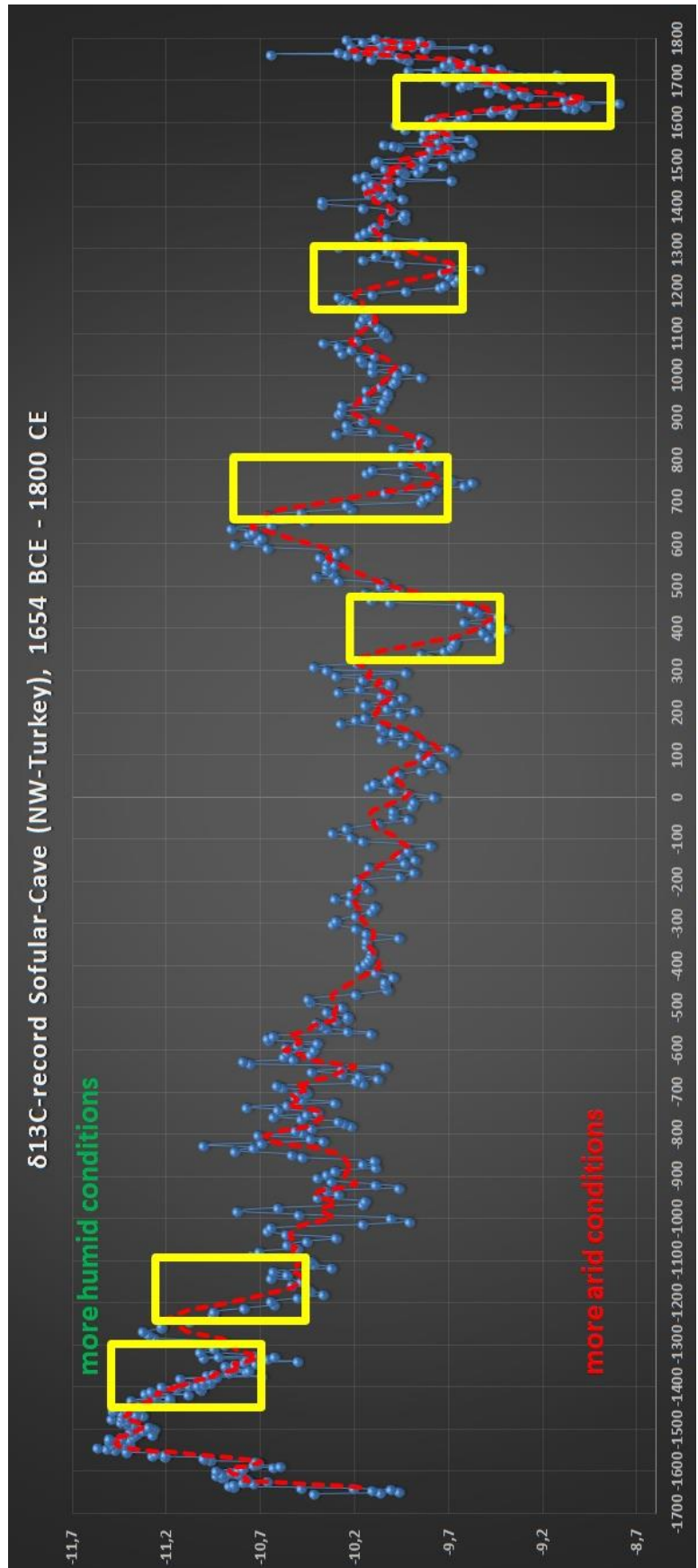

**Fig. 25:** Carbonate isotopes data from speleothems in the Sofular Cave (Turkey) as precipitation proxy, 1654 BCE–1800 CE (data: FLEITMANN *et alii*, Sofular Cave, Turkey 50KYr Stalagmite Stable Isotope Data); yellow bars mark the periods of most rapid change from more humid to more arid conditions.

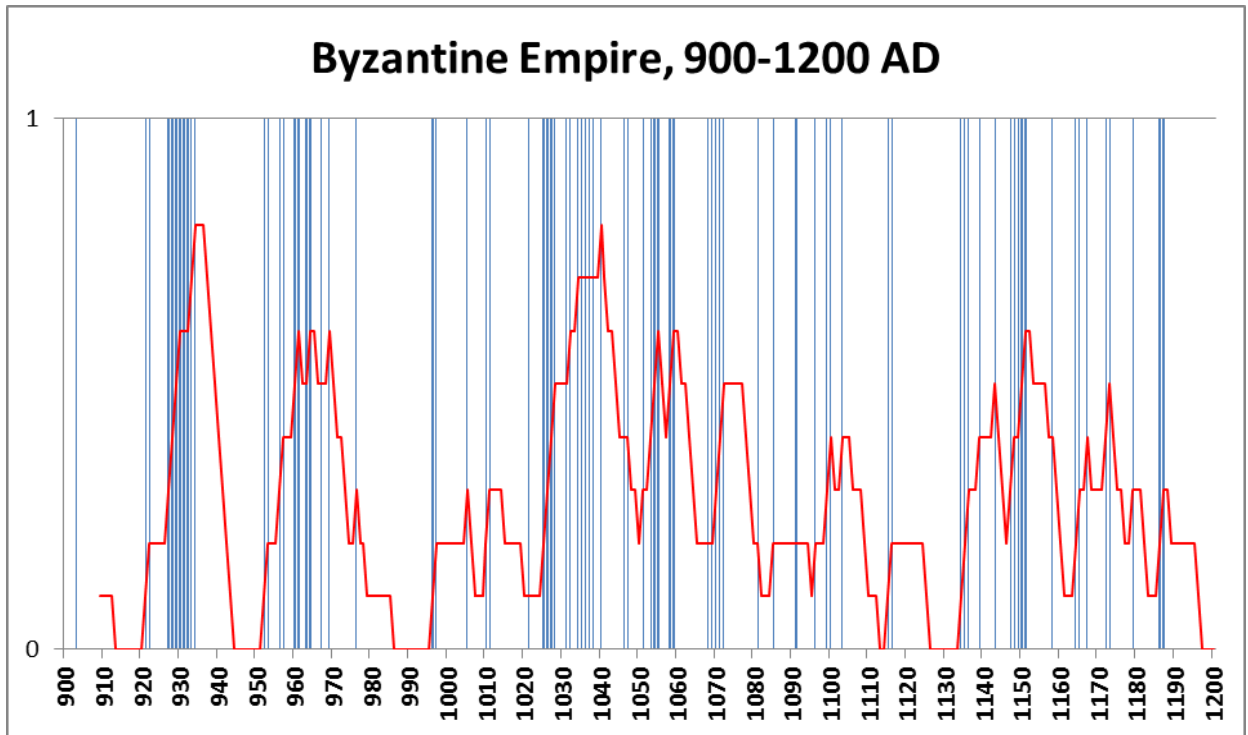

**Fig. 26:** Frequency of years with extreme weather events in the areas of the Byzantine Empire as documented in written sources, 900–1200 AD (red: 10 years moving average; data: TELELIS, Μετεωρολογικά φαινόμενα; HALDON *et alii*, The Climate and Environment of Byzantine Anatolia)

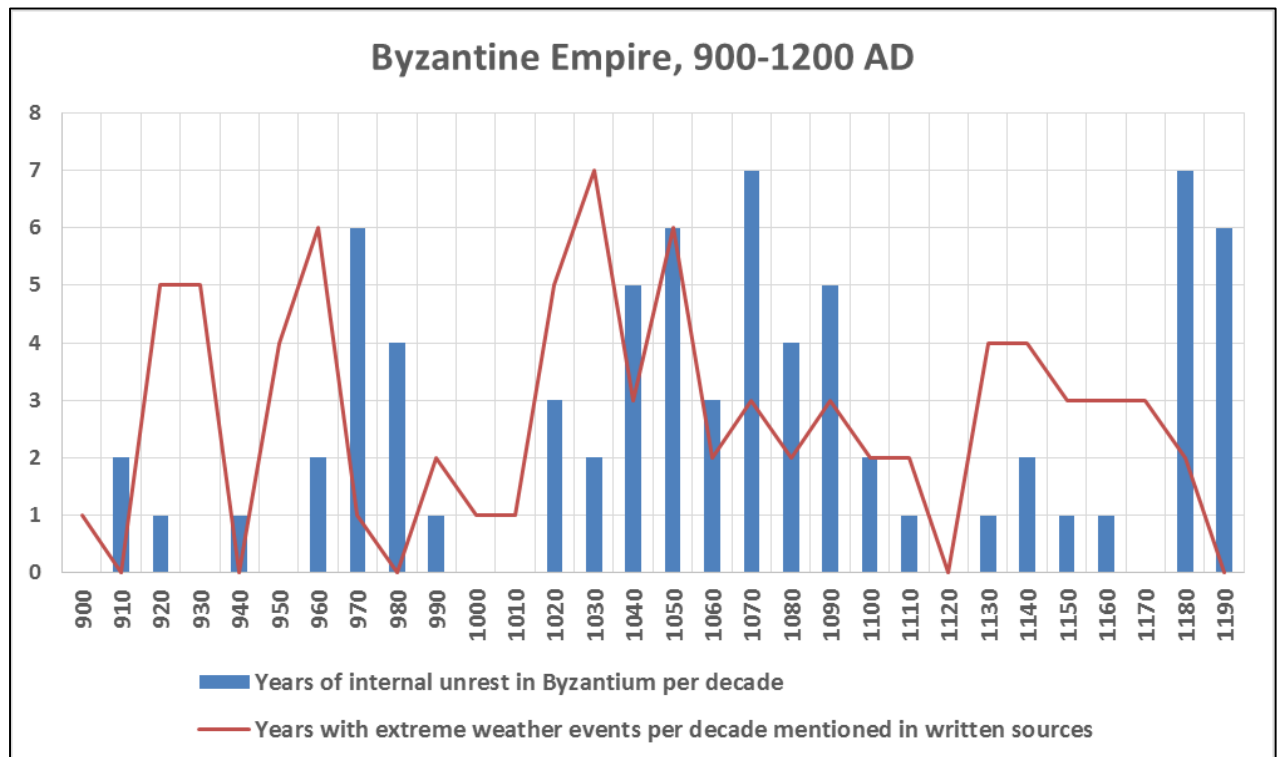

**Fig. 27:** Frequency of years of internal unrest and with extreme weather events per decade in the areas of the Byzantine Empire as documented in written sources, 900–1200 AD (data: CHEYNET, Pouvoir et contestations à Byzance; TELELIS, Μετεωρολογικά φαινόμενα; HALDON *et alii*, The Climate and Environment of Byzantine Anatolia)

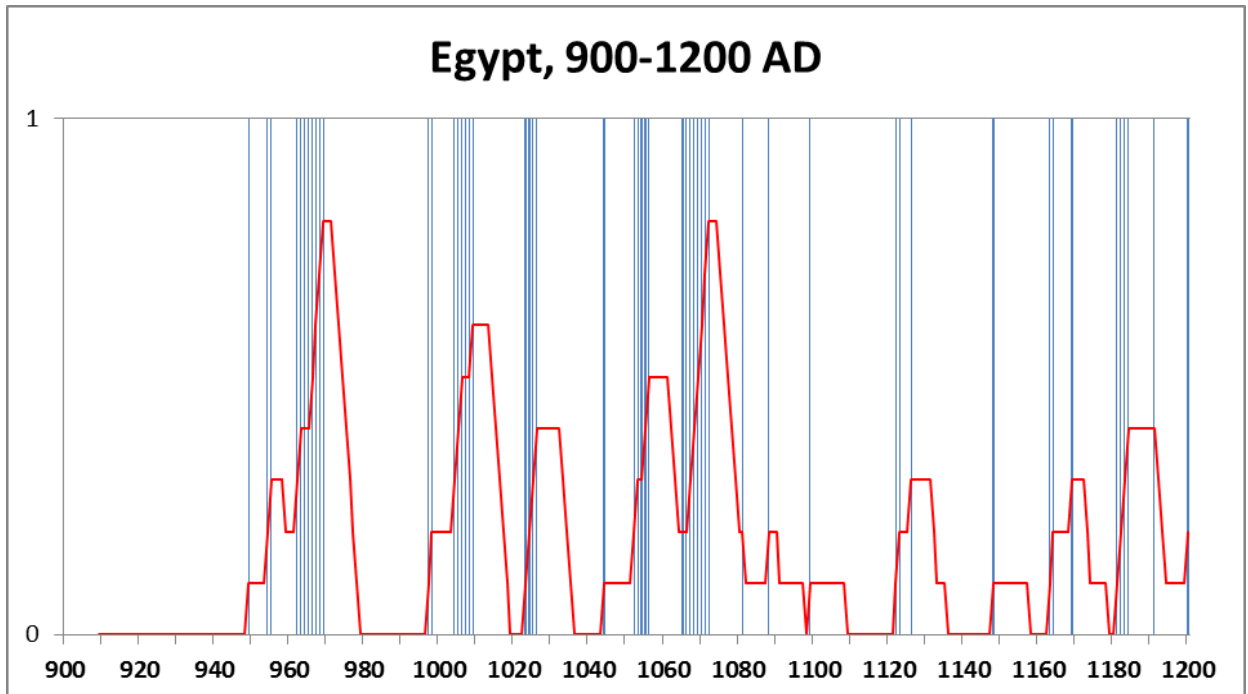

**Fig. 28:** Frequency of years with extreme (high or low) Nile floods in Egypt, 900–1200 AD (red: 10 years moving average; data: HASSAN, Extreme Nile floods and famines in Medieval Egypt; ELLENBLUM, The Collapse of the Eastern Mediterranean)

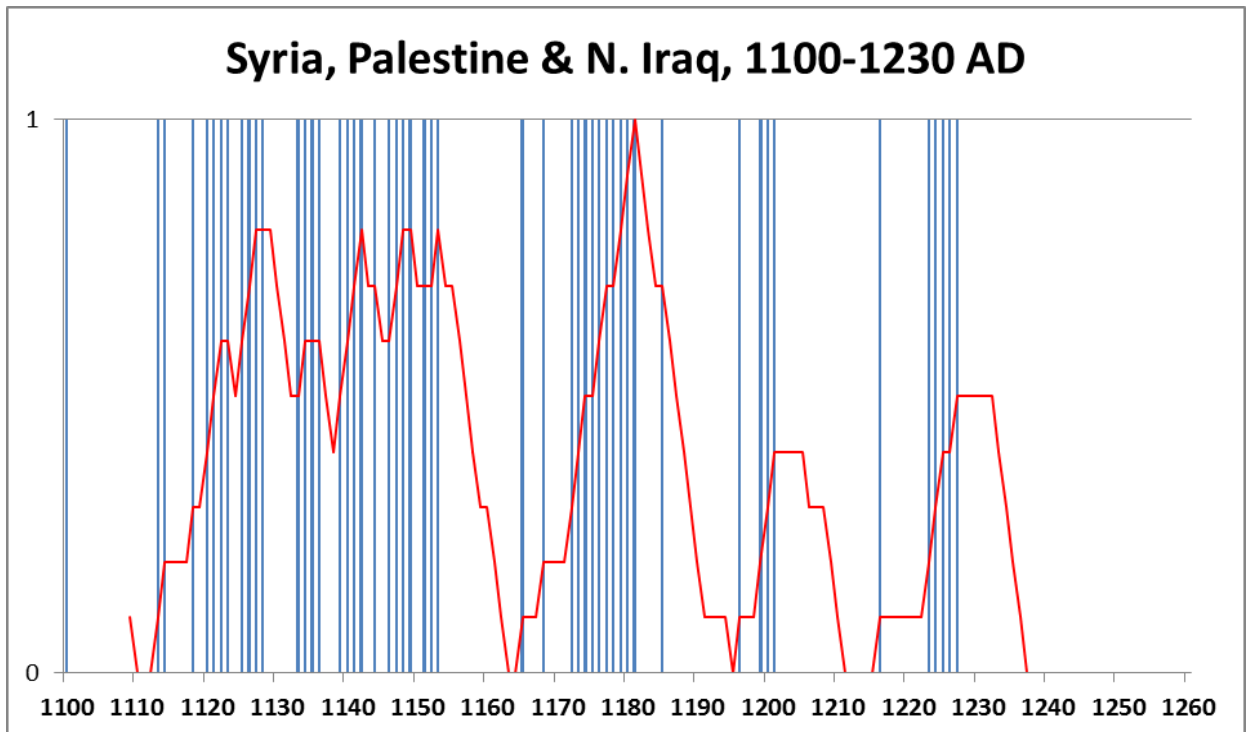

**Fig. 29:** Frequency of years with extreme weather events in the areas of Syria, Palestine and Northern Iraq as documented in written sources, 1100–1230 AD (red: 10 years moving average; data: RAPHAEL, Climate and Political Climate; WIDELL, Historical Evidence for Climate Instability)

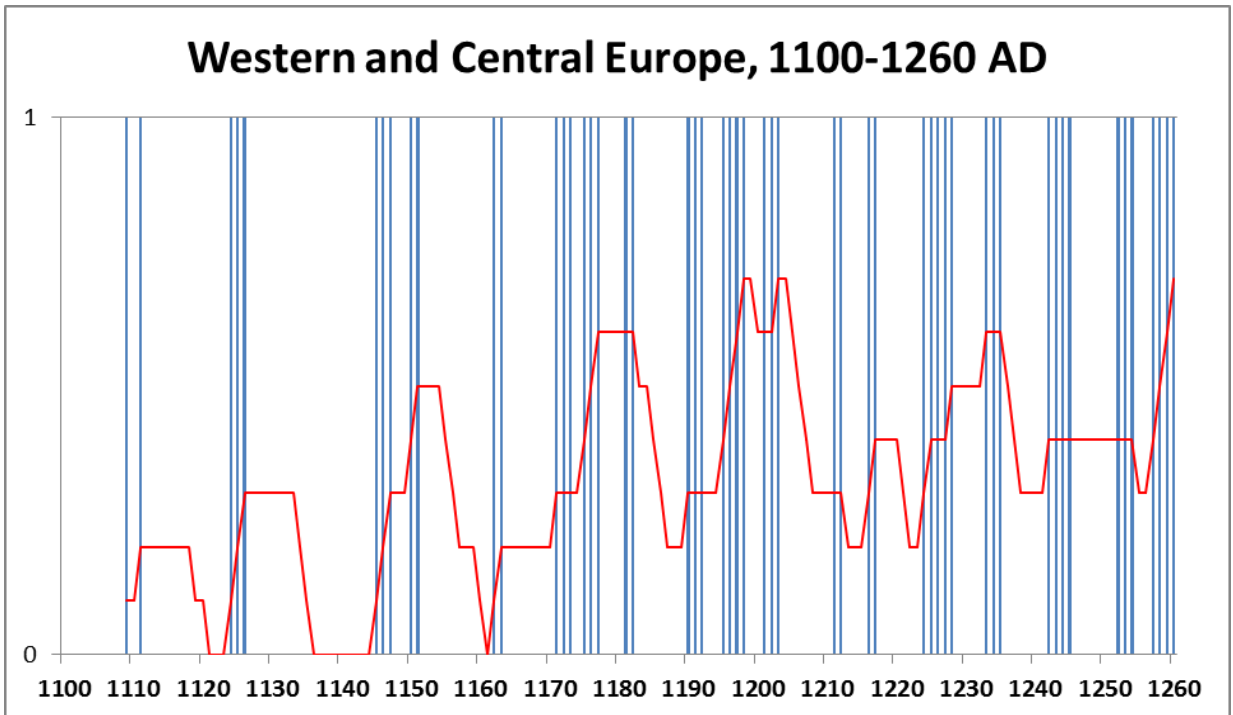

**Fig. 30:** Frequency of years with over-regional famines in Western and Central Europe as documented in written sources, 1100–1260 AD (red: 10 years moving average; data: I MONCLÚS, Famines sans frontiers en Occident avant la "conjuncture de 1300").

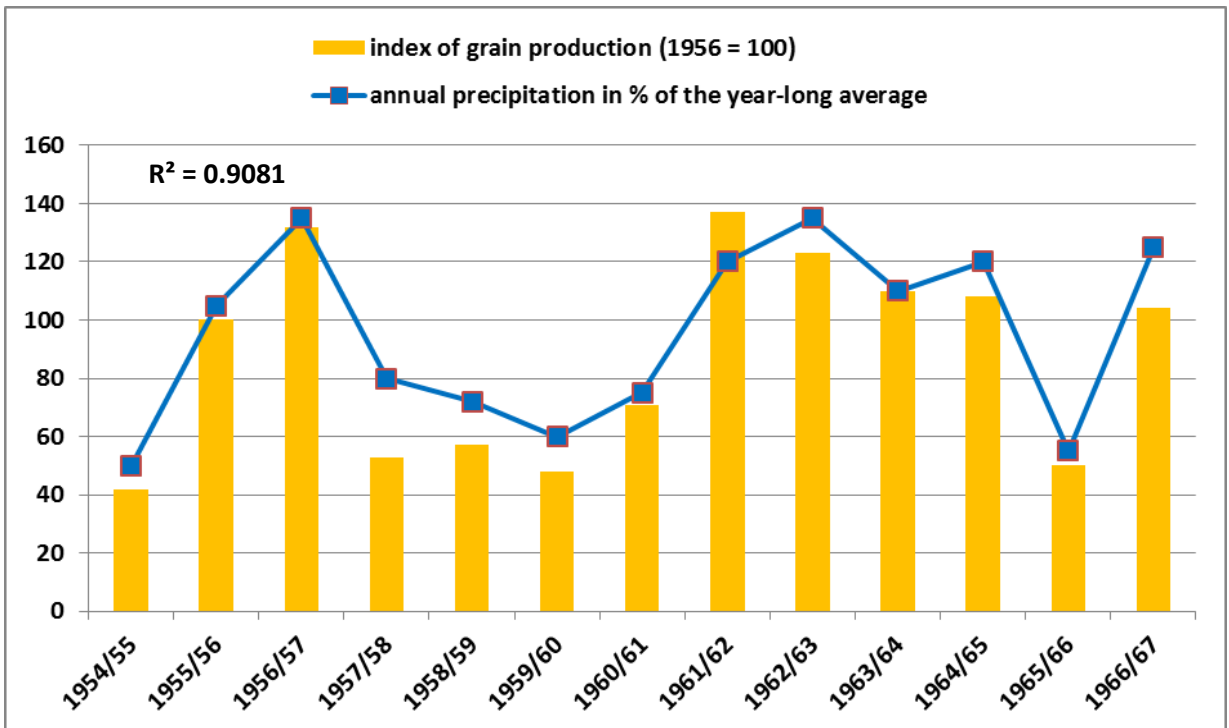

**Fig. 31:** Index of grain production (1956 = 100) and annual precipitation in percentage of the year-long average in Syria, 1954–1967 (data: E. WIRTH, Syrien. Eine geographische Landeskunde [*Wissenschaftliche Länderkunden* 4/5]. Darmstadt 1971, 13).

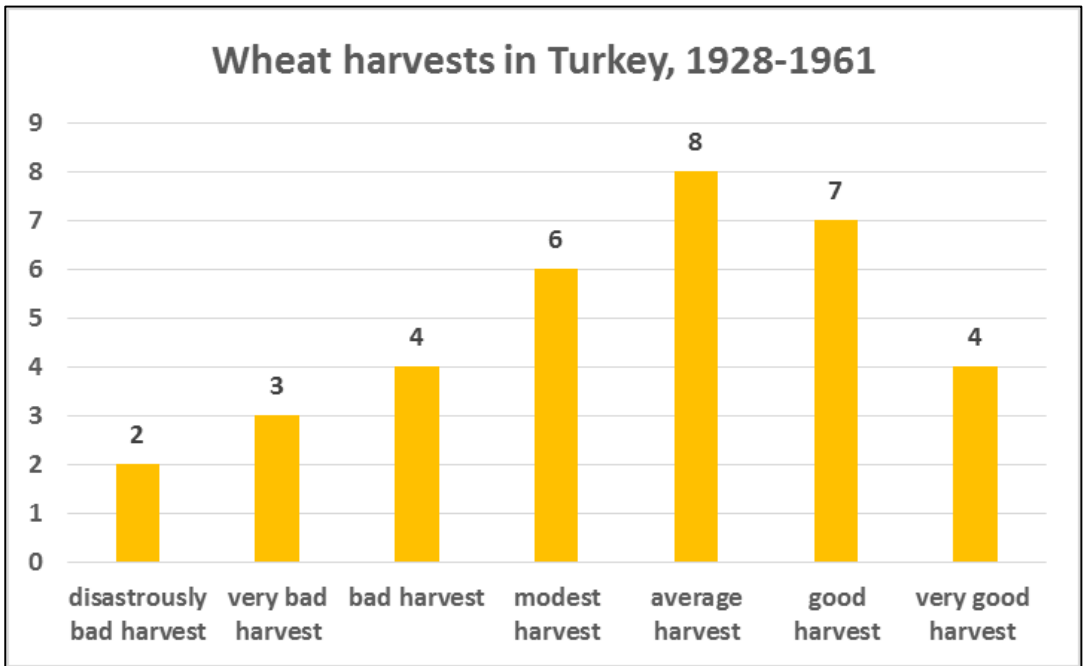

**Fig. 32:** Frequency of the quality of wheat harvests in Turkey, 1928–1961 (data: HÜTTEROTH, Türkei 126)

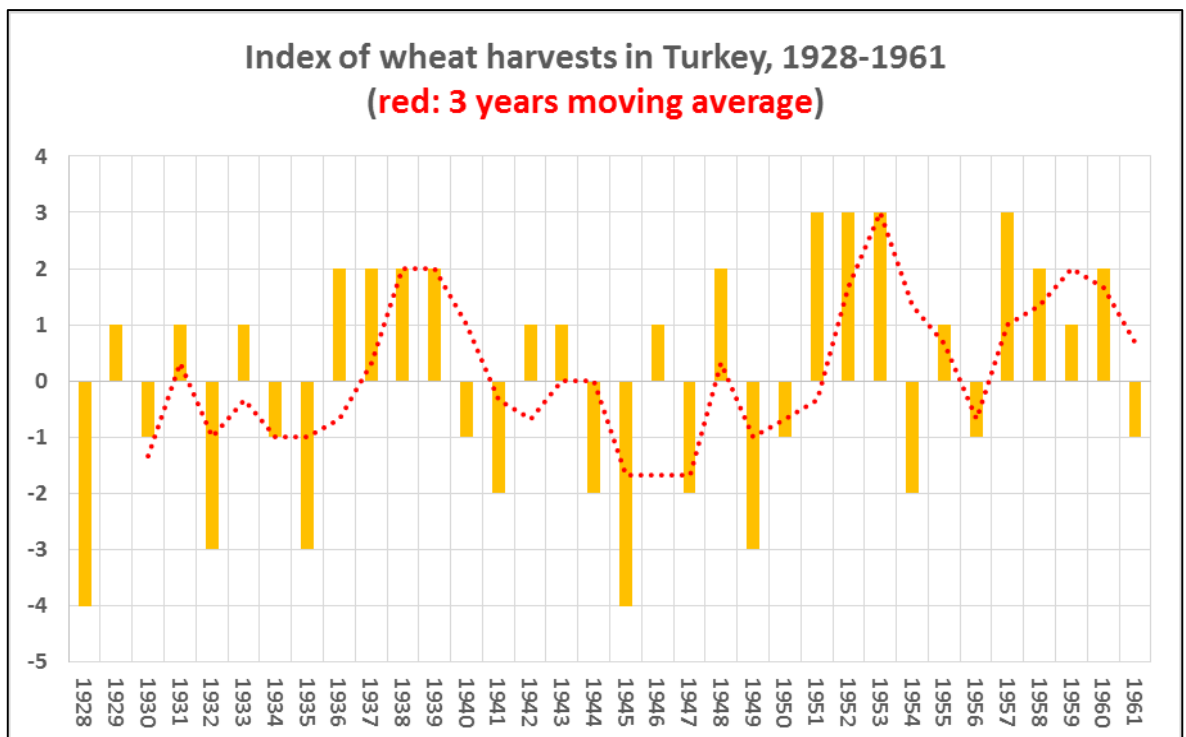

**Fig. 33:** Trajectory of the quality of wheat harvests in Turkey, 1928–1961 (data: HÜTTEROTH, Türkei 126)